\documentclass[twocolumn]{emulateapj}
\usepackage{lscape}

\def\Sec{\hbox{${}^{\prime\prime}$\llap{.}}}
\def\deg{\hbox{${}^\circ$}}
\def\min{\hbox{${}^{\prime}$}}
\def\sec{\hbox{${}^{\prime\prime}$}}
\def\lae{\mathrel{<\kern-1.0em\lower0.9ex\hbox{$\sim$}}} 
\def\gae{\mathrel{>\kern-1.0em\lower0.9ex\hbox{$\sim$}}} 
\newcommand{\be}{\begin{equation}} 
\newcommand{\ee}{\end{equation}}

\shorttitle{A Cepheid Distance to NGC 5128}
\shortauthors{Ferrarese \etal}

\begin{document}

\title{The Discovery of Cepheids and a  Distance to
NGC~5128\altaffilmark{1}}

\author{Laura Ferrarese\altaffilmark{2}, 
Jeremy R. Mould\altaffilmark{3}, 
Peter B. Stetson\altaffilmark{2}, 
John L. Tonry\altaffilmark{4}, 
John P. Blakeslee\altaffilmark{5}, \& 
Edward A. Ajhar\altaffilmark{6}}

\altaffiltext{1}{Based on observations with the NASA/ESA {\it  Hubble
Space Telescope} obtained at the Space Telescope Science Institute,
which is operated by the association  of Universities for Research in
Astronomy, Inc., under NASA contract NAS 5-26555.}
\altaffiltext{2}{Herzberg Institute of Astrophysics, National Research
Council of Canada, 5071 West Saanich Road, Victoria, BC, V8X 4M6,
Canada;  {\sf laura.ferrarese@nrc-cnrc.gc.ca,
  peter.stetson@nrc-cnrc.gc.ca}}
\altaffiltext{3}{National Optical Astronomy Observatory, P.O. Box
  26732, Tucson, AZ 85726, USA; {\sf jrm@noao.edu}}
\altaffiltext{4}{Institute for Astronomy, University of
Hawaii, 2680 Woodlawn Drive, Honolulu, HI 96822, USA; {\sf jt@ifa.hawaii.edu}}
\altaffiltext{5}{Department of Physics and Astronomy, PO Box 642814, Washington State University,
Pullman, WA 99164, USA; {\sf jblakes@wsu.edu}}
\altaffiltext{6}{Department of Natural Sciences, Mathematics, and Computer Science, St. Thomas
  University, 16401 NW 37th Avenue, Miami Gardens, FL 33054, USA; {\sf ajhar@stu.edu}}

\def\spose#1{\hbox to 0pt{#1\hss}}
\def\simlt{\mathrel{\spose{\lower 3pt\hbox{$\mathchar"218$}}
     \raise 2.0pt\hbox{$\mathchar"13C$}}}
\def\simgt{\mathrel{\spose{\lower 3pt\hbox{$\mathchar"218$}}
     \raise 2.0pt\hbox{$\mathchar"13E$}}}
\def\eg{{\rm e.g. }}
\def\ie{{\rm i.e. }}
\def\etal{{\rm et~al. }}

\begin{abstract}
We discuss a new distance to NGC~5128 (Centaurus A) based on Cepheid
variables observed with  the {\it Hubble Space Telescope.} Twelve
F555W ($V$) and six F814W ($I$) epochs of cosmic-ray-split {\it Wide
Field  Planetary Camera 2} observations were obtained.  A total of 56
{\it bona-fide} Cepheids were discovered, with periods ranging from 5
to $\sim50$ days; five of these are likely Population II Cepheids of
the W Virginis class,  associated with the bulge or halo of NGC 5128.
Based on the period and $V$ and $I-$band luminosities of  a sub-sample
of 42 classical (Pop I) Cepheids, and adopting a Large Magellanic
Cloud distance modulus and extinction of $18.50\pm 0.10$ mag and
E($B-V$)=0.10 mag, respectively, the true  reddening-corrected
distance modulus to NGC 5128 is $\mu_\circ=27.67\pm 0.12~({\rm
random})\pm 0.16~ ({\rm systematic})$ mag, corresponding to a
distance of $3.42 \pm 0.18~({\rm random})\pm 0.25~ ({\rm
systematic})$ Mpc. The random uncertainty in the distance is dominated
by the error on the assumed value for the  ratio of total to selective
absorption, $R_V$, in NGC 5128, and by the possible metallicity
dependence of the Cepheid Period-Luminosity relation at $V$ and
$I$. This represent the first determination of a Cepheid distance to
an early-type galaxy.
\end{abstract}


\section{Introduction}
\label{introduction}
The HST Key Project on the Extragalactic Distance Scale (Mould et al.\
2000a; Freedman et al.\ 2001) and the HST project on the ``Calibration
of Nearby Type Ia Supernovae'' (Sandage et al.\ 1992) have greatly
improved our knowledge of  the Hubble Constant by providing a solid
zero point for distance indicators applicable to Population I (Pop I)
stellar systems, in particular the Tully- Fisher (TF) relation (Sakai
et al.\ 2000) and Type Ia Supernovae (SNIa; Saha et al.\ 2001; Gibson
et al.\ 2000). This goal was achieved by measuring accurate Cepheid
distances to over two dozen nearby spiral galaxies, in which Cepheid,
TF and/or SNIa distances could be compared directly.  The calibration
of distance  estimators applicable to early-type galaxies cannot
benefit from such direct comparison, leading to considerable
disagreement in the Population II (Pop II) distance scale (Ferrarese
et al.\ 2000; Blakeslee et al.\ 2002; Ciardullo et al.\ 2002).
Resolving such discrepancies  is a must, not only as a means of
checking for systematic errors in the  Pop I distance scale ladder,
but also for the practical reason that early-type galaxies are found
in all types of environments, that they are more abundant than spirals
in rich clusters, and that they enjoy a privileged position in a
cluster's gravitational center.

Pop II distance indicators include several classes of variable stars
(most notably RR-Lyrae, e.g. Dolphin et al.\ 2001 and references
therein, and Mira Variables, e.g. Feast et al.\ 1989), the Tip of the
Red Giant Branch (TRGB; e.g. Mould \& Kristian 1986), the red clump
(Paczynski \& Stanek 1998; Udalski 2000), the Surface Brightness
Fluctuation method (SBF; Tonry \& Schneider 1988), the Planetary
Nebula Luminosity Function (PNLF; Jacoby et al.\ 1999), the Globular
Cluster Luminosity Function (GCLF; e.g. Richtler 2003), the Globular
Cluster Sizes (GCS, Jord\'an et al.\ 2005), and  the Fundamental Plane
(FP; Dressler et al.\ 1987; Djorgovski \& Davis 1987).
PNLF and TRGB can be used with equal success across the entire
Hubble sequence and provide, at least in local galaxies, a direct link
between the Pop II and the Pop I (Cepheids in particular) distance
scales. Although all methods can be used  to study local deviations
from the Hubble flow (see M\'endez et al.\ 2002 for an application
using the TRGB), SBF and the FP have the additional advantage of being
the only estimators, besides TF and SNIa, to extend far enough into
the unperturbed Hubble flow to allow a direct measurement of the
Hubble constant. SBF in particular  benefits from an intrinsic scatter
which only SNIa can rival (Tonry et al.\ 2001), making it an order of
magnitude less susceptible than TF to Malmquist biases and systematic
error.

As a  step in securing a Cepheid-based calibration of Pop II distance
indicators, we present in this paper the first determination of a
Cepheid distance to an early-type galaxy. Our  target is the giant
elliptical NGC~5128. The galaxy is the dominant member of the
Centaurus group, a powerful radio galaxy (Centaurus A), the host of
the very well sampled (but highly reddened) SNIa SN1986G (Phillips et
al.\ 1987), and an archetypal merger  (Malin, Quinn \& Graham
1983). Its Pop I component, including the Cepheids, is supported by
the reservoir of gas (Oosterloo et al.\ 2002) deposited in the merger;
star formation is triggered partly by the merger and partly by
interaction between the  gas cloud and the radio jet (Mould et al.\
2000b).  Distances to NGC 5128 have been measured using the TRGB in
the halo (Soria et al.\ 1996; Harris et al.\ 2002; Rejkuba et al.\
2005), SBF (Tonry et al.\ 2001), PNLF (Hui et al.\ 1995), and GCLF
(Harris et al.\ 1988). In future contributions, we will discuss a
Cepheid distance to NGC 4647, a spiral galaxy interacting with the
giant elliptical NGC 4649, strategically located in the Virgo cluster
core. A revision of the zero points for Pop II distance indicators,
and SBF in particular, will also be presented.

The paper is organized as follows. In \S \ref{photometry} we present
the multi-epoch {\it Hubble Space Telescope} (HST) {\it Wide Field
Planetary Camera 2} (WFPC2) observations on which our distance
determination is based.  The photometric analysis is discussed in \S
\ref{dophot}, while the identification of Cepheids and their derived
properties are discussed in \S \ref{cepheids}.  The derived distance
to NGC~5128 is presented in \S \ref{distance}. The result is discussed
and contrasted with previous distance determinations for NGC~5128 and
the Centaurus group in \S \ref{discussion}, where conclusions can also
be found.

\section{Observations and Initial Data Reduction}
\label{photometry}

NGC 5128 was observed between 2001 July 8 and 2001 August 20 as part
of program GO-9043. The 44-day length of the observing window was
imposed  by the need to observe the field at a single orientation,
i.e., without changing the  roll angle of the telescope. Within this
window, WFPC2 F555W (similar to Johnson $V$) and F814W ($\sim$
Kron-Cousins $I$) images were taken at 12 and six epochs respectively,
spaced in such a way to optimize the discovery of variable stars with
period between $\sim 10$ and $50$ days, as described in Freedman et
al.\ (1994).  Each image was split in a pair of consecutive exposures,
referred to as a ``cosmic ray-split pair'', of duration between 1100
and 1400 s.  Details of the observing procedure, including the
calendar and Heliocentric Julian Date (HJD) of the observations, the
HST archive rootnames, the filter employed, and exposure times are
listed in Table \ref{tbl:observations}.

\begin{figure}
\centering\plotone{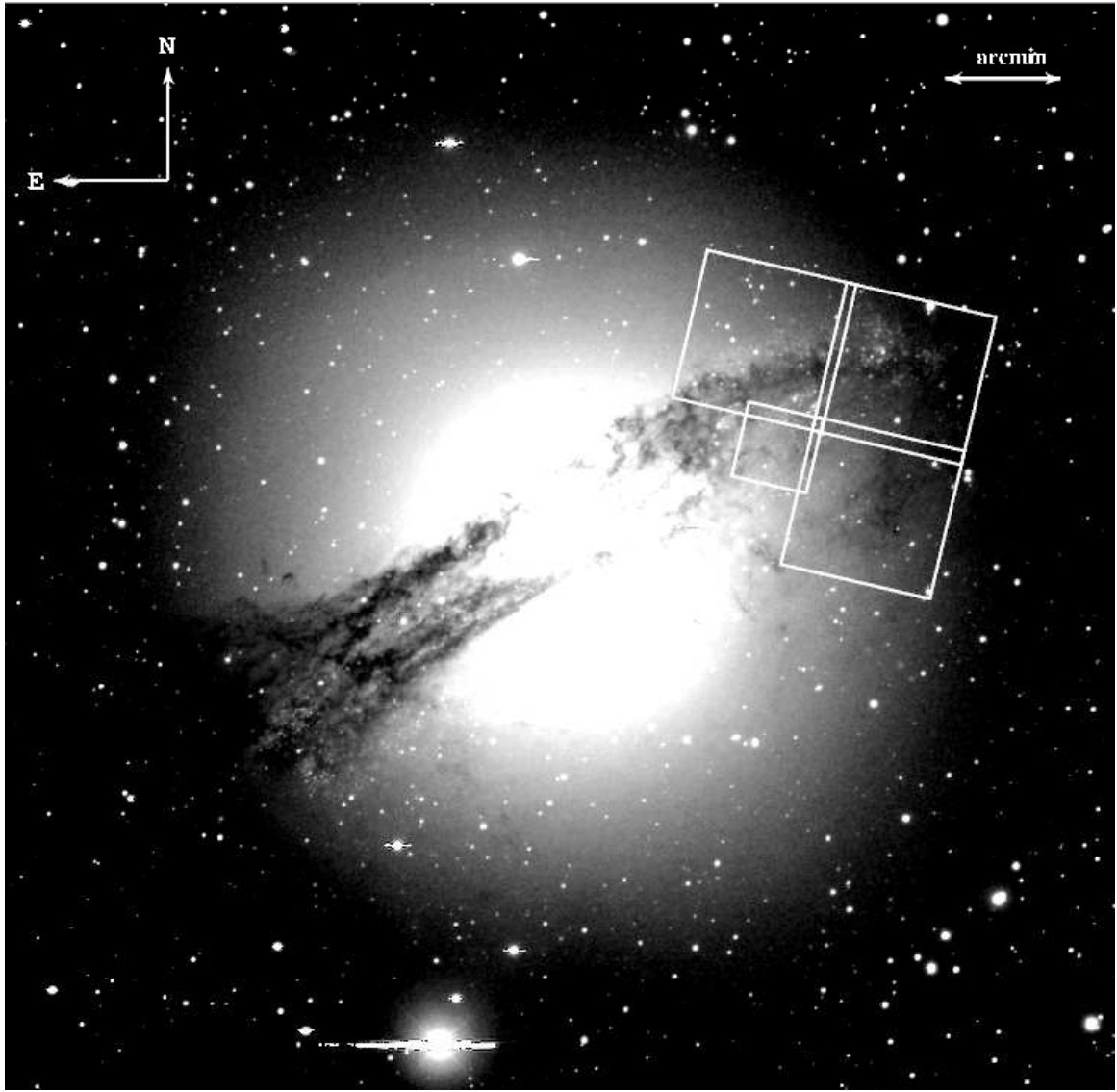}
\caption{A ground based image of NGC 5128, taken with the Prime Focus
  8K Mosaic CCD Imager at the CTIO 4.0 meter Victor M. Blanco
  telescope (courtesy of Holland Ford and Eric Peng). The footprint of
  the HST/WFPC2 detectors is shown, with the PC having the smaller
  field of view, and the WF2, WF3 and WF4 detectors arranged in a
  counter-clockwise fashion  following the PC. The image is shown with
  a logarithmic gray-scale.
\label{ground}}
\end{figure}

The WFPC2 field of view (FOV) is divided between three {\it Wide Field
Cameras} (WFC), each with pixel scale of 0\Sec1, and one {\it
Planetary Camera} (PC), with pixel scale of 0\Sec046. The FOV of each
WFC is 80\sec$\times$80\sec, while the FOV of the PC is
36\Sec8$\times$36\Sec8.  Figure~\ref{ground} shows the position of the
WFPC2 FOV  superimposed to a ground-based image of NGC 5128. The
region targeted purposely avoided the high background associated with
the galaxy core, and included the western edge of the equatorial dust
lane, where the abundance of recent star formation should prove
conducive to the detection of Cepheids.  To minimize the impact of
dead pixels and other CCD imperfections, at each epoch the telescope
pointing was offset slightly (up to  2\Sec6 relative to the first
epoch) following a spiral hexagonal pattern.  All observations were
obtained with the telescope pointing in fine lock, giving a nominal
pointing stability of 5 mas (RMS) or better over 60-second intervals.

Basic data reduction (bias removal, dark subtraction and flat
fielding) was performed by the IRAF task CALWP2 when downloading the
images from the STScI data archive, selecting the ``On the Fly
Reprocessing'' option. Details of the calibration procedures executed
by CALWP2 can be found in the WFPC2 Data Handbook (Heyer et al.\ 2005).

\section{Photometric Analysis}
\label{dophot}

The photometric reduction was performed independently using a variant
of DoPHOT (Schechter, Mateo \& Saha 1993; Saha et al.\ 1994) and DAOPHOT
II/ALLFRAME (Stetson 1994).  A detailed description of the
photometric reduction process can be found in Ferrarese et al.\ (1996)
and Stetson et al.\ (1998). In what follows, we only provide a brief
summary of the key steps.

\subsection{DoPHOT Procedure}

Unlike ALLFRAME, DoPHOT works most effectively on images which have
been cleaned of cosmic ray (CR) hits.  CR-split pairs were therefore
combined using a sigma detection algorithm which takes into account
the problems of Point Spread Function (PSF) under-sampling (Saha et
al.\ 1996).  The DoPHOT reduction of the resulting  12 F555W and six
F814W images was performed using as first guess a parametric
representation of the PSF which best represents  point sources in the
frames. To aid source detection, the photometric reduction of each
epoch adopts, as a starting point,   an input star-list generated by
running DoPHOT on deep, CR-free F555W and F814W images, respectively
created by combining all 24 F555W and 12 F814W  frames listed in
Table~\ref{tbl:observations}.  In combining the frames, single images
are first brought into alignment with the first epoch frames, using
shifts calculated by matching positions for a list of objects in
common; rotations between frames were found to be negligible in all
cases.  To insure photometric stability, in aligning the images
sub-pixel interpolation was not performed and images were not
corrected for geometric distortion. Although this unavoidably leads to
a degradation in the PSF of the deep frames relative to that of the
single exposures, this has no detrimental effects on the photometry.

DoPHOT  magnitudes are defined as $m_{Do} = -2.5$log(DN/$t$)$+30.0$,
where DN is the number of counts within an aperture of radius equal to
5 pixels, and $t$ is the exposure time. The raw output from DoPHOT is
converted to DoPHOT magnitudes by applying an aperture correction,
calculated by performing aperture photometry on  bright isolated
stars. Telescope jitter or focus changes between subsequent epochs can
result in a slight time dependence of the aperture corrections,
however, rather than deriving independent aperture corrections for
each epoch/filter, we instead derived aperture corrections from the
deep images (using between 35 and 160 stars depending on the
filter/chip), applied them to each epochs' photometry, matched the
deep frame photometry to that of the first epoch (both of which have
been aperture corrected), and finally matched each epoch's
aperture-corrected photometry to the aperture-corrected deep
photometry. With this procedure, the first epoch is used to tie the
photometry, the aperture corrections are derived from the high
signal-to-noise deep images, and small changes in aperture corrections
are accounted for by (effectively) matching the photometry of several
thousand stars in each epoch to that of the first. The deep frames are
not used directly to tie the photometry because of possible
complications introduced by the degradation in the PSF as a result of
the procedure used to combine the single exposures. Variations in
aperture corrections between epochs are found to be 0.03 mag or less.

\subsection{ALLFRAME}
\label{allframe}

ALLFRAME is run directly on the single exposures, without combining
CR-split  pairs or rejecting cosmic rays. An input star list to
ALLFRAME was generated from median-averaged  cosmic-ray-free F555W and
F814W deep frames.  Iterative application of DAOPHOT and ALLSTAR led
to a single final master star list, including stars detected in either
(or both) filters. This star list was input to ALLFRAME, and used to
extract profile-fitting stellar photometry from the 36 individual
frames.  The adopted PSFs were derived from public domain HST WFPC2
observations of the globular clusters Pal~4 and NGC~2419.

Aperture photometry was performed on a set of bright isolated
stars. The program DAOGROW was then employed to generate growth curves
out to 0\Sec5,  allowing an aperture correction to be derived for each
chip and filter,  using the approach of Holtzman et al.\ (1995).
As seen for the DoPHOT photometry, variations in aperture 
corrections for the same filter/chip and different epochs are found to
be, on average, 0.03 mag or less.

\subsection{Absolute Photometric Calibration}
\label{zeropt}

The continuing degradation in the WFPC2 performance demands a frequent
updating of the coefficients needed to convert DoPHOT and ALLFRAME
magnitudes to standard-system magnitudes, $m_{F555W}$ and $m_{F814W}$,
and ultimately, through the adoption of a color term, to Johnson $V$
and Kron-Cousin $I$. As part of this project, we obtained  F555W and
F814W WFPC2 single-epoch images of the Draco dwarf spheroidal,  and
three globular clusters, NGC 5262, NGC 6341 and NGC 2419. Comparison
of ground-based and HST ALLFRAME photometry for secondary standard
stars in these fields has led to the updated ALLFRAME photometric
calibration equations listed in  Table~\ref{tbl:zpall}. The
transformations assume a color correction as in Holtzman et al.\
(1995), so that only the zero points are derived when matching the HST
to the ground-based photometry; a full description of the procedure
will be provided in an upcoming paper (Stetson, in preparation). These
transformations were derived using 40 s F555W and F814W exposures, and
are therefore equivalent to the ``short-exposure zero points"
presented in Hill et al.\ (1998). Comparing the Hill et al.\ to the
new  coefficients implies a 10 to 15\% degradation in the WFPC2
performance (slightly worse in F555W than in F814W) between May 1994
(when the fields analyzed by Hill et al.\ were  observed) and July
2001. Comparing the new  coefficients to those presented by Dolphin
(2000, for a ``cold'' operating temperature and gain = 7), which were
derived from data taken between January  1994 and May 2000, implies a
5 to 10\% degradation in the WFPC2 performance (again slightly worse
in F555W than in F814W).

There is some debate in the literature as to whether the photometric
zero points differ depending on whether short ($\lesssim$ 60 s) or
long exposure ($\gtrsim$1300 s) observations are used in their
derivation. Hill et al., as well as previous investigations, find
that  zero points obtained from the long exposure observations are
systematically fainter, on average by 0.05 mag, with no significant
chip/filter dependence, while Dolphin (2000) find no evidence of such
effect. In addition to the 40 s exposures mentioned above, exposures
between 230 and 300 s were obtained for all standard fields (Draco,
NGC 5262, NGC 6341 and NGC 2419) at the same time as the NGC 5128
frames; comparing these longer exposures (analyzed with ALLFRAME as
described in the previous paragraph) to the 40 s exposures yields zero
points which are on average 0.025 to 0.030 mag fainter than those
reported in Table~\ref{tbl:zpall} (again with no significant
dependence on chip/filter), in the same sense noted by Hill et al.,
i.e., the detector quantum efficiency appears to be reduced in the
shorter exposure. In light of this,  we assume a 0.05 mag difference,
in both $V$ and $I$, between the zero points listed in
Table~\ref{tbl:zpall} and those applicable to the longer NGC 5128
exposures. {\it In the rest of this paper, all tables and Figures use
the short exposure zero points given in Table~\ref{tbl:zpall},
however, we will add 0.05 mag to the final distance modulus to account
for the fact that the NGC 5128 frames used   longer exposures time
than the standard fields from which the photometric calibration is
derived.} We will include the uncertainty in the photometric zero
points in the discussion of the error budget in \S 5.3.

Analogous transformation equations applicable to the DoPHOT photometry
were obtained by comparing the DoPHOT and ALLFRAME photometry for a
set of bright, isolated secondary standard stars (listed in the
Appendix) in the NGC 5128 field.  Figure~\ref{comp1} shows a
chip-by-chip comparison of the $V-$ and $I-$band ALLFRAME and DoPHOT
photometry for these secondary standards, the former adopting the
transformations listed in Table~\ref{tbl:zpall}, and the latter using
the short exposure zero points from Hill et al.\ (1998). Mean
differences between the DoPHOT and ALLFRAME photometry are listed in
Table~\ref{tbl:comp} for each chip/filter combination.  The slightly
larger offsets noted for the PC (and more so in $I$ than in $V$) could
be due to several causes, including a slight degradation in the PSF
(the size of the aperture, in arcsec,  used in measuring DoPHOT
magnitudes is smaller in the PC than in the WF chips) between 1994 and
2001, and/or a change in the Charge-Transfer Inefficiency (CTE) which is
known to affect WFPC2 data (Stetson 1998).

Bringing the DoPHOT photometry into agreement with the ALLFRAME photometry by
correcting for the observed offsets leads to the DoPHOT photometric
transformations listed in Table~\ref{tbl:zpdo}. Figure~\ref{comp2}
shows the match between the DoPHOT and ALLFRAME photometry for the
same secondary standard stars shown in Figure~\ref{comp1} once these
final zero points are adopted. It is worth noticing that a small scale
error appears to be present in the $I-$band comparison. Regression
fits, accounting for errors in both variables, to the data shown in
Figure~\ref{comp2} give:

$$
 \; V_{\rm DoPHOT} - V_{\rm ALLFRAME} =  - (0.0023 \pm  0.0080) -
 $$
 \be
- (0.006 \pm 0.012) \times (V_{\rm DoPHOT} - 22.72)
\ee

$$
 I_{\rm DoPHOT} - I_{\rm ALLFRAME} =  - (0.0047 \pm  0.0030) - 
$$
\be
- (0.0202 \pm 0.0049) \times (I_{\rm DoPHOT} - 21.43)
\ee

Scale errors of this magnitude are not uncommon when comparing DoPHOT
and ALLFRAME photometry (e.g. Hill et al. 1998), and are likely to
reflect the inherent limitations of PSF fitting procedures in crowded
fields. In the worst case scenario, i.e. assuming that the error
resides entirely with the DoPHOT photometry (which will be used in \S
5 to measure the distance to NGC 5128), the  $I-$band 0.02 mag/mag
scale error would lead to overestimate the magnitudes of the faintest
Cepheids (mean $I-$band magnitude $\sim 23.5$) by $\sim 0.04$ mag
($I_{\rm DoPHOT}$ and $I_{\rm ALLFRAME}$ agree at the mean magnitudes
of the brightest Cepheids).  Correcting all DoPHOT photometry
according to equation (2) would increase all distances (derived using
the same procedure discussed in \S 5) by 0.04 mag. However, given the
uncertain nature of this scale error, and its limited impact compared
to all other sources of error which will be discussed in \S 5.3,  we
feel justified in neglecting its effects in the remainder of this
paper.

Finally, we note that the zero points listed in Tables \ref{tbl:zpall}
and \ref{tbl:zpdo} do not include a correction for CTE. The electron
loss due to CTE is dependent on position within the chip, as well as
on the brightness of the star and underlying sky background.  The zero
points used in this paper are translated to the chips' centers;
therefore we expect that (uncorrected) CTE losses will add scatter (at
the few hundredths of a magnitude level) in the photometry, but not
produce any systematic biases in the photometry of a sample of stars
distributed uniformly throughout the chip.  We note that the
combination of $V = 20$ mag standards and 80 second exposures with $V
= 25$ mag Cepheids and 1100 second exposures leads to equal CTE
corrections for both Cepheids and standards, according to the
coefficients of Stetson (1998). Although we assume that this remained
true for our 2001 data, we expect to test this assumption in ongoing
calibration work.

\section{Cepheid Identification}
\label{cepheids}

The search for variable stars was conducted using three independent
methods: 1) a variant of Stellingwerf's (1978) phase dispersion
minimization routine, as described in Saha \& Hoessel (1990), applied
to the DoPHOT photometry; 2) the template light curve fitting
algorithm TRIAL (Stetson 1996) applied to the ALLFRAME photometry; and
3) a photometry-independent  image subtraction method, following Alard
\& Lupton (1998). The first two methods produce both a list of likely
variables and an estimate of the variability period, while the third
method was implemented to simply flag objects undergoing
luminosity variations, with no attempt to construct light curves.

Every star that satisfies the variability criteria imposed by the
previous methods was visually inspected by  blinking the images
against each other. This step is necessary since random errors in the
photometry,  uncorrected CR hits, or crowding, do occasionally result
in spurious light curves which look credible based on the photometry
alone. Of the 126 variables identified based on one or more of the
methods above, a total of 56 were judged to be {\it bona-fide} Cepheid
variables based on the character of the light curve, the
visually-established reliability of the light variations, and some
visual assurance that the star does not appear to be a blend or is
found in an unusually crowded region. The remaining 70 stars are
likely genuine variables, although either their photometry is
considered suspect because of crowding, or  their light curves do not
appear Cepheid-like. 

For the remainder of this paper, we will adopt the DoPHOT photometry
for all variables. The periods quoted are therefore determined based
on method 1) above; a comparison of these periods and the TRIAL
periods best fitting the ALLFRAME light curves for the 36 {\it
bona-fide} Cepheids in common shows excellent agreement, with  a mean
difference of $(-0.5 \pm 5.4)$\%.

For each variable star, we calculated both intensity-averaged
magnitudes, defined as

\be
m = -2.5 \log_{10}\sum_{i=1}^n {1 \over n} 10^{-0.4\times m_i},
\ee

\noindent and phase-weighted magnitudes (Saha and Hoessel 1990),
defined as

\be
m = -2.5\log_{10}\sum_{i=1}^n 0.5(\phi_{i+1} - \phi_{i-1})10^{-0.4
\times m_i},
\ee

\noindent where the phase $\phi$ varies between 0 and 1.  In the above
equations, $n$ is the total number of observations, and $m_i$ and
$\phi_i$ are the magnitude and phase of the $i-$th observation in
order of increasing phase. Phase-weighted magnitudes are more robust
than intensity-averaged magnitudes for variables with non-uniform
temporal sampling (Saha and Hoessel, 1990).

$X$ and $Y$ positions within the chip (in the coordinate frame of the
first epoch), right ascension and declination, period,
intensity-averaged and phase-weighted $V$ and $I$ magnitudes for all
Cepheids are listed in Table~\ref{tbl:ceph}, while parameters for the
suspected variable stars are listed in the Appendix, where single
epoch F555W and F814W magnitudes of all Cepheids are also
reported. Note that for Cepheids which vary on timescales close to or
longer than the 44-day contiguous observing window, often only a lower
limit on the period can be placed.

The spatial distribution of the Cepheids and suspected variable stars
in each chip is shown in Figure~\ref{PC}. Detailed
finding charts are given in Figure~\ref{FC1} for the
Cepheids, and in the Appendix for the suspected variable stars.

DoPHOT light curves for each Cepheid, phased to the appropriate
period, are presented in Figure~\ref{lc1} -- the F555W and F814W
magnitudes are shown by solid and open dots, respectively.  Figure
\ref{cmd} shows the location of the 56 Cepheids (using phase-weighted
$V$ and $I$ magnitudes) listed in Table \ref{tbl:ceph} in a $V,V-I$
color-magnitude diagram  (CMD), constructed using the DoPHOT
photometry of the deep F555W and F814W frames; mean photometric
errors, calculated in bins of width equal to 0.5 mag, are shown in
Figure \ref{cmderr} as a function of magnitude and color. Note that in
Figure \ref{cmd}, most Cepheids  lie outside the instability strip,
which is shown in the Figure assuming a $V-$band distance modulus of
28.85 mag and reddening $E(V-I)=0.55$ mag (see \S 5). As will be
discussed in \S 5, this is a consequence of the large and highly
position dependent reddening of the NGC 5128 field.

\section{The Distance to NGC~5128}
\label{distance}

\subsection{Standard Procedure}

The DoPHOT  $V-$ and $I-$band Period-Luminosity (PL) relations for
NGC~5128 are shown in the upper panels of Figure \ref{pl1}, using
phase-weighted magnitudes. Cepheids identified by a symbol with an
arrow are those for which only a lower limit on the period could be
placed.  In the lower two panels of Figure \ref{pl1}, the $<V>$ and
$<I>$ phase-weighted magnitudes are combined to give the Wesenheit
function $W = V - (V-I)\times A(V)/E(V-I)$ (Madore 1982). By
construction, $W$ is unaffected by line-of-sight reddening, therefore
the extrinsic scatter in the $W-\log P$ plane should be smaller than
in either the  $<I>-\log P$ or (especially) $<V>-\log P$ planes. The
lower right panel of Figure \ref{pl1} shows the same data plotted in
the lower left panel, but Cepheids are color-coded according to the
chip to which they belong.

The apparent $V-$ and $I-$band distance moduli (\ie $\mu_{\rm V}$ and
$\mu_{\rm I}$) to NGC~5128 are derived relative to those of the Large
Magellanic Cloud. We adopt the LMC PL relations of Udalski et al.\
(1999), scaled to a true LMC distance modulus of
$\mu_{\circ,LMC}=18.50\pm 0.10$ mag and reddening
E($B-V$)$_{LMC}=0.10$ mag as in Freedman et al.\ (2001):

$$
\; M_V  = -2.760[\pm 0.03](\log P - 1.0) - 4.218[\pm 0.02] 
$$
\be
(\sigma_V = \pm 0.16)
\ee

\noindent and

$$
\;M_I = -2.962[\pm 0.02](\log P - 1.0) - 4.904[\pm 0.01] 
$$
\be
(\sigma_I = \pm 0.11). 
\ee

Once a ratio of total to selective absorption $R_V = A(V)/E(B-V)$ and
a reddening law are adopted, a reddening corrected true distance
modulus $\mu_0$ can be calculated {\it for each} of the NGC 5128
Cepheids. In the absence of a metallicity dependence of the Cepheid PL
relation, and for a reddening law as derived by Cardelli, Clayton \&
Mathis (1989) with $R_V = 3.3$\footnotemark, the true distance modulus
for each individual Cepheid becomes:

\footnotetext{$R_V = 3.3$ was adopted in all Key Project papers
  (e.g. Freedman et al.\ 2001), and is in agreement with the recent
  estimate by McCall (2004). Under these conditions, $A$(F814W)$/A(V)=0.599$,
  $A$(F555W)$/A(V)=0.999$, and $A$(F555W)$/[A($F555W$)-A($F814W$)]=2.497$}

\begin{eqnarray} 
\begin{array}{l}
\mu_0 = \mu_V - A(V) = \mu_I - A(I) = \\ 
\\
\phantom{\mu_0 }= \mu_V - A(V)/E(V-I)(\mu_V - \mu_I) = \\
\\
\phantom{\mu_0 }= 2.497\mu_I - 1.497\mu_V = \\
\\
\phantom{\mu_0 }= W + 3.27[\pm 0.01](\log P-1) + 5.94[\pm 0.01] \\
\\
\phantom{\mu_0 = } (\sigma_0=\pm0.08). 
\end{array}
\end{eqnarray} 

The true distance modulus to NGC 5128 is then taken as the average of
the distance moduli to the individual Cepheids. In the process,
Cepheids which deviate by more than 0.48 mag (two times the width of
the instability strip) from the main ridge-line of the $W - \log P$
plot are excluded from the fit. These Cepheids are shown as solid dots
surrounded by larger circles in Figure~\ref{pl1}; note that the
rejection is not performed based on the $V$ and $I$ PL relations, for
which large deviations could be due to differential reddening. In
addition, Cepheids with period less than 8 days, which could be
overtone pulsators (open circles in Figure~\ref{pl1}), and Cepheids
for which only a lower limit on the period could be determined, were
also excluded from the fits.  The procedure described above is
equivalent to 1) finding the  sample-mean $V$ and $I$ magnitudes at
$\log P = 1$ by fitting $V$ and $I$ PL relations to the NGC 5128 data,
with slopes fixed to those of equations (5) and (6); 2) subtracting
them from the zero points of equations (5) and (6) respectively, to
find sample-mean distance moduli $\mu_V$ and $\mu_I$; and 3) applying
the middle expression of equation (7) to derive $\mu_0$.

Table~\ref{tbl:dist} lists the distances obtained by means of the
method described above. Using Cepheids from all chips, with period $>
8$ days and excluding outliers (case 1 in the Table), the resulting
apparent distance moduli are $\mu_{V}=28.85\pm 0.10$ mag and
$\mu_{I}=28.30\pm 0.06$  mag (short exposure zero points). The
corresponding true distance modulus is $\mu_0 = 27.48 \pm 0.05$ mag,
giving a linear distance of $3.4 \pm 0.1$ Mpc, where the quoted
uncertainties are 1$\sigma$ fitting uncertainties.  These fits are
shown  by  the solid lines in Figure~\ref{pl1}, while the dotted lines
represent 2$\sigma$ deviations from the mean of the LMC relations (\ie
0.32 mag in $V$, and 0.22 mag in $I$, equations 5 and 6).

Previous papers have noted the importance of testing the sample for
incompleteness bias (\eg Lanoix, Paturel, \& Garnier 1999) by imposing
a long pass filter on the period distribution.  Table~\ref{tbl:dist}
lists distance moduli obtained by applying lower period cutoffs of 15,
20, 25 and 30 days (cases 2 to 5). In all cases, the distances agree
well within the quoted uncertainties. The same is true if
distances are derived independently for the four chips
(cases 6 to 9). Finally, if all the suspected variable stars listed
in the Appendix with period $> 8$ days are added to the Cepheid
sample, the distance remains virtually unchanged (case 10).

Using all Cepheids with period $> 8$ days gives a total (foreground
plus intrinsic) reddening to NGC 5128 of $E(V-I)=0.55\pm0.05$  mag.
The uncertainty in this estimate is not simply the quadrature sum of
the uncertainties in the apparent moduli (Stetson et al.\ 1998):
because $A(V) \propto \mu_V - \mu_I$, the intrinsic scatter in the PL
relation does not propagate into the reddening uncertainty.  The
DIRBE/IRAS dust maps of Schlegel, Finkbeiner \& Davis (1998) show a
foreground reddening component of $E(B-V)=0.115$ or $E(V-I)=0.152$ mag
along the sight-line to NGC 5128, implying an average $A(V) = 1.0$ mag
of internal extinction for the NGC 5128 Cepheids.

\subsection{The Effect of Reddening on the Distance to NGC 5128}

One complication in the analysis described in \S 5.1 is that it relies
on an assumed value for $R_V$ which, in the case discussed above, was
set to 3.3 for both NGC 5128 and the LMC. For a different $R_V$, the
coefficients in equation (7) will change, and the distance will be
affected. Furthermore, as demonstrated in Ferrarese et al.\ (1996), if
$R_V$ differs in the LMC and NGC 5128, $\mu_0$ becomes mildly
dependent on the absolute absorption $A(V)$ to the LMC:

$$
\; \mu_0 = \mu_V - \left[{A(V)\over E(V-I)}\right]^{N5128}(\mu_V -
\mu_I) + 
$$
\be
+ 0.582~ \left[{E(B-V)\over
E(V-I)}\right]^{N5128}\left[{{R_V^{N5128}} \over
{R_V^{LMC}}}-1\right]~A(V)^{LMC}\\ \ee

In their study of the wavelength of maximum polarization of SN1986G,
Hough et al.\ (1987)  found $R_V = 2.4 \pm 0.13$ for the dust affected
regions of NGC 5128, significantly  different from the standard value
of 3.3 adopted in \S5.1. The generalized reddening law of Cardelli,
Clayton \& Mathis (1989) applied to equation (8) then yields the
following expression for the true distance modulus, to be compared
with equation (7):

\be
\mu_0 = 2.143\mu_I - 1.143\mu_V  - 0.047
\ee 

In the last two columns of Table~\ref{tbl:dist}, distance moduli are
calculated assuming $R_V=3.3$ for the LMC and $R_V=2.4$  for NGC 5128.
The distance modulus increases by 0.14 mag relative to the case in
which $R_V=3.3$ for both galaxies, from $\mu_0 = 27.48 \pm 0.05$ to
$\mu_0 = 27.62 \pm 0.04$ for the complete sample (case 1).

The issue is explored further in Figure \ref{rvfits}, where dereddened
distance moduli are shown as a function of the value of $R_V$ assumed
for NGC 5128 (in all cases, $R_V = 3.3$ for the LMC). The upper panel
of the Figure shows the standard deviation around the best-fit line in
the $P-W$ plot as a function of $R_V$, while the lower panel shows the
corresponding variation in $\mu_0$. The scatter in the $P-W$ relation
is minimized around $R_V \sim 1.8$; at this value $\mu_0 = 27.76 \pm
0.04$, corresponding to a linear distance of $3.56 \pm 0.07$ Mpc. We
refrain from being guided by considerations of the scatter in the
$P-W$ relation in choosing $R_V$; further insight into the nature of
the reddening in NGC 5128 will require infrared photometry following
Macri et al.\ (2001).

\subsection{Error Estimates and a Final Distance}

In view of the discussion above, and given the existence of an
independent estimate of $R_V$ in NGC 5128, we adopt as our final
distance to NGC 5128 that obtained for $R_V^{NGC 5128} = 2.4$ and
$R_V^{LMC} = 3.3$. In what follows, however, we will take a
conservative approach in quoting the uncertainty in the final
distance, specifically to account for possible differences between the
reddening to SN1986G and that of our Cepheid sample.

The errors listed in \S 5.1 and \S 5.2 reflect internal errors alone,
arising from  scatter in the NGC~5128 PL relations.  A more complete
assessment of the  associated uncertainty, incorporating other
currently identified random and systematic  errors, is presented in
Table \ref{tbl:error}. Uncertainties due to  metallicity, LMC distance
modulus ($\pm 0.10$ mag, Madore \& Freedman 1991; Westerlund 1996;
Freedman et al.\ 2001), reddening and photometric calibration all
contribute  to the NGC~5128 distance modulus error budget. Of these,
we identify as ``systematic" those sources of error that affect
equally all   Cepheid distances derived using the same instrumental
setup and calibration of the LMC PL relations as those used in this
paper.

Errors related to reddening estimates in NGC 5128 contribute 0.08 mag
to the (random) error budget. This is calculated following equation
(8), assuming $R_V^{NGC5128} = 2.4 \pm 0.3$ and $R_V^{LMC}=3.3$. The
error on $R_V^{NGC5128}$ assumes that the difference between $R_V=2.4$
and 3.3 represents a $3\sigma$ uncertainty on the true value of
$R_V$. The $1\sigma$ uncertainty on $A_V^{LMC}$ is assumed to be 0.05
mag.

The remaining random uncertainty in Table \ref{tbl:error} which should
be  noted here is that due to a possible metallicity dependence of the
Cepheid PL relation at $V$ and $I$.  Sakai et al.\ (2004)  find a
metallicity dependence of the form  d$\mu_\circ$/d[O/H] =
$-0.24\pm0.05$ mag/dex. This is in agreement with Groenewegen et
al. (2004), although we note that  Romaniello et al.\ (2005)  find an
[Fe/H] dependence of approximately the same  size, but opposite
sign. If the NGC~5128 Cepheids  differ substantially in metal
abundance from those of the LMC  Cepheids which calibrate the PL
relation, a significant systematic error could  be present in the
derived distance. Since the source of the neutral hydrogen  gas for
the Pop I in NGC~5128 is a galaxy like the LMC, one might expect
parity in metal abundance. The H~I masses of NGC 5128 and the LMC are
7.2 $\times$  10$^8$ M$_\odot$ and 3.1 $\times$ 10$^8$ M$_\odot$
respectively (van Gorkom  et al.\ 1990; Luks \& Rohlfs 1992). On the
other hand, Sutherland, Bicknell and Dopita (1994) were able to fit
spectra of the jet-excited knots in Cen A  with solar abundances. In
these circumstances, we are unable to correct the NGC 5128 distance
for the LMC/NGC 5128 metallicity difference, however we can account
for it in our error estimate. Adopting  [O/H]$=-3.5$ for the LMC
Cepheids (Kennicutt et al. 1998) and assuming solar metallicity for
NGC 5128, the metallicity trend of Sakai et al.\ (2004) leads to a
0.35 $\times$ 0.24 = 0.08 mag uncertainty in the distance modulus,
which we include in our final error budget.

In light of the uncertainties listed in Table~\ref{tbl:error}, {\it our
final quoted Cepheid-based true distance modulus to NGC 5128 is
(adding 0.05 mag to bring the photometry on the long exposure zero
points, see \S 3.3) $\mu_\circ=27.67\pm 0.12\,({\rm random})\pm 0.16\,
({\rm systematic})$ mag, with a reddening of $E(V-I)=0.55\pm0.05$
(internal+foreground).  The corresponding distance is $3.42 \pm
0.18\,({\rm random})\pm 0.25\,({\rm systematic})$ Mpc.}

\section{Discussion and Conclusions}
\label{discussion}

The most extensive study of the structure of the Centaurus A group has
been published by Karachentsev et al.\ (2002). Based on HST/WFPC2 TRGB
distances to 17 dwarf galaxies, the authors conclude that the group is
composed of two spatially distinct substructures, with NGC 5128 and
NGC 5236 (M 83) as dominant members. The mean TRGB distances to the
two substructures are found to be 3.63 $\pm$ 0.07 Mpc ($\mu_0$ = 27.80
$\pm$ 0.04 mag), and 4.57 $\pm$ 0.05 Mpc ($\mu_0$ = 28.30 $\pm$ 0.02
mag), respectively; the latter agrees with the Cepheid distance
modulus ($\mu_0=28.25 \pm 0.15$ mag) measured for NGC 5236 itself by
Thim et al.\ (2003).  A Cepheid distance exists for an additional
member of the Cen A group, NGC 5253, host of the Type Ia SN1972E,
although its value is controversial ($27.61 \pm 0.11 \pm 0.16$ mag
according to Gibson et al.\ 2000, but $28.08 \pm 0.2$ mag according to
Saha et al.\ 1995).

A comprehensive summary of published distances to NGC 5128 is given by
Rejkuba (2004). Determinations have been made using Mira variable
stars, the TRGB, PNLF and SBF; although NGC 5128 was host to the Type
Ia SN1986G, unfortunately the large reddening towards this supernova
(Phillips et al.\ 1999; $E(B-V) = 0.50$) makes a distance from the
standard candle estimator unreliable\footnotemark.  Using $K-$band
data, Rejkuba (2004) derived $\mu_0 = 27.96 \pm 0.11$ mag based on the
PL relation for Mira variables (calibrated using a distance modulus to
the LMC of 18.50 mag, as in this paper), and $\mu_0 = 27.87 \pm 0.16$
mag based on the TRGB.  The latter has been established as a solid
standard candle for stars with [Fe/H] $< -0.7$ (Mould \& Kristian
1986; Lee et al.\ 1993; Madore \& Freedman 1995; Sakai et al.\ 1996);
the $I-$band TRGB magnitude for NGC 5128 halo stars has been placed,
based on HST data, at  $m_{TRGB} = 23.88 \pm 0.1$ mag (Soria et al.\
1996;  Harris et al.\ 1999, corrected assuming a foreground extinction
$A_I = 0.22$ mag, Schlegel, Finkbeiner \& Davis 1998) and $m_{TRGB} =
23.83 \pm 0.05$ mag (Rejkuba et al. 2005 -- the value of 24.05 mag
quoted in the original paper is uncorrected for extinction). These
TRGB magnitudes can be converted to distances using the empirical
calibration of the $I-$band tip absolute magnitude, $M^{TRGB}$, as a
function of  the stars' metallicity, [Fe/H], derived by Bellazzini et
al.\ (2001) from Galactic globular clusters, with zero point anchored
by observations of the globular cluster $\omega$Cen. Harris et
al. (1999) consider only stars with [Fe/H] $< -0.7$ in measuring
$m_{TRGB}$. Soria et al.\ (1996) use stars for which $(V-I) > 1.5$ mag
(corresponding to [Fe/H] $ > -1.68$ according to Bellazzini et al.\
2001), while the $m_{TRGB}$ determination of Rejkuba et al. (2005) is
based on stars with $(V-I) < 1.8$ ([Fe/H] $ < -1.28$). Based on the
$(V-I)$ range spanned by the TRGB in the published color magnitude
diagrams, we estimate a mean [Fe/H] of $-1.2 \pm 0.5$ for both the
Soria et al. (1996) and Harris et al.\ (1999) data, and $-1.4 \pm 0.1$
for the Rejkuba et al. (2005) data. The Bellazzini et al.\ calibration
then leads to TRGB  distances to NGC 5128 of $\mu_0 = 27.91 \pm 0.44$
mag (Soria et al.\ 1996; Harris et al. 1999), and $\mu_0 = 27.89 \pm
0.16$ mag (Rejkuba et al. 2005), where the errors include a 0.12 mag
uncertainty in the zero point of the Bellazzini et al.\ calibration.

A PNLF distance of $\mu_0^{PNLF} = 27.73^{+0.03}_{-0.05}$ was
published by Hui et al.\ (1995) based on a sample of 224 planetary
nebulae. The authors use an empirical calibration of the PNLF based on
the Cepheid distance to M31; using the updated calibration by
Ferrarese et al.\ (2000), based on Cepheid distances to six galaxies
(including M31) with PNLF measurements, gives $\mu_0^{PNLF} =
27.83^{+0.03}_{-0.05}$.  The SBF distance to NGC 5128 is, at the
moment, less firmly constrained.  In their SBF survey, Tonry et al.\
(2001) report $\mu_0^{SBF} = 28.12 \pm 0.14$, but two factors have
surfaced since which require this distance to be revised. First, the
zero point of the SBF distance scale (based on Cepheid distances to
the bulges of six spirals with SBF measurements) has been updated
(Blakeslee et al.\ 2002). Adopting the new zero point gives a revised
distance modulus $\mu_0^{SBF} = 28.06 \pm 0.14$ mag.  Second, new
wide-field mosaic observations (Peng et al.\ 2004;  E.~Peng, private
communication) indicate that the original $(V{-}I)$ color measured by
Tonry et al.\ (2001) for the bulge of NGC 5128 was significantly
underestimated, by approximately 0.07 mag.  The revised color is more
reasonable for an early-type galaxy of this luminosity.  The reason
for the unexpectedly large error was the inadequate area available for
sky estimation on the small format CCDs used to observe this large
galaxy in the ground-based SBF survey.  Because the slope of the
linear SBF--$(V{-}I)$ calibration is 4.5, the Tonry et al.\ SBF
distance must be revised further by $4.5\times0.07 = 0.32$ mag.  With
these corrections, the SBF distance modulus becomes $\mu_\circ =
27.74$ mag.  The SBF calibration will be revisited in a future
contribution comparing SBF and Cepheid distances to NGC4647/NGC4649.
    
To summarize, distance estimates for NGC 5128 are confined in the very
narrow range $\mu_0 = 27.74 \pm 0.14$ (SBF) to $\mu_0 = 27.96 \pm
0.11$ mag (Mira variables, Rejkuba 2004).  Our Cepheid distance
modulus, $\mu_\circ=27.67\pm 0.12\,({\rm random})\pm 0.16\, ({\rm
systematic})$ is consistent with both estimates within the quoted
uncertainties.

\footnotetext{Note that the GCLF distance published by Harris et
  al.\ (1988) is to be considered tentative since based on a luminosity
  function which did not reach the turnover; this distance was in fact not
  used in subsequent papers by some of the authors (e.g. Harris et
  al.\ 1999).}

We conclude by returning briefly to the five {\it bona-fide} Cepheids
which appeared under-luminous in the $P-W$ plot of Figure
\ref{pl1}. These objects are identified by the large squares in Figure
\ref{cmd} and correspond to Cepheids  C43, C50, C52, C54, and C56 in
Table \ref{tbl:ceph}. Given the presence of an older stellar
population associated with the bulge and halo of NGC 5128, it is not
unreasonable to expect the detection of Population II variables, such
as RR-Lyrae stars, Anomalous Cepheids, RV Tauri and Pop II Cepheids
(themselves divided in BL Herculis for periods between 1 and 8 days,
and W Virginis for longer periods) within our WFPC2 field.  RR-Lyrae
are fainter (by 4 to 6 $V-$band magnitudes) and have shorter
variability timescales ($< 0.8$ day) than any of the variables
discussed here; Anomalous Cepheids, which have been detected in a
number of dwarf spheroidals (Nemec et al.\ 1994; Wallerstein 2002) are
also not known for periods longer than 1.6 days. RV Tauri stars have
characteristic light curves showing alternating deep and shallow
minima (see Alcock et al.\ 1998 for several exceptionally well sampled
light curves in the LMC), which are not characteristic of any of the
variables in NGC 5128.

The remaining, and indeed likely possibility is that the five
under-luminous variables in NGC 5128 are Population II Cepheids of the
W Virginis class. These are known to have regular light curves, which
compared to those of classical (Pop I) Cepheids, often display a flat
maximum, a symmetric minimum and/or a hump during the decline phase
(Schmidt et al.\ 2004; Alcock et al.\ 1998). Although such differences
cannot be easily appreciated in sparsely sampled light curves such as
the ones available for NGC 5128,   all of the under-luminous variables
in NGC 5128, with the exception of C50, show a broad maximum (C43,
C52), a symmetric light curve (C54) and/or a slow ascent (C56): in
other words, their light curves are consistent with those seen for Pop
II Cepheids. In a CMD, Pop II Cepheids occupy the instability strip,
and indeed Figure \ref{cmd} shows that the five under-luminous
variables in NGC 5128 do not appear to have abnormal colors (the fact
that four of them are among the bluest of the Cepheids discovered in
NGC 5128 could simply reflect the fact that they suffer from lower
internal extinction, as expected for objects belonging to the bulge or
halo).  Finally, the under-luminous Cepheids have mean $V-$band
magnitudes consistent with those expected based on the PL relation
observed by Alcock et al.\ (1998) for Pop II Cepheids and RV Tauri
stars in the LMC (Figure \ref{pl1})\footnotemark.

\footnotetext{Unfortunately, the PL relation for Pop II Cepheids is
not characterized in the $I-$band.}

Beyond the Milky Way globular clusters, bulge and halo, Pop II
Cepheids have been detected only in the LMC (Alcock et al.\ 1998), the
Fornax dwarf spheroidal (Bersier \& Wood 2002) and, possibly, NGC 6822
(Antonello et al.\ 2002), IC 1613 (Antonello et al.\ 1999) and the
And I and And III dwarf spheroidal companions of M31 (Pritzl et al.\
2005).  If our assessment of the under-luminous variables in NGC 5128
is correct,  to the best of our knowledge this would represent the
first detection of Pop II Cepheids outside the Local group.

\acknowledgments

The work presented in this paper is based on observations with the NASA/ESA 
Hubble Space Telescope, obtained by the Space Telescope Science Institute, 
which is operated by AURA, Inc. under NASA contract No. 5-26555. 
Support for this work was provided by NASA through grant GO-09043.02 from 
the Space Telescope Science Institute (STScI).

\clearpage 

\begin{deluxetable}{cclcc}
\tabletypesize{\scriptsize}
\tablecaption{Log of Observations.\label{tbl:observations}}
\tablewidth{0pt}
\tablehead{
\colhead{Date Obs.} &
\colhead{HJD} &
\colhead{Rootname} &
\colhead{Filter} &
\colhead{Exp. Time}\\
\colhead{} &
\colhead{(days)} &
\colhead{} &
\colhead{} &
\colhead{(s)}
}
\startdata
2001-07-08 &  2452099.00 &  u6dm2101r,u6dm2102r &  F555W &  1200+1100\\
           &             &  u6dm2103r,u6dm2104r &  F814W &  1300+1100\\
2001-07-14 &  2452105.50 &  u6dm2201r,u6dm2202r &  F555W &  1300+1300\\
2001-07-22 &  2452112.50 &  u6dm2301r,u6dm2302r &  F555W &  1300+1300\\
           &             &  u6dm2303r,u6dm2304r &  F814W &  1300+1400\\
2001-07-24 &  2452114.50 &  u6dm2401r,u6dm2402r &  F555W &  1300+1300\\
2001-07-26 &  2452116.75 &  u6dm2501r,u6dm2502r &  F555W &  1300+1300\\
           &             &  u6dm2503r,u6dm2504m &  F814W &  1300+1400\\
2001-07-28 &  2452119.25 &  u6dm2601r,u6dm2602r &  F555W &  1300+1300\\
2001-08-01 &  2452123.00 &  u6dm2701m,u6dm2702r &  F555W &  1300+1300\\
           &             &  u6dm2703r,u6dm2704r &  F814W &  1300+1400\\
2001-08-03 &  2452125.25 &  u6dm2801r,u6dm2802r &  F555W &  1300+1300\\
2001-08-07 &  2452128.75 &  u6dm2901m,u6dm2902m &  F555W &  1300+1300\\
           &             &  u6dm2903m,u6dm2904m &  F814W &  1300+1400\\
2001-08-11 &  2452133.00 &  u6dm3001m,u6dm3002m &  F555W &  1300+1300\\
2001-08-16 &  2452137.75 &  u6dm3101m,u6dm3102m &  F555W &  1300+1300\\
           &             &  u6dm3103m,u6dm3104m &  F814W &  1300+1400\\
2001-08-20 &  2452142.00 &  u6dm3201m,u6dm3202m &  F555W &  1300+1300\\
\enddata
\end{deluxetable}

\begin{deluxetable}{lc}
\tabletypesize{\scriptsize}
\tablecaption{Adopted ALLFRAME Photometric Zero Points (short-exposure).\label{tbl:zpall}}
\tablewidth{0pt}
\tablehead{
\colhead{Chip} &
\colhead{Transformation equations}
}
\startdata
PC & F555W = $-2.5\log_{10}({\rm DN}/t)$ + 22.328 \\
        & F814W = $-2.5\log_{10}({\rm DN}/t)$ + 21.483 \\
WF2 & F555W = $-2.5\log_{10}({\rm DN}/t)$ + 22.388 \\
        & F814W = $-2.5\log_{10}({\rm DN}/t)$ + 21.529 \\
WF3 & F555W = $-2.5\log_{10}({\rm DN}/t)$ + 22.387 \\
        & F814W = $-2.5\log_{10}({\rm DN}/t)$ + 21.505 \\
WF4 & F555W = $-2.5\log_{10}({\rm DN}/t)$ + 22.361 \\
        & F814W = $-2.5\log_{10}({\rm DN}/t)$ + 21.498 \\
All & V = F555W $-$ 0.052($V-I$) + 0.027($V-I$)$^2$\\
        & I = F814W $-$ 0.063($V-I$) + 0.025($V-I$)$^2$  \\
\enddata
\end{deluxetable}

\begin{deluxetable}{llcc}
\tabletypesize{\scriptsize}
\tablecaption{Comparison of DoPHOT and ALLFRAME Photometry
for the  Secondary Standard Stars.\label{tbl:comp}}
\tablewidth{0pt}
\tablehead{
\colhead{Chip} &
\colhead{$\Delta V$(DoP.$-$ALL.)} &
\colhead{$\Delta I$(DoP.$-$ALL.)} &
\colhead{No. Stars}\\
\colhead{} &
\colhead{(mag)} &
\colhead{(mag)} &
\colhead{}
}
\startdata
PC  & \phantom{$-$}0.059 $\pm$ 0.002 & \phantom{$-$}0.031 $\pm$ 0.004 & 36\\
WF2 & \phantom{$-$}0.032 $\pm$ 0.002 & $-$0.105 $\pm$ 0.002 & 23\\
WF3 & $-$0.008 $\pm$ 0.001 & $-$0.084 $\pm$ 0.001 & 58\\
WF4 & $-$0.008 $\pm$ 0.001 & $-$0.080 $\pm$ 0.001 & 42\\
\enddata
\end{deluxetable}

\begin{deluxetable}{lc}
\tabletypesize{\scriptsize}
\tablecaption{Adopted DoPHOT Photometric Zero Points (short-exposure).\label{tbl:zpdo}}
\tablewidth{0pt}
\tablehead{
\colhead{Chip} &
\colhead{Transformation equations}
}
\startdata
PC & F555W  = $-2.5\log_{10}({\rm DN}/t)$ + 1.202 \\
    & F814W  = $-2.5\log_{10}({\rm DN}/t)$ + 0.324 \\
WF2 & F555W  = $-2.5\log_{10}({\rm DN}/t)$ + 1.323 \\
    & F814W  = $-2.5\log_{10}({\rm DN}/t)$ + 0.615 \\
WF3 & F555W  = $-2.5\log_{10}({\rm DN}/t)$ + 1.345 \\
    & F814W  = $-2.5\log_{10}({\rm DN}/t)$ + 0.539 \\
WF4 & F555W  = $-2.5\log_{10}({\rm DN}/t)$ + 1.340 \\
    & F814W  = $-2.5\log_{10}({\rm DN}/t)$ + 0.530 \\
All & V = F555W $-$ 0.052($V-I$) + 0.027($V-I$)$^2$\\
    & I = F814W $-$ 0.063($V-I$) + 0.025($V-I$)$^2$\\
\enddata
\end{deluxetable}

\clearpage

\begin{deluxetable}{lcrcccccccl}
\tabletypesize{\scriptsize}
\tablecaption{Cepheid Variable Stars in NGC 5128.\label{tbl:ceph}}
\tablewidth{0pt}
\tablehead{
\colhead{ID} &
\colhead{$X$} &
\colhead{$Y$} &
\colhead{RA (J2000)} &
\colhead{Dec (J2000)} &
\colhead{Period} &
\colhead{$V^{Int}$} &
\colhead{$I^{Int}$} &
\colhead{$V^{ph}$} &
\colhead{$I^{ph}$} &
\colhead{Chip}\\
\colhead{} &
\colhead{(pixel)} &
\colhead{(pixel)} &
\colhead{(h:m:s)} &
\colhead{(\deg:\min:\sec)} &
\colhead{(days)} &
\colhead{(mag)} &
\colhead{(mag)} &
\colhead{(mag)} &
\colhead{(mag)} &
\colhead{}
}
\startdata
C1  & 147.2 & 305.5 & 13:25:13.534 & $-$43:00:17.05 &  \phantom{$\geq$}5.0 & 25.14 & 24.12 & 25.08 & 24.06 & WF2\\
C2  & 268.4 & 201.4 & 13:25:14.711 & $-$43:00:26.29 &  \phantom{$\geq$}5.3 & 24.61 & 23.46 & 24.66 & 23.52 & WF2\\
C3  & 129.8 & 651.9 & 13:25:13.659 & $-$42:59:04.99 &  \phantom{$\geq$}7.0 & 24.12 & 23.45 & 24.08 & 23.44 & WF3\\
C4  & 616.0 & 621.5 & 13:25:09.448 & $-$42:59:19.04 &  \phantom{$\geq$}7.3 & 23.94 & 23.18 & 23.94 & 23.20 & WF3\\
C5  & 687.9 & 358.4 & 13:25:14.224 & $-$43:01:10.44 &  \phantom{$\geq$}7.4 & 24.98 & 23.97 & 24.95 & 23.92 & WF2\\
C6  & 547.4 & 411.9 & 13:25:18.163 & $-$43:00:11.80 &  \phantom{$\geq$}8.2 & 24.86 & 23.66 & 24.86 & 23.67 & PC\\
C7  & 603.4 & 611.7 & 13:25:09.578 & $-$42:59:19.68 &  \phantom{$\geq$}8.6 & 23.99 & 22.94 & 24.04 & 22.94 & WF3\\
C8  & 590.3 & 683.2 & 13:25:18.589 & $-$43:00:23.32 &  \phantom{$\geq$}8.8 & 24.93 & 23.38 & 24.88 & 23.36 & PC\\
C9  & 775.3 & 498.0 & 13:25:13.184 & $-$43:01:22.05 &  \phantom{$\geq$}9.4 & 24.16 & 22.94 & 24.11 & 22.93 & WF2\\
C10 & 410.4 & 552.9 & 13:25:19.519 & $-$42:59:14.54 & \phantom{$\geq$}10.6 & 23.98 & 22.95 & 23.98 & 22.96 & WF4\\
C11 & 126.2 & 587.3 & 13:25:13.827 & $-$42:59:11.10 & \phantom{$\geq$}11.0 & 23.84 & 23.01 & 23.84 & 23.00 & WF3\\
C12 & 440.9 & 478.7 & 13:25:18.802 & $-$42:59:13.22 & \phantom{$\geq$}11.2 & 23.60 & 22.70 & 23.60 & 22.70 & WF4\\
C13 & 713.1 & 536.5 & 13:25:18.945 & $-$43:00:15.57 & \phantom{$\geq$}12.7 & 25.43 & 23.54 & 25.38 & 23.54 & PC\\
C14 & 317.6 & 562.9 & 13:25:19.794 & $-$42:59:23.31 & \phantom{$\geq$}12.7 & 23.99 & 22.93 & 23.92 & 22.92 & WF4\\
C15 & 473.8 & 318.7 & 13:25:17.778 & $-$43:00:08.44 & \phantom{$\geq$}12.8 & 24.64 & 23.22 & 24.66 & 23.21 & PC\\
C16 & 422.6 & 403.3 & 13:25:18.171 & $-$42:59:16.66 & \phantom{$\geq$}13.9 & 23.27 & 22.40 & 23.31 & 22.41 & WF4\\
C17 & 478.9 & 162.1 & 13:25:15.929 & $-$42:59:16.53 & \phantom{$\geq$}13.9 & 23.64 & 22.61 & 23.64 & 22.60 & WF4\\
C18 & 158.0 & 229.3 & 13:25:14.227 & $-$43:00:16.31 & \phantom{$\geq$}14.3 & 24.08 & 22.88 & 24.03 & 22.85 & WF2\\
C19 & 294.3 & 754.6 & 13:25:21.524 & $-$42:59:21.32 & \phantom{$\geq$}14.9 & 23.41 & 22.36 & 23.41 & 22.42 & WF4\\
C20 & 646.7 & 305.9 & 13:25:18.462 & $-$43:00:06.09 & \phantom{$\geq$}15.1 & 23.86 & 22.80 & 23.87 & 22.83 & PC\\
C21 & 336.9 & 589.5 & 13:25:19.990 & $-$42:59:20.85 & \phantom{$\geq$}15.1 & 23.73 & 22.80 & 23.68 & 22.79 & WF4\\
C22 & 715.0 & 716.8 & 13:25:08.390 & $-$42:59:12.20 & \phantom{$\geq$}15.5 & 23.76 & 22.58 & 23.77 & 22.55 & WF3\\
C23 & 487.8 & 491.5 & 13:25:10.845 & $-$42:59:28.62 & \phantom{$\geq$}16.1 & 25.03 & 23.15 & 25.05 & 23.16 & WF3\\
C24 & 170.5 & 267.1 & 13:25:13.922 & $-$43:00:18.39 & \phantom{$\geq$}16.5 & 23.91 & 22.80 & 23.91 & 22.79 & WF2\\
C25 & 139.5 & 600.2 & 13:25:10.924 & $-$43:00:23.27 & \phantom{$\geq$}16.5 & 24.73 & 22.97 & 24.74 & 22.98 & WF2\\
C26 & 503.2 & 215.1 & 13:25:16.346 & $-$42:59:13.00 & \phantom{$\geq$}16.6 & 23.35 & 22.36 & 23.34 & 22.35 & WF4\\
C27 & 123.3 & 617.8 & 13:25:10.736 & $-$43:00:22.13 & \phantom{$\geq$}17.3 & 24.74 & 23.24 & 24.73 & 23.25 & WF2\\
C28 & 429.5 & 755.4 & 13:25:21.260 & $-$42:59:08.25 & \phantom{$\geq$}17.9 & 23.71 & 22.62 & 23.71 & 22.61 & WF4\\
C29 & 382.8 & 574.6 & 13:25:11.667 & $-$43:00:46.11 & \phantom{$\geq$}20.5 & 23.94 & 22.45 & 23.85 & 22.44 & WF2\\
C30 & 657.0 & 337.8 & 13:25:18.533 & $-$43:00:07.39 & \phantom{$\geq$}20.9 & 24.21 & 22.81 & 24.23 & 22.83 & PC\\
C31 & 421.0 & 589.1 & 13:25:11.230 & $-$42:59:17.66 & \phantom{$\geq$}21.3 & 23.57 & 22.44 & 23.51 & 22.42 & WF3\\
C32 & 422.6 & 123.5 & 13:25:12.191 & $-$43:00:02.65 & \phantom{$\geq$}22.2 & 24.05 & 22.62 & 24.10 & 22.63 & WF3\\
C33 & 728.9 &  50.2 & 13:25:17.005 & $-$43:01:07.04 & \phantom{$\geq$}22.4 & 23.38 & 22.27 & 23.30 & 22.25 & WF2\\
C34 &  65.7 & 233.4 & 13:25:15.097 & $-$42:59:43.83 & \phantom{$\geq$}22.5 & 24.59 & 22.86 & 24.63 & 22.90 & WF3\\
C35 & 698.1 & 198.6 & 13:25:18.565 & $-$43:00:00.84 & \phantom{$\geq$}22.6 & 24.93 & 22.89 & 24.91 & 22.91 & PC\\
C36 & 474.8 & 288.4 & 13:25:17.050 & $-$42:59:14.13 & \phantom{$\geq$}23.1 & 23.01 & 22.10 & 22.96 & 22.07 & WF4\\
C37 & 423.5 & 369.9 & 13:25:17.875 & $-$42:59:17.31 & \phantom{$\geq$}23.1 & 22.98 & 21.77 & 23.02 & 21.76 & WF4\\
C38 & 295.2 & 696.4 & 13:25:12.113 & $-$42:59:04.46 & \phantom{$\geq$}24.1 & 23.27 & 22.22 & 23.34 & 22.26 & WF3\\
C39 &  94.6 & 539.4 & 13:25:14.205 & $-$42:59:15.00 & \phantom{$\geq$}26.4 & 23.32 & 22.22 & 23.30 & 22.21 & WF3\\
C40 & 320.7 & 632.0 & 13:25:12.024 & $-$42:59:11.23 & \phantom{$\geq$}27.1 & 23.35 & 22.28 & 23.33 & 22.27 & WF3\\
C41 & 755.1 & 554.5 & 13:25:08.370 & $-$42:59:28.69 & \phantom{$\geq$}27.8 & 23.67 & 22.33 & 23.67 & 22.32 & WF3\\
C42 & 476.9 &  78.9 & 13:25:15.202 & $-$42:59:18.55 & \phantom{$\geq$}30.5 & 23.34 & 22.14 & 23.35 & 22.15 & WF4\\
C43 & 222.6 & 215.3 & 13:25:13.761 & $-$42:59:49.21 & \phantom{$\geq$}34.1 & 24.77 & 24.02 & 24.77 & 24.02 & WF3\\
C44 & 500.9 & 439.6 & 13:25:10.838 & $-$42:59:33.94 & \phantom{$\geq$}34.7 & 24.26 & 22.24 & 24.25 & 22.23 & WF3\\
C45 & 257.6 & 387.2 & 13:25:13.095 & $-$42:59:33.41 & \phantom{$\geq$}41.2 & 23.74 & 21.92 & 23.73 & 21.91 & WF3\\
C46 & 474.9 & 106.9 & 13:25:15.981 & $-$43:00:43.99 & \phantom{$\geq$}42.8 & 24.43 & 22.33 & 24.41 & 22.31 & WF2\\
C47 & 475.5 & 338.9 & 13:25:17.804 & $-$43:00:09.32 & $\geq$44.0 & 23.37 & 21.59 & 23.35 & 21.58 & PC\\
C48 & 500.2 & 114.7 & 13:25:15.967 & $-$43:00:46.62 & $\geq$44.0 & 24.44 & 22.42 & 24.42 & 22.42 & WF2\\
C49 &  85.1 &  98.6 & 13:25:16.166 & $-$42:59:55.92 & \phantom{$\geq$}44.4 & 23.12 & 21.70 & 23.13 & 21.69 & WF4\\
C50 & 275.9 & 341.2 & 13:25:13.495 & $-$43:00:30.28 & \phantom{$\geq$}47.0 & 24.32 & 23.32 & 24.27 & 23.29 & WF2\\
C51 & 348.4 & 695.1 & 13:25:11.648 & $-$42:59:05.80 & $\geq$48.0 & 25.11 & 22.92 & 25.10 & 22.92 & WF3\\
C52 & 341.9 & 450.9 & 13:25:12.669 & $-$43:00:39.25 & $\geq$48.5 & 25.06 & 24.05 & 24.96 & 24.02 & WF2\\
C53 & 544.7 & 440.4 & 13:25:18.254 & $-$42:59:04.00 & \phantom{$\geq$}48.5 & 22.27 & 21.34 & 22.33 & 21.37 & WF4\\
C54 & 439.2 & 668.6 & 13:25:17.968 & $-$43:00:24.26 & $\geq$50.4 & 23.59 & 22.70 & 23.55 & 22.68 & PC\\
C55 & 599.0 & 480.1 & 13:25:18.496 & $-$42:58:57.88 & $\geq$74.0 & 24.98 & 22.75 & 25.02 & 22.76 & WF4\\ 
C56 & 604.1 & 670.5 & 13:25:20.163 & $-$42:58:53.25 & $\geq$80.0 & 24.11 & 22.88 & 24.12 & 22.89 & WF4\\
\enddata
\tablecomments{Cepheids are numbered in  order of increasing
  period. $X$ and $Y$ positions are in the reference frame of 
the first exposure, u6dm2101r (see Table~\ref{tbl:observations}). 
Columns 7 and 8 give intensity-averaged
$V$ and $I$ magnitudes respectively (equation 3), while columns 9 and
10 give phase-weighted magnitudes (equation 4) assuming the period 
listed in column 6. The last column lists the chip in which the
Cepheid 
is located.}
\end{deluxetable}

\clearpage

\LongTables
\begin{landscape}
\begin{deluxetable}{llcccccccc}
\tabletypesize{\scriptsize}
\tablecaption{Cepheid Distance Estimates to NGC 5128.\label{tbl:dist}}
\tablewidth{0pt}
\tablehead{
\colhead{} &
\colhead{} &
\colhead{} &
\colhead{} &
\colhead{} &
\colhead{} &
\multicolumn{2}{c}{$R_V^{NGC5128} = 3.3$} &
\multicolumn{2}{c}{$R_V^{NGC5128} = 2.4$}\\
\colhead{Case} &
\colhead{Sample} &
\colhead{No. Cepheids} &
\colhead{$ \mu_V$} &
\colhead{$ \mu_I$} &
\colhead{$E(V-I)$} &
\colhead{$ \mu_0$} &
\colhead{d}&
\colhead{$ \mu_0$} &
\colhead{d}\\
\colhead{} &
\colhead{} &
\colhead{} &
\colhead{(mag)} &
\colhead{(mag)} &
\colhead{(mag)} &
\colhead{(mag)} &
\colhead{(Mpc)} &
\colhead{(mag)} &
\colhead{(Mpc)}
} 
\startdata
1 & All Chips, Ceph. only, P$\geq$8d, Ph.Mag. & 42              &  28.85 $\pm$ 0.10 & 28.30 $\pm$ 0.06 &  0.55 $\pm$ 0.05 & 27.48 $\pm$ 0.05 &  3.1  $\pm$  0.1 & 27.62 $\pm$ 0.04 &  3.4  $\pm$  0.1 \\ 
2 &  All Chips, Ceph. only, P$\geq$15d, Ph.Mag. & 28            &  29.02 $\pm$ 0.12 & 28.43 $\pm$ 0.06 &  0.59 $\pm$ 0.07 & 27.54 $\pm$ 0.07 &  3.2  $\pm$  0.1 & 27.70 $\pm$ 0.05 &  3.5  $\pm$  0.1 \\  
3 &  All Chips, Ceph. only, P$\geq$20d, Ph.Mag. & 19            &  29.09 $\pm$ 0.15 & 28.47 $\pm$ 0.07 &  0.61 $\pm$ 0.09 & 27.55 $\pm$ 0.09 &  3.2  $\pm$  0.1 & 27.72 $\pm$ 0.06 &  3.5  $\pm$  0.1 \\ 
4 &  All Chips, Ceph. only, P$\geq$25d, Ph.Mag. & \phantom{0}9  &  29.22 $\pm$ 0.21 & 28.57 $\pm$ 0.08 &  0.65 $\pm$ 0.14 & 27.59 $\pm$ 0.14 &  3.3  $\pm$  0.2 & 27.77 $\pm$ 0.10 &  3.6  $\pm$  0.2 \\ 
5 &  All Chips, Ceph. only, P$\geq$30d, Ph.Mag. & \phantom{0}6  &  29.41 $\pm$ 0.29 & 28.63 $\pm$ 0.11 &  0.78 $\pm$ 0.19 & 27.45 $\pm$ 0.19 &  3.1  $\pm$  0.3 & 27.68 $\pm$ 0.13 &  3.4  $\pm$  0.2 \\ 
6 & PC, P$\geq$8d, Ph.Mag. & \phantom{0}7                       &  29.34 $\pm$ 0.21 & 28.51 $\pm$ 0.10 &  0.83 $\pm$ 0.13 & 27.27 $\pm$ 0.14 &  2.9  $\pm$  0.2 & 27.52 $\pm$ 0.11 &  3.2  $\pm$  0.2 \\  
7 & WF2, P$\geq$8d, Ph.Mag. & \phantom{0}8                      &  29.14 $\pm$ 0.24 & 28.30 $\pm$ 0.15 &  0.84 $\pm$ 0.12 & 27.05 $\pm$ 0.12 &  2.6  $\pm$  0.1 & 27.30 $\pm$ 0.10 &  2.9  $\pm$  0.1 \\ 
8 & WF3, P$\geq$8d, Ph.Mag. & 13                                &  28.99 $\pm$ 0.18 & 28.31 $\pm$ 0.09 &  0.67 $\pm$ 0.10 & 27.30 $\pm$ 0.11 &  2.9  $\pm$  0.1 & 27.50 $\pm$ 0.08 &  3.2  $\pm$  0.1 \\ 
9 & WF4, P$\geq$8d, Ph.Mag. & 14                               &  28.35 $\pm$ 0.09 & 27.98 $\pm$ 0.07 &  0.37 $\pm$ 0.04 & 27.42 $\pm$ 0.07 &  3.0  $\pm$  0.1 & 27.50 $\pm$ 0.07 &  3.2  $\pm$  0.1 \\ 
10 &  All Chips, Ceph.+Var., P$\geq$8d, Ph.Mag. & 70            &  28.93 $\pm$ 0.10 & 28.39 $\pm$ 0.07 &  0.55 $\pm$ 0.05 & 27.56 $\pm$ 0.07 &  3.3  $\pm$  0.1 & 27.71 $\pm$ 0.07 &  3.5  $\pm$  0.1 \\  
\enddata
\tablecomments{All distance moduli are based on short-exposure zero
  points (see \S~\ref{zeropt})}
\end{deluxetable}
\clearpage
\end{landscape}

\begin{deluxetable}{llc}
\tabletypesize{\scriptsize}
\tablecaption{Error Budget in the Distance to NGC 5128.\label{tbl:error}}
\tablewidth{0pt}
\tablehead{
\colhead{Systematic Uncertainties} &
\colhead{$\delta\mu_0$ (mag)} &
\colhead{Notes} 
}
\startdata
\hskip.1in a) WFPC2 $V$ band zero point                                                                         & $\pm 0.02$       &\\
\hskip.1in b) WFPC2 $I$ band zero point                                                                         & $\pm 0.01$       &\\
\hskip.1in c) $V-$band Short to Long Exposure Zero Point Uncertainty                                              & $\pm 0.05$       &\\
\hskip.1in d) $I-$band Short to Long Exposure Zero Point Uncertainty                                                 & $\pm 0.05$       &\\
\hskip.1in e) Charge Transfer Efficiency                                                                        & $\pm 0.01$       &\\
\hskip.3in A) Cumulative Error on $\mu_V$:   $\sqrt{a^2 + c^2 + e^2}$   & \phantom{$\pm 0.00$}$\pm 0.05$ & (1)\\
\hskip.3in B) Cumulative Error on $\mu_I$:   $\sqrt{b^2 + d^2 + e^2}$     & \phantom{$\pm 0.00$}$\pm 0.05$ & (1)\\
\hskip.3in C) Error on $\mu_0$ due to A and B: $\sqrt{A^2\times(1-R)^2 + B^2\times R^2}$& \phantom{$\pm 0.00$}$\pm 0.13$ & (2,3)\\
\hskip.1in i) Error on LMC Distance Modulus                       & $\pm 0.10$ &\\
\hskip.3in {\bf Final Systematic Error on $\mu_0$ due to C and i:  $\sqrt{C^2 + i^2}$} &    \phantom{$\pm 0.00$}${\bf \pm 0.16}$ & (1)\\
& & \\
\hline
& & \\
\multicolumn{1}{c}{Random Uncertainties} &
\multicolumn{1}{c}{$\delta\mu_0$ (mag)} &
\multicolumn{1}{c}{Notes} \\
\hline
& & \\
\hskip.1in f) Error on $\mu_0$ due to $R_V$ for NGC 5128: & $\pm 0.077$ & (3)\\
\hskip.1in g) Error on $\mu_0$ due to $A_V$ of LMC: & $\pm 0.007$ & (3)\\
\hskip.3in D) Error on $\mu_0$ due to f and g: $\sqrt{f^2+ g^2}$& \phantom{$\pm 0.00$}$\pm 0.08$ & (1)\\
\hskip.1in h) Error on $\mu_0$ due to Metallicity:                & $\pm 0.08$ &\\
\hskip.1in j) Random Error on $\mu_0$ from fitting the PL relation (from Table~\ref{tbl:dist}):                & $\pm 0.04$ &\\
\hskip.3in {\bf Final Random Error on $\mu_0$ due to D, h, and j:  $\sqrt{D^2 + h^2 + j^2}$} &    \phantom{$\pm 0.00$}${\bf \pm 0.12}$ & (1)\\
\enddata
\tablecomments{(1): the errors are uncorrelated, and therefore
summed in quadrature. (2): $R$ is defined as $A(V)/E(V-I) = 2.143$, according to the extinction law by Cardelli, Clayton and Mathis (1989) for $R_V(NGC 5128)=2.4$. (3) The expression for the error is derived from equation (8), assuming that the errors are uncorrelated.}
\end{deluxetable}

\clearpage

\clearpage

\begin{figure}
\centering\plotone{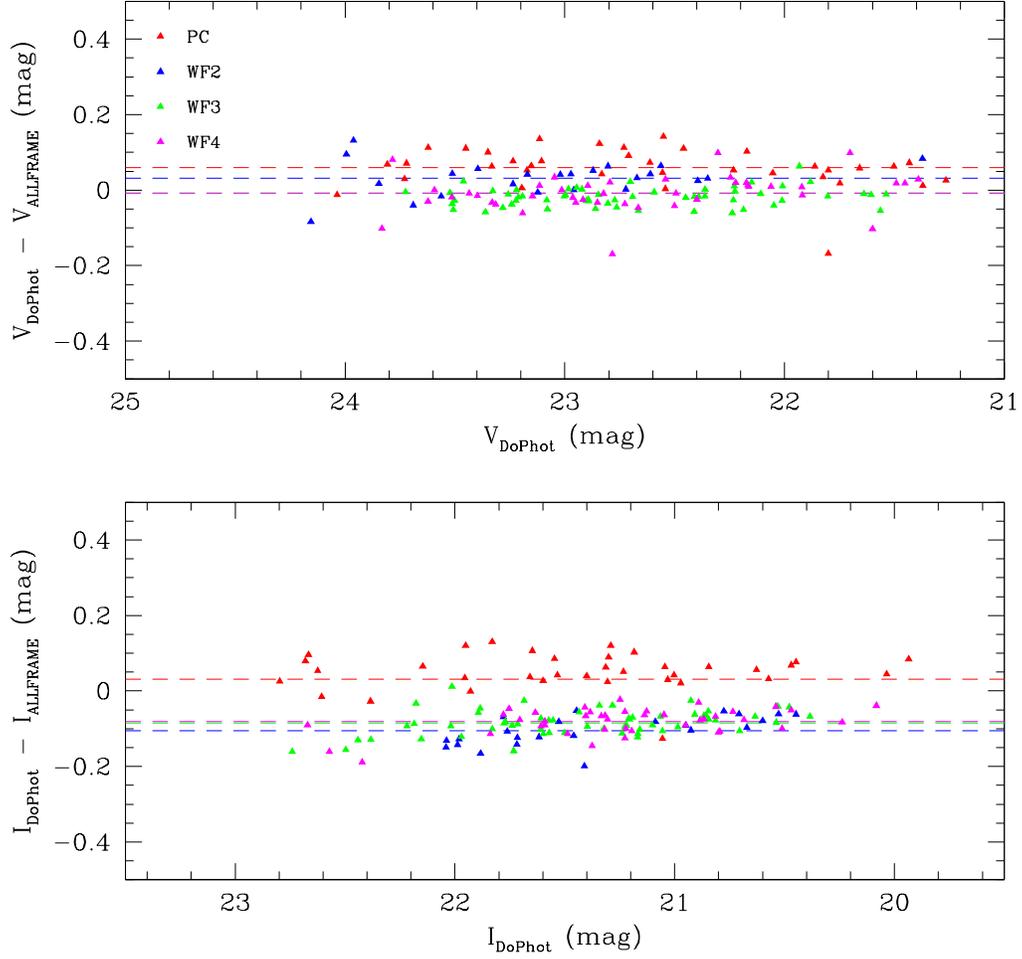}
\caption{Comparison between the $V$ and $I$ DoPHOT and ALLFRAME
  photometry for the secondary standard stars listed in Table 7. The
  photometric calibration listed in Table~\ref{tbl:zpall} has been
  adopted for ALLFRAME, while the DoPHOT calibration follows the short
  exposure zero points of  Hill et al.\ (1998).  The mean differences
  between the DoPHOT and ALLFRAME photometry (shown for each chip by
  the dashed lines) are listed in Table~\ref{tbl:comp}. See text for
  further details.
\label{comp1}}
\end{figure}

\clearpage

\begin{figure}
\centering\plotone{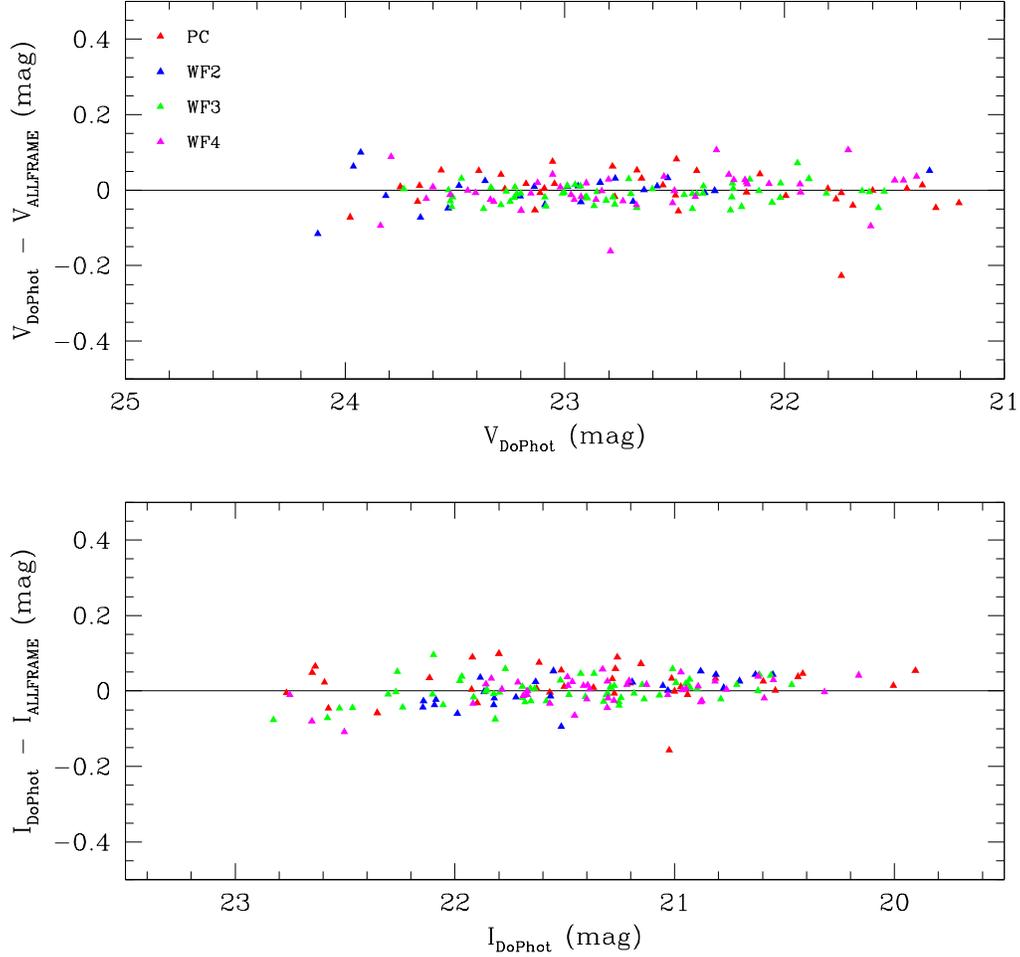}
\caption{Comparison between the $V$ and $I$ DoPHOT and ALLFRAME
  photometry for the secondary standard stars listed in Table 7, using
  the photometric calibrations given in Tables~\ref{tbl:zpall}
  and~\ref{tbl:zpdo}.
\label{comp2}}
\end{figure}

\begin{figure}
\caption{NOTE: Figure 4 can be found on the version of the paper available at http://astrowww.phys.uvic.ca/~lff/publications.html.
A deep image, obtained by combining all F555W exposures,
  showing the PC field of view. To give a sense of the dynamic range
  spanned by the four chips, the  logarithmic gray-scale  is kept the
  same in this figure and the next three figures, showing the WF2, WF3
  and WF4 fields of view respectively.  All images are shown with the
  center of the WFPC2 pyramid on the lower left corner; the direction
  of the North is also marked  (as customary, East is 90\deg~
  counter-clockwise from North).  Cepheids are identified by open
  circles and labeled with the IDs listed in Table 5. Squares identify
  probable variable stars and are labeled with the IDs listed in Table
  10.
\label{PC}}
\end{figure}

\clearpage

\begin{figure}
\centering\plotone{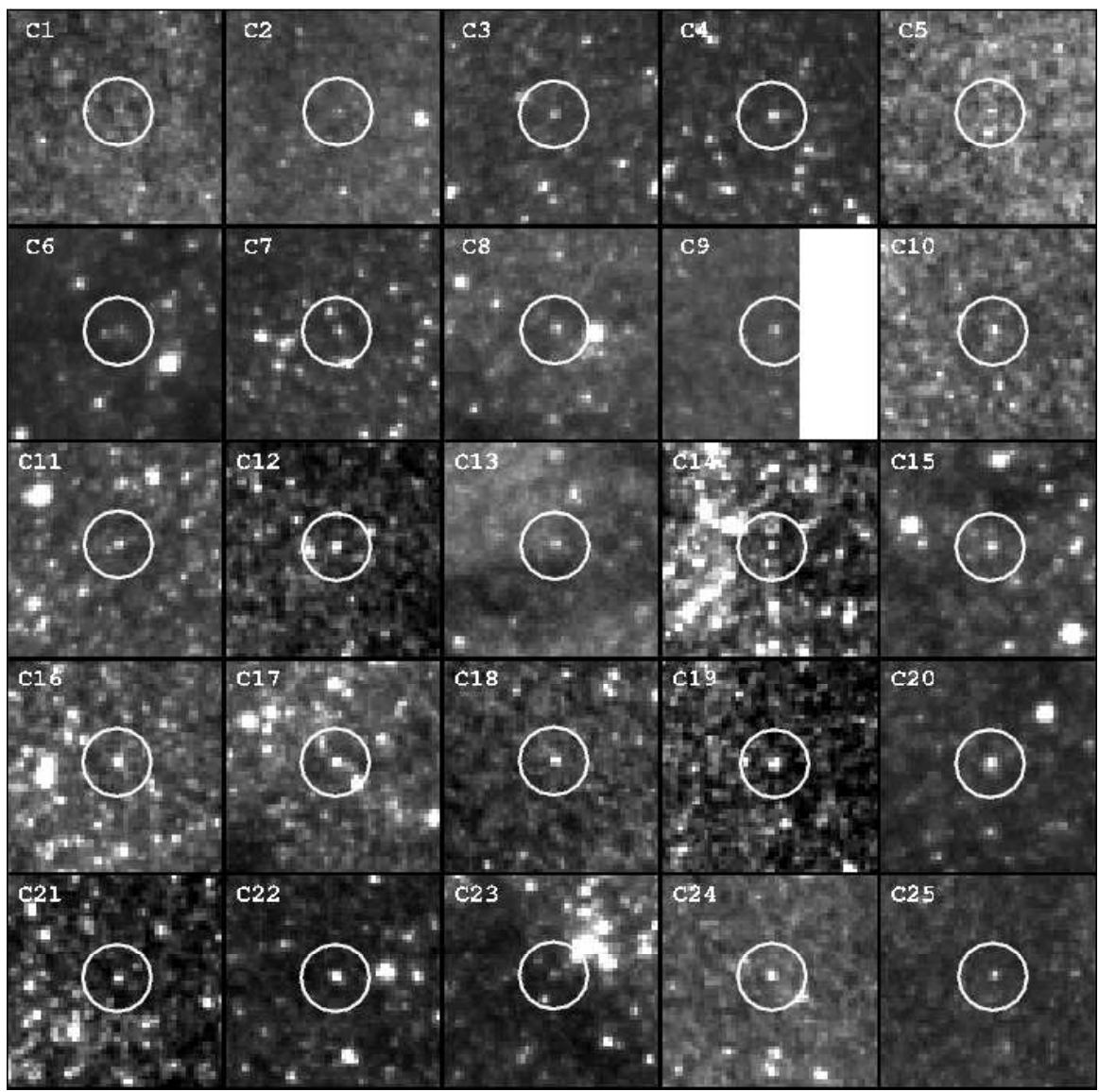}
\caption{Finding charts for the confirmed Cepheids in NGC 5128. The
  fields shown are 5\sec$\times$5\sec~ for the WF chips, and
  2\Sec5$\times$2\Sec5 for the PC; the orientation of each finding
  chart is the same as that of the chip to which it belongs, as
  shown in Figure~\ref{PC}. The
  gray-scale is different for each finding chart. The Cepheids are
  numbers following the IDs listed in Table 5.
\label{FC1}}
\end{figure}

\clearpage

\addtocounter{figure}{-1}
\begin{figure}
\centering\plotone{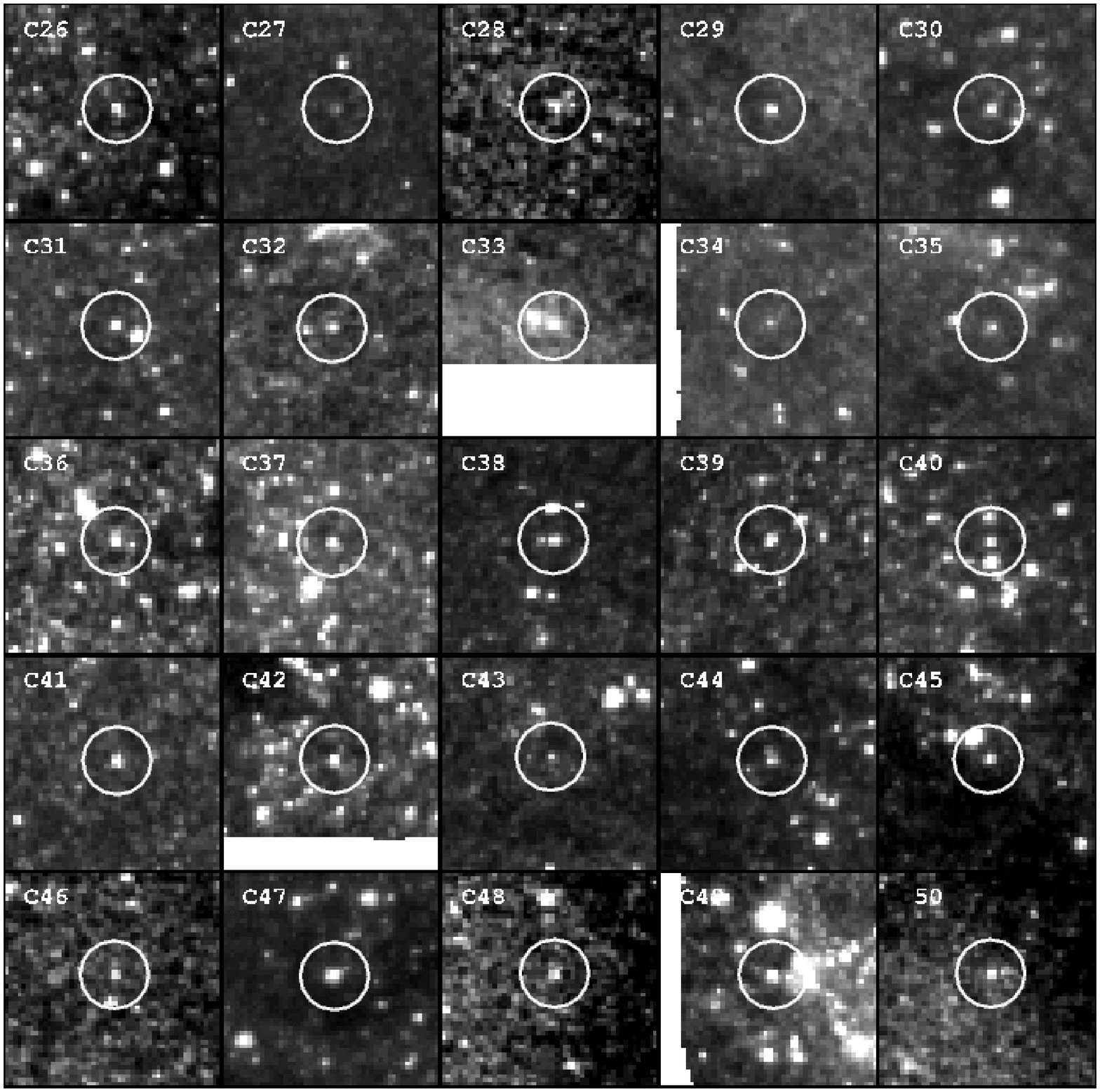}
\caption{Continued.
\label{FC2}}
\end{figure}

\clearpage

\addtocounter{figure}{-1}
\begin{figure}
\centering\plotone{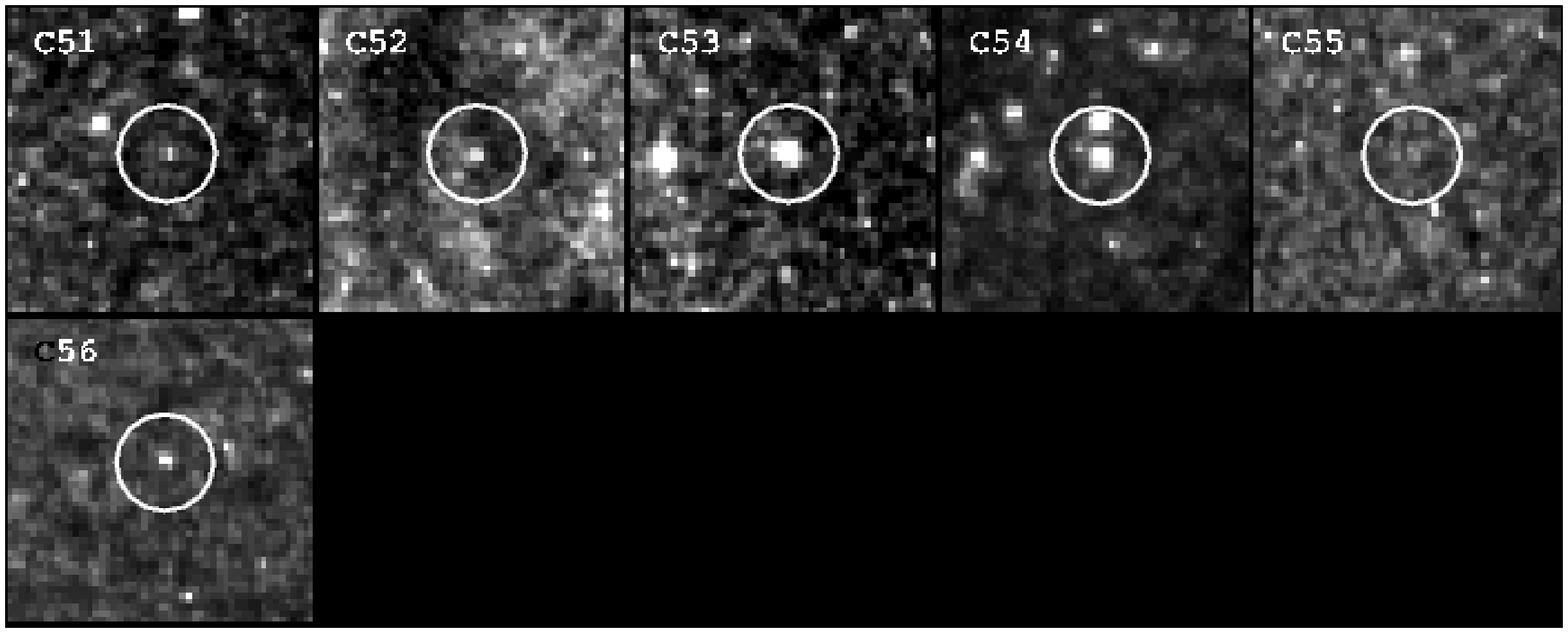}
\caption{Continued.
\label{FC3}}
\end{figure}

\begin{figure}
\centering\plotone{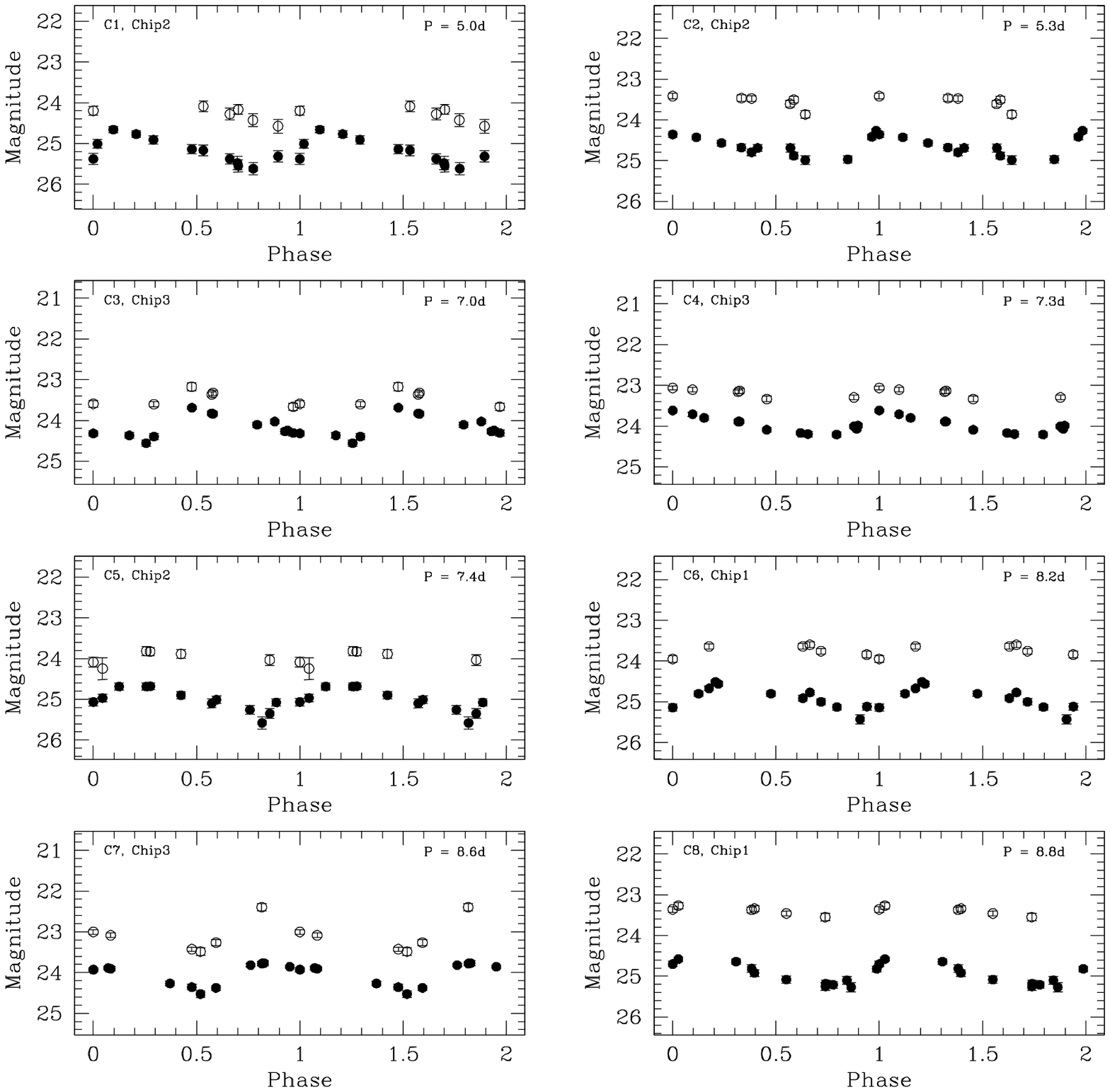}
\caption{Light curves for the NGC 5128 Cepheids, numbered as in
  Figure~\ref{PC} and Table
  5. The DoPHOT photometry is shown; solid and open circles represent
  the F555W and F814W data respectively. The period used in phasing
  the light curves is listed in the upper right corner of each panel as
  well as in Table 5.
\label{lc1}}
\end{figure}

\clearpage

\addtocounter{figure}{-1}
\begin{figure}
\centering\plotone{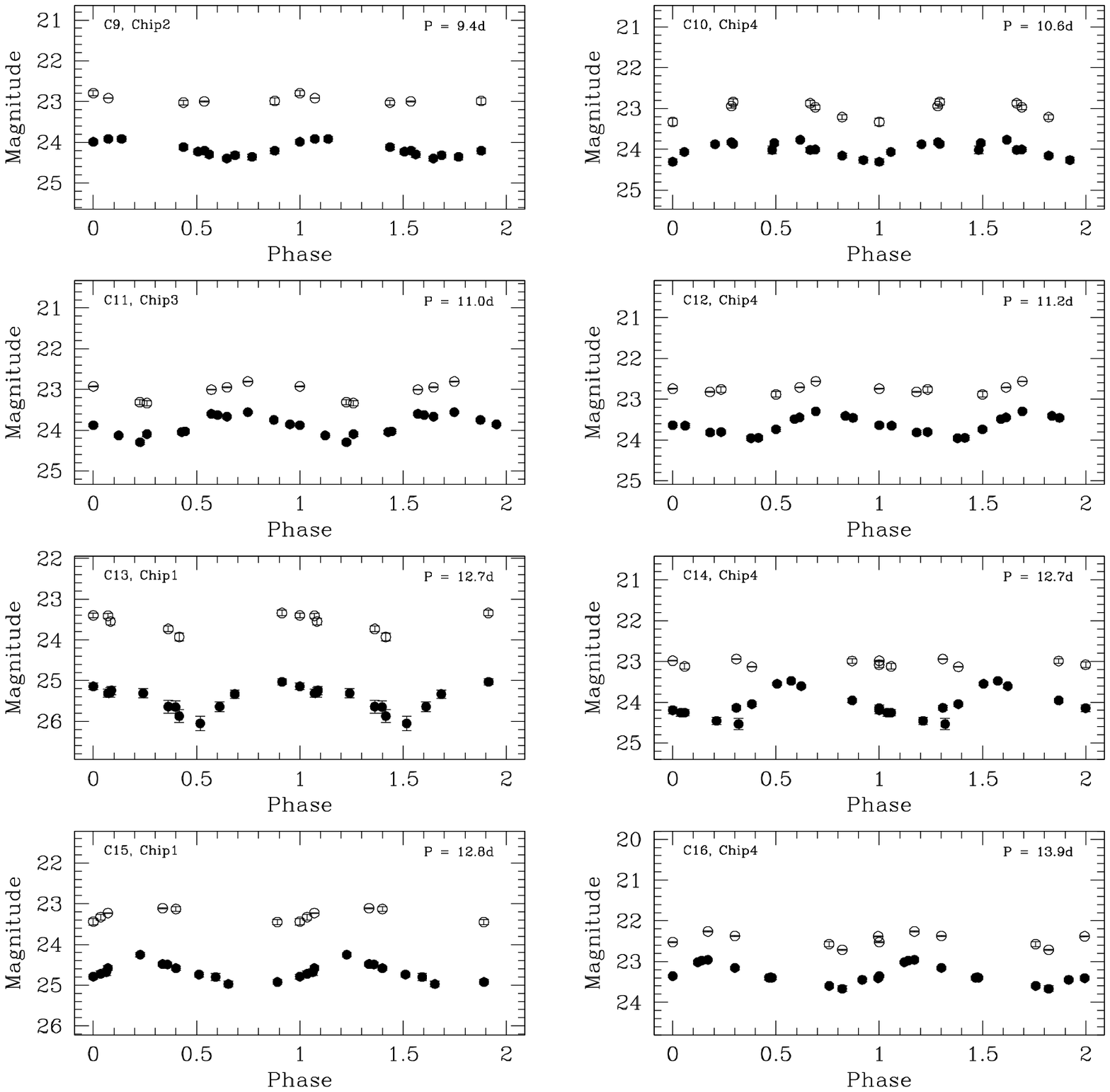}
\caption{Continued.
\label{lc2}}
\end{figure}

\clearpage

\addtocounter{figure}{-1}
\begin{figure}
\centering\plotone{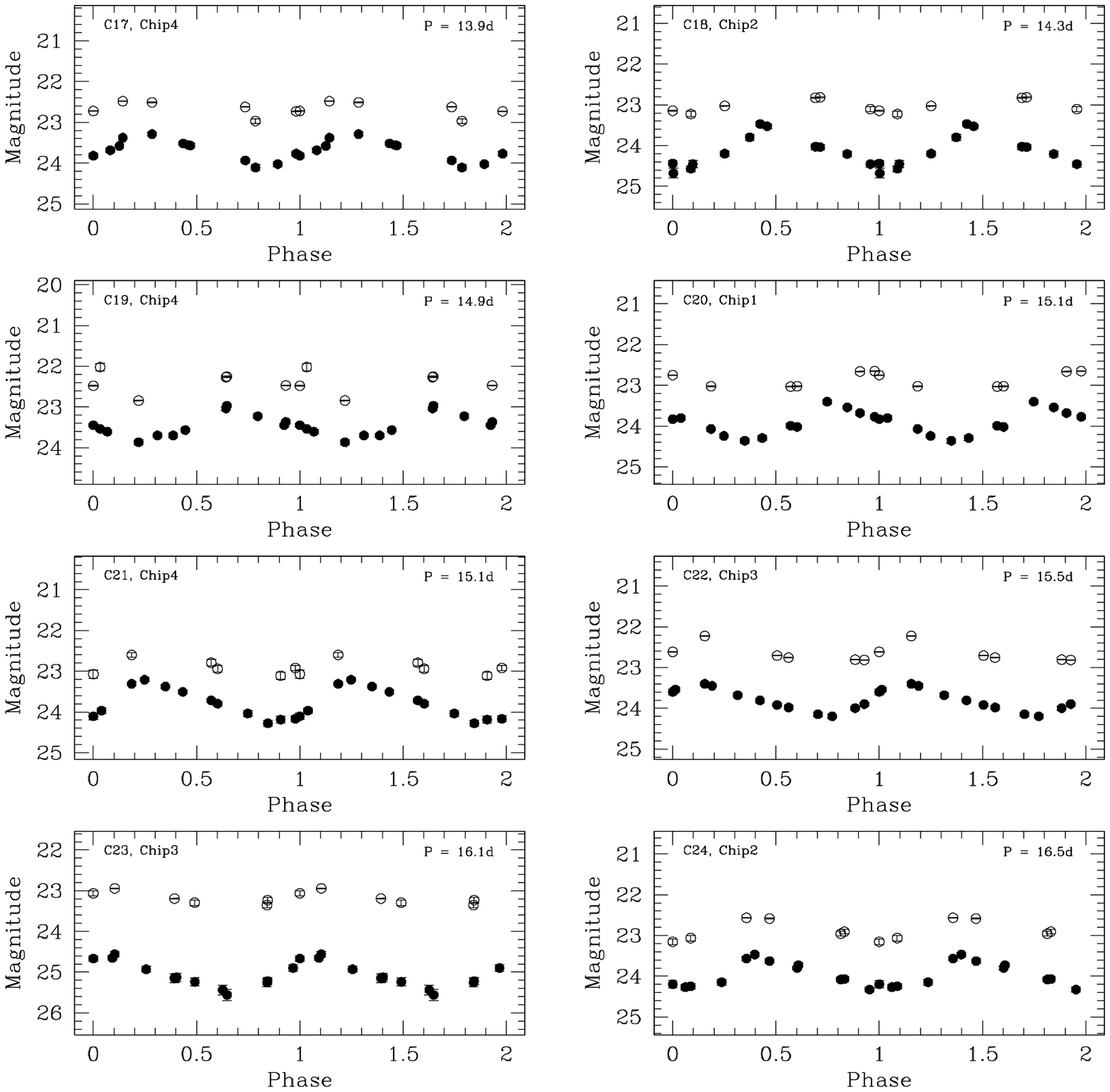}
\caption{Continued.
\label{lc3}}
\end{figure}

\clearpage

\addtocounter{figure}{-1}
\begin{figure}
\centering\plotone{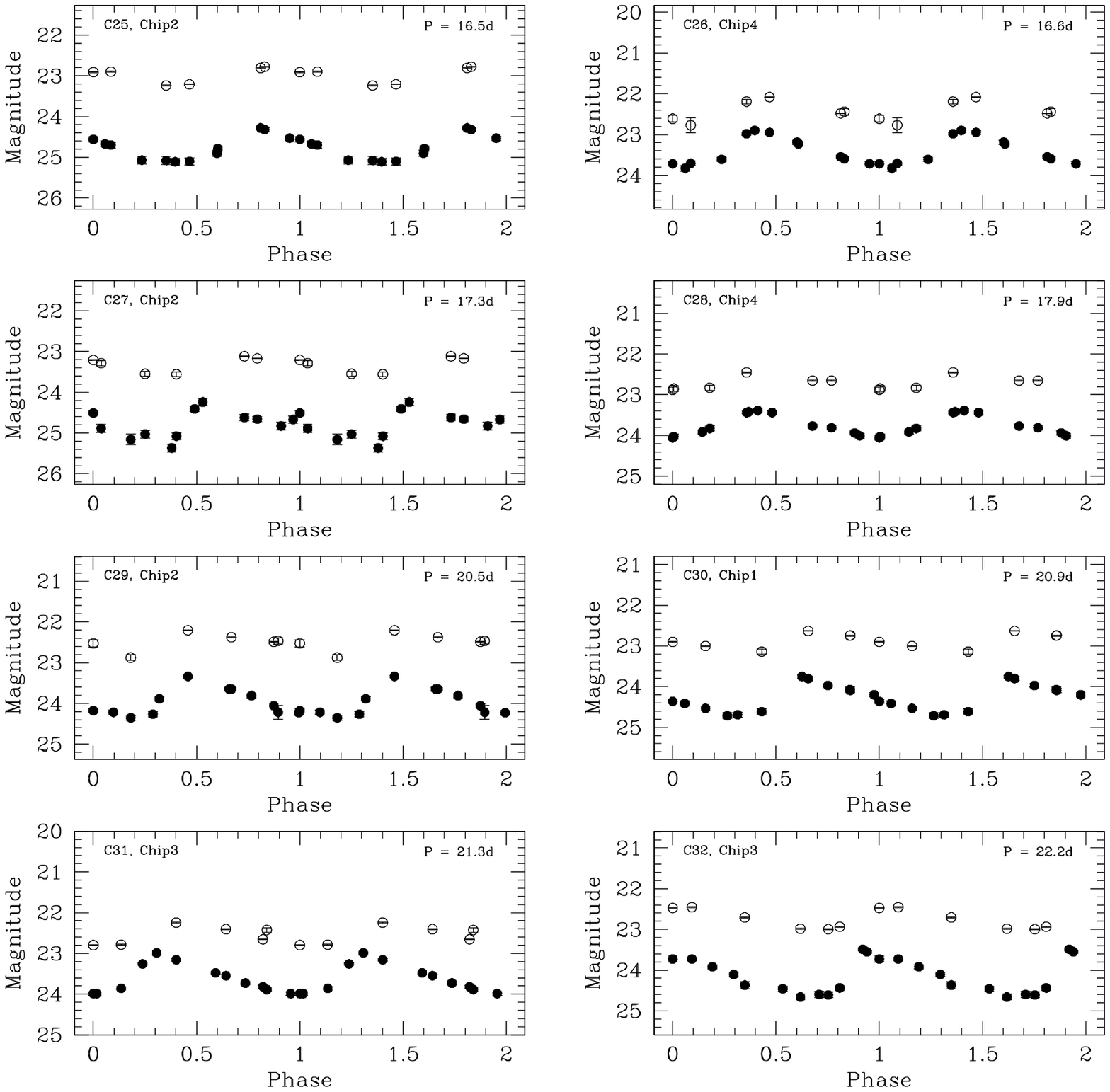}
\caption{Continued.
\label{lc4}}
\end{figure}

\clearpage

\addtocounter{figure}{-1}
\begin{figure}
\centering\plotone{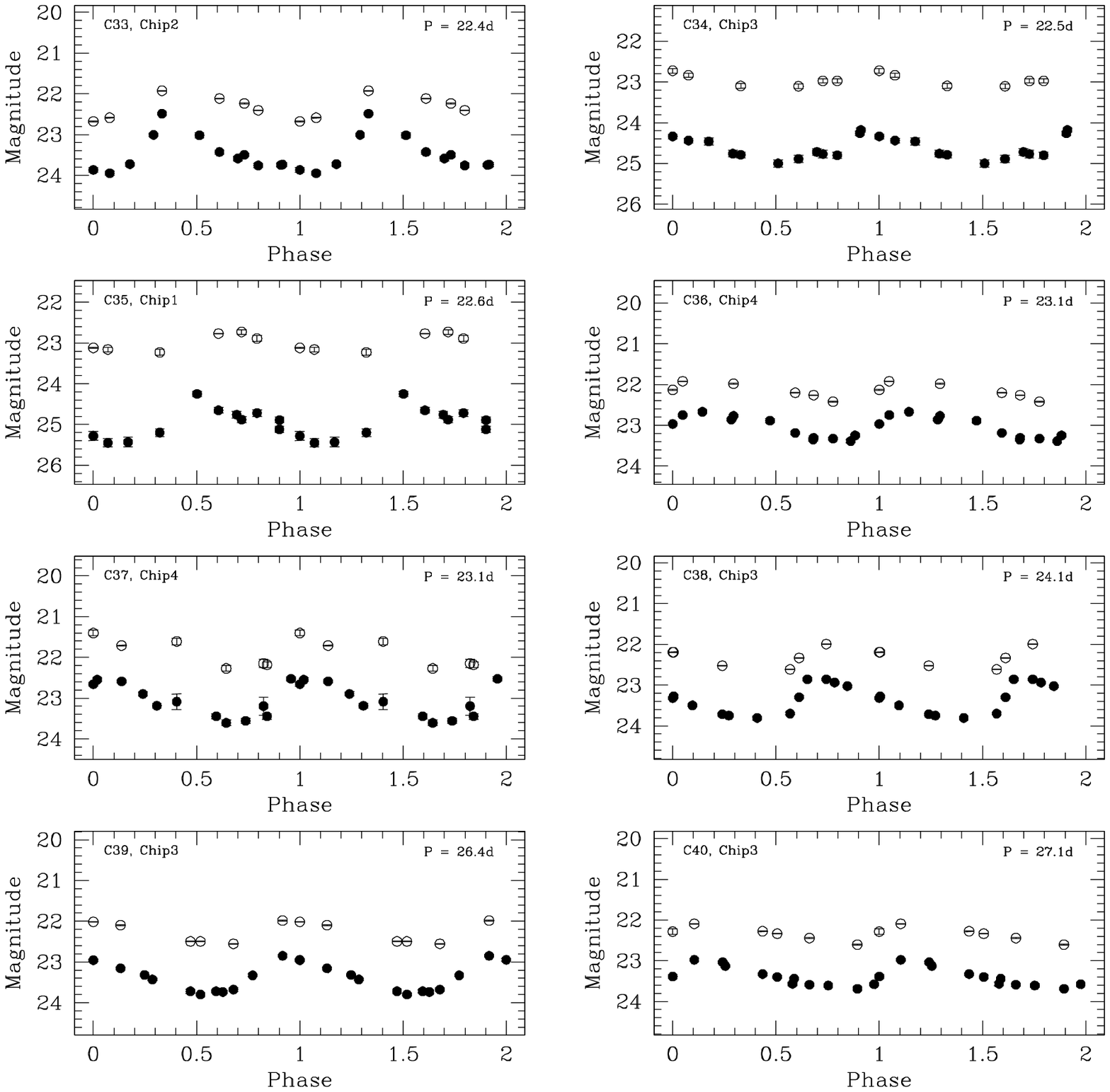}
\caption{Continued.
\label{lc5}}
\end{figure}

\clearpage

\addtocounter{figure}{-1}
\begin{figure}
\centering\plotone{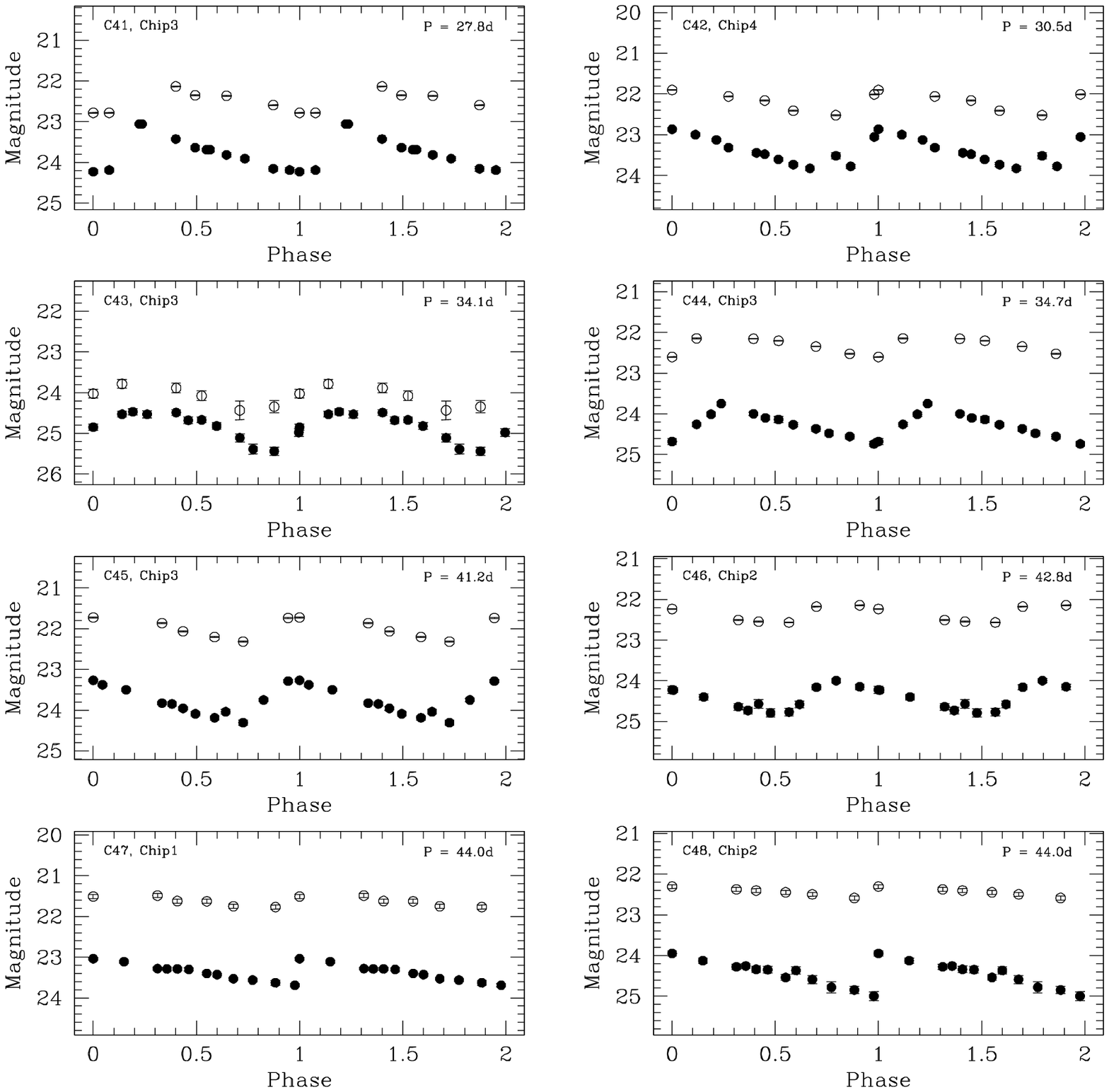}
\caption{Continued.
\label{lc6}}
\end{figure}

\clearpage

\addtocounter{figure}{-1}
\begin{figure}
\centering\plotone{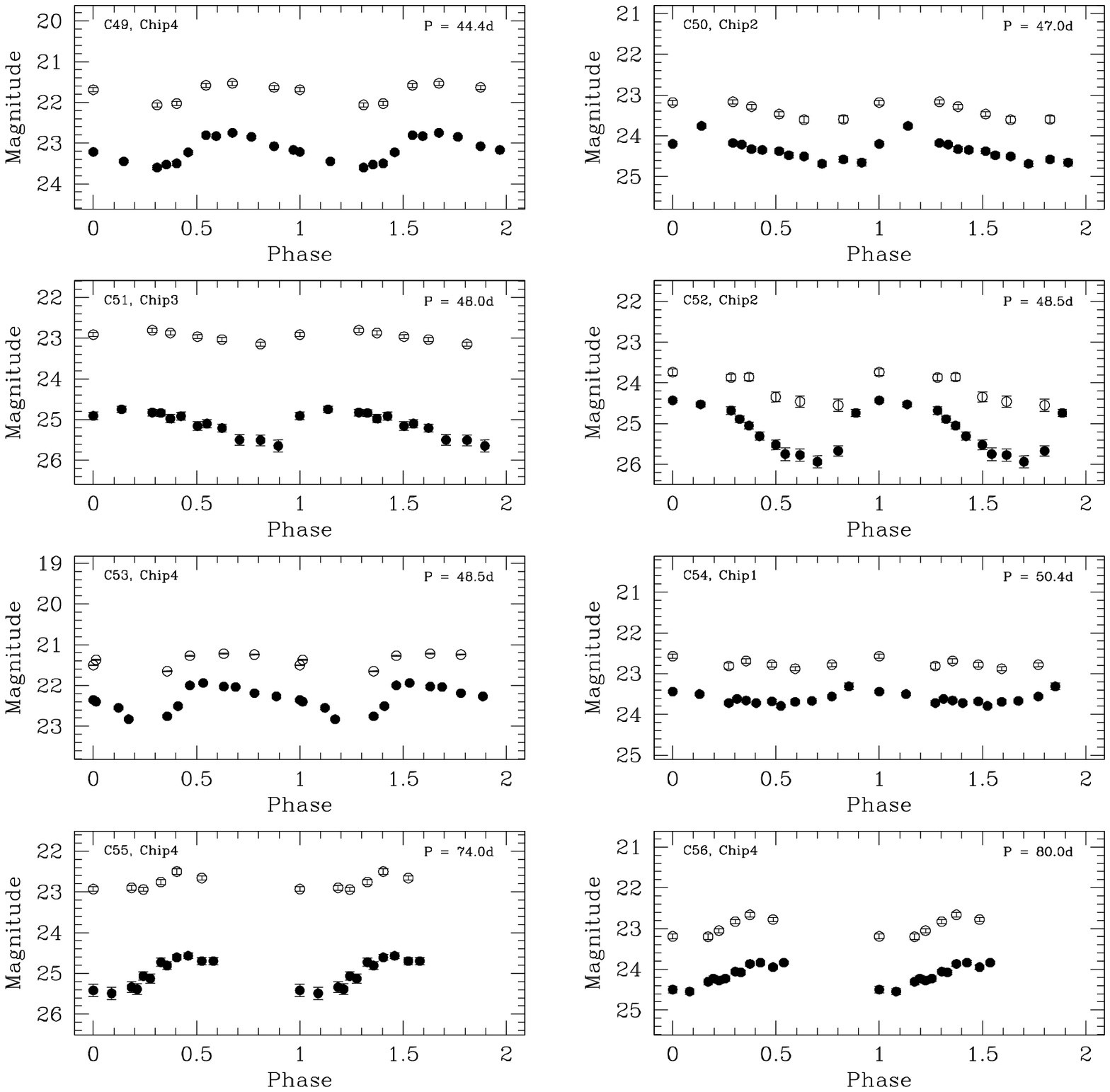}
\caption{Continued.
\label{lc7}}
\end{figure}

\clearpage

\renewcommand{\thefigure}{\arabic{figure}}

\begin{figure}
\centering\plotone{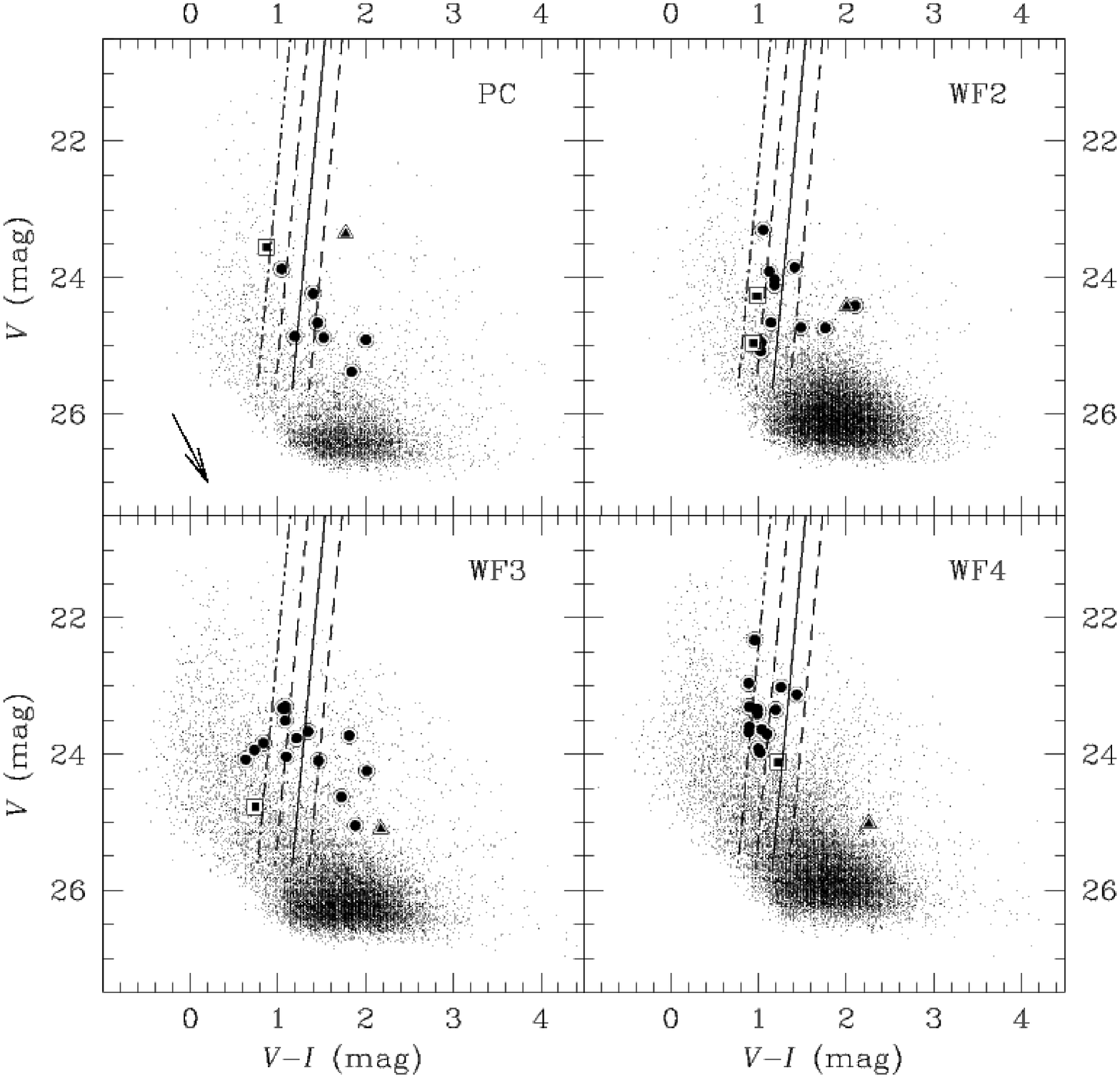}
\caption{$V$, $V-I$ Color-Magnitude diagrams for NGC 5128, shown
  separately for each WFPC2 chip. Cepheids used in fitting the PL
  relation are identified by large circles, while Cepheids for which
  only a lower limit on the period could be placed are plotted as
  triangles. The squares identify the Population II Cepheids discussed
  in the text. The solid line shows the ridge-line of the Cepheid
  instability strip, with width marked by the dashed lines, assuming a
  $V-$band distance modulus of 28.85 mag and reddening $E(V-I)=0.55$
  mag (see \S 5). The dot-dashed line shows the main ridgeline of the
  instability strip assuming $E(V-I)=0.152$ mag which, being equal to
  the Galactic reddening along the line of sight to NGC 5128,
  represents a lower limit to the amount of total reddening to the
  Cepheids. The arrow in the upper right panel shows where a star would
  move if subject to one magnitude of visual extinction (for $R_V =
  3.3$). The data are consistent with the assumption that Cepheids lie
  outside the instability strip as a consequence of the large and
  highly position dependent reddening of the  NGC 5128 field.
\label{cmd}}
\end{figure}

\clearpage 

\begin{figure}
\centering\plotone{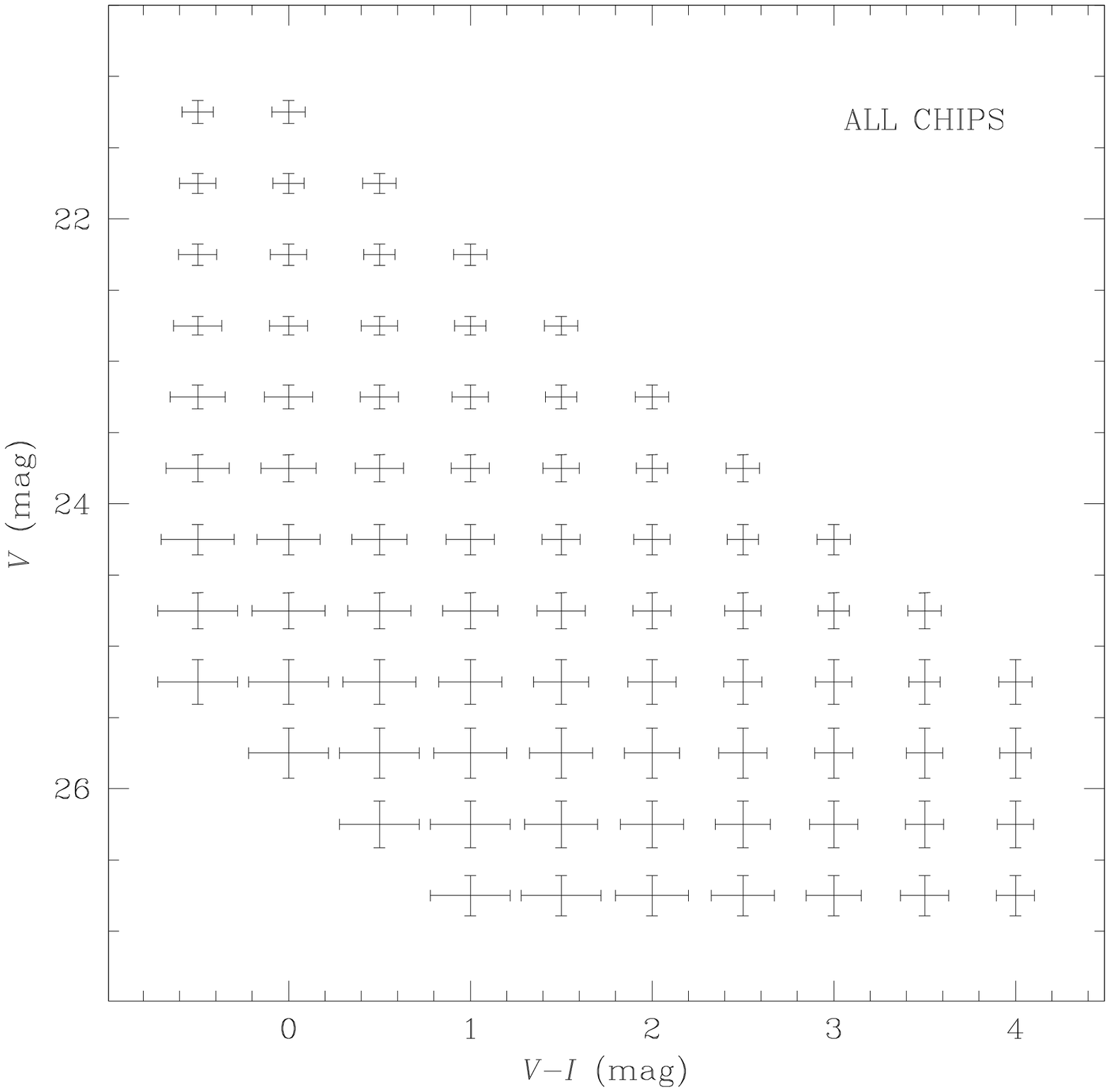}
\caption{Mean photometric errors associated with the DoPHOT photometry, as a
  function of $V$, $V-I$. The mean is calculated in 0.5 magnitudes
  bins, for both $V$ and $I$.
\label{cmderr}}
\end{figure}

\clearpage 
\begin{figure}
\centering\plotone{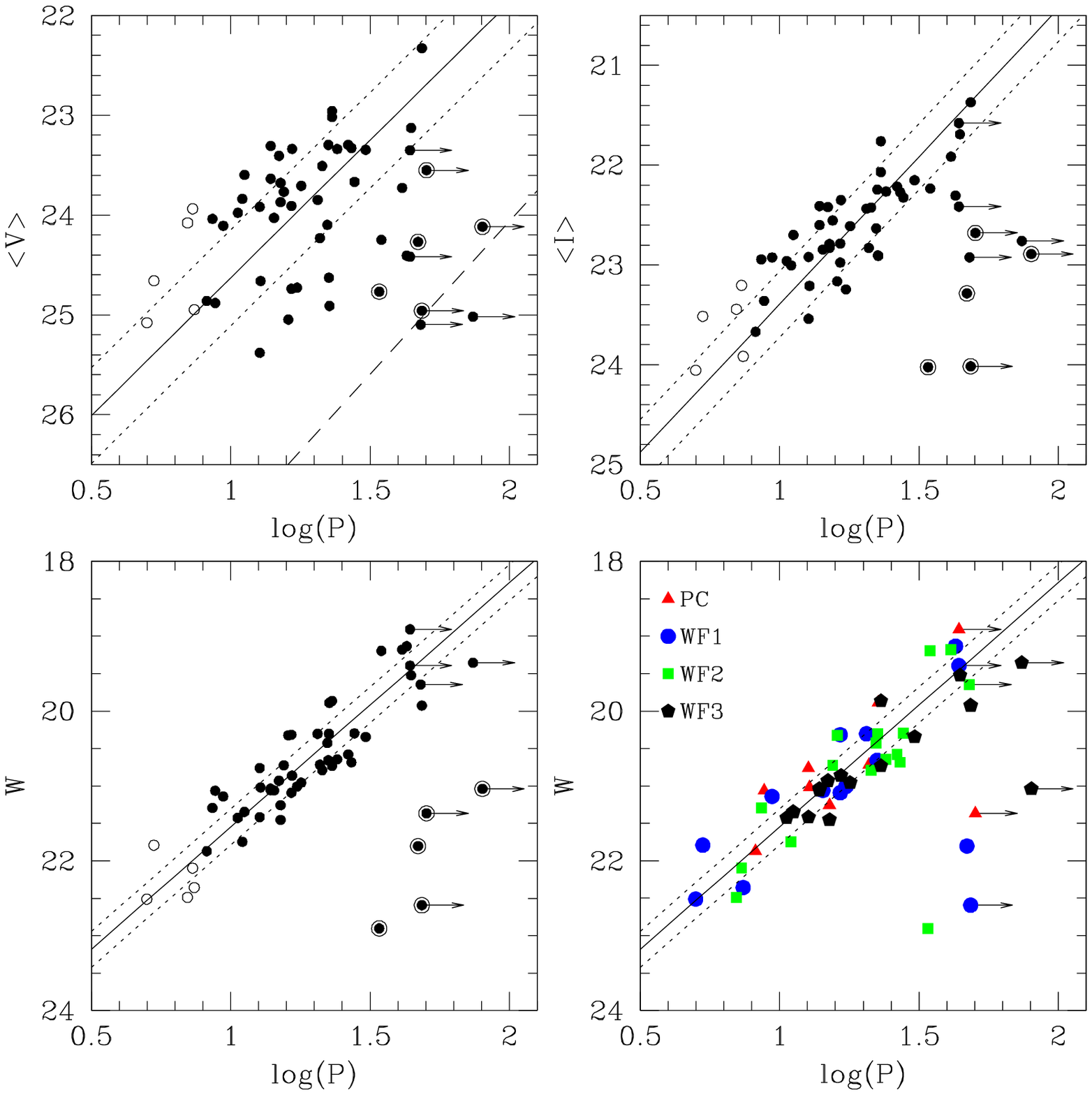}
\caption{$V$, $I$, and $W$ Period-Luminosity relations for the NGC
  5128 Cepheids. Magnitudes are DoPHOT phase-weighted mean
  magnitudes. Cepheids plotted as open circles have period less than 8
  days, and were excluded from the fits. Cepheids plotted as solid
  circles surrounded by a larger open circle, which are identified in
  the text as Pop II Cepheids,  deviate by more than 6$\sigma$ (two
  times the width of the instability strip) from the main ridge-line,
  and were also excluded from the fits. Periods of Cepheids plotted as
  a circle plus an arrow should be considered lower limits. In the
  bottom right panel, Cepheids are separated according to the chips in
  which they were detected. $W$ is calculated assuming
  $R_V$(LMC)=$R_V$(N5128)=3.3. The solid lines represent the Cepheids
  PL relations with slopes fixed at the LMC value, and scaled to
  $\mu_V=28.85$ mag;  $\mu_I=28.30$ mag, and  $\mu_0=27.48$ mag (Table
  6). The 3$\sigma$ confidence limit on these PL relations are shown by the
  dotted lines. The dashed line in the $V-$band PL relation shows the
  PL relation for Pop II Cepheids in the LMC from Alcock et al.\
  (1998), scaled to $\mu_V=28.85$ mag. The 3$\sigma$ confidence limit
  on this relation (not shown) is 1.32 mag.
\label{pl1}}
\end{figure}

\clearpage

\begin{figure}
\centering\plotone{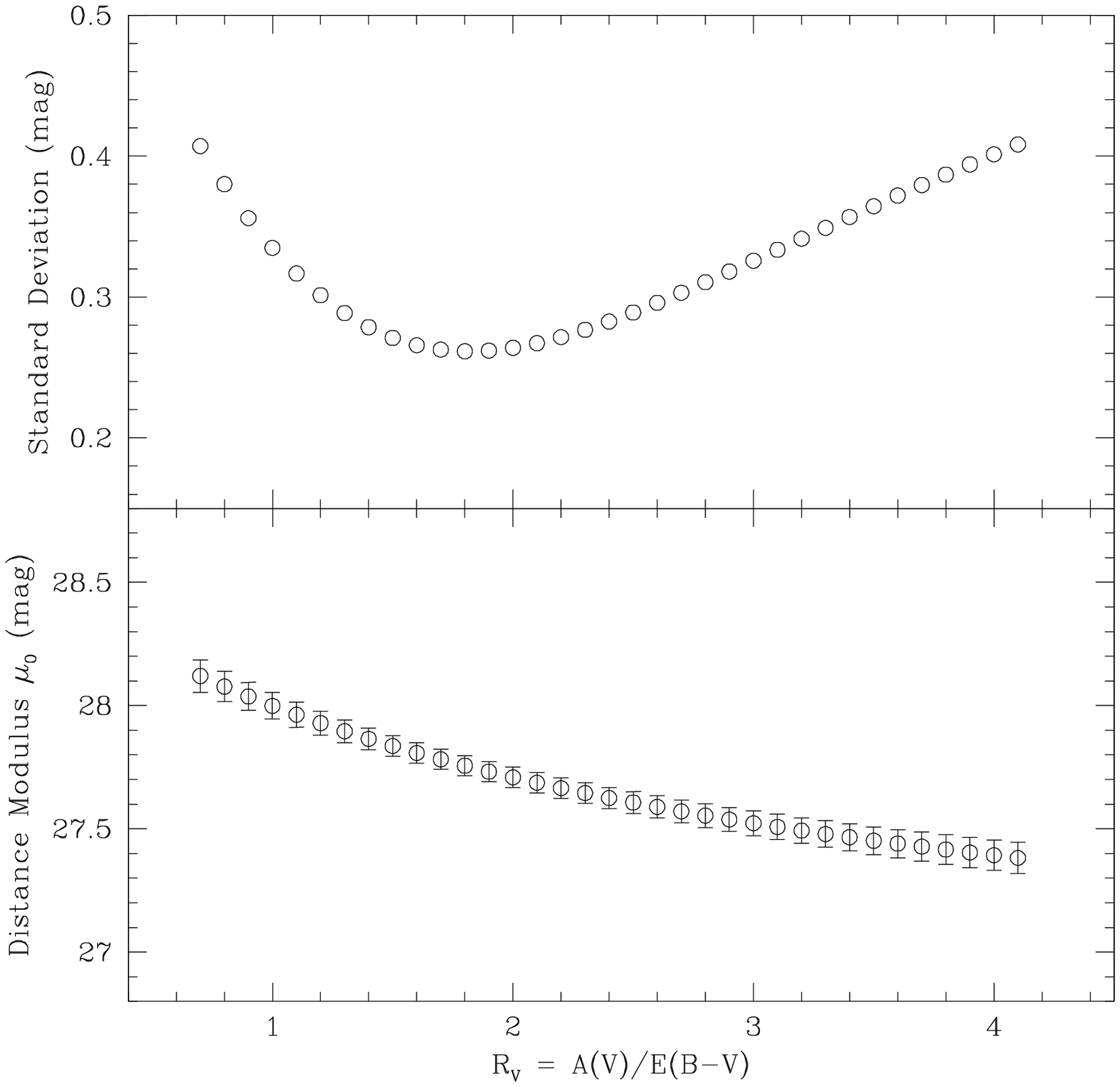}
\caption{Distance modulus (bottom panel) and standard deviation from
  the best fitting $W$ PL relation (upper panel) as a function of the
  $R_V$ value assumed for NGC 5128. See text for further details.
\label{rvfits}}
\end{figure}

\clearpage

\centerline{\bf{Appendix}}

\begin{deluxetable}{lrrcc}
\tabletypesize{\scriptsize}
\tablecaption{Secondary Standard Stars in NGC 5128.\label{tbl:secstd}}
\tablewidth{0pt}
\tablehead{
\colhead{Chip} &
\colhead{$X$} &
\colhead{$Y$} &
\colhead{$V$} &
\colhead{$I$}\\
\colhead{} &
\colhead{(pix)} &
\colhead{(pix)} &
\colhead{(mag)} &
\colhead{(mag)}
}
\startdata
PC & 673.4 & 236.3 & 21.24$\pm$ 0.01 & 19.99$\pm$ 0.02 \\ 
PC & 340.5 & 500.6 & 21.44$\pm$ 0.01 & 20.37$\pm$ 0.02 \\ 
PC & 331.1 & 540.7 & 21.36$\pm$ 0.01 & 20.40$\pm$ 0.02 \\ 
PC & 565.5 & 164.0 & 21.36$\pm$ 0.01 & 20.54$\pm$ 0.02 \\ 
PC & 150.0 & 415.8 & 22.41$\pm$ 0.01 & 19.85$\pm$ 0.02 \\ 
PC & 515.7 & 703.4 & 21.75$\pm$ 0.01 & 20.98$\pm$ 0.02 \\ 
PC & 715.9 & 300.4 & 22.62$\pm$ 0.01 & 20.57$\pm$ 0.02 \\ 
PC & 511.4 & 745.1 & 22.07$\pm$ 0.01 & 21.17$\pm$ 0.02 \\ 
PC & 279.3 & 142.1 & 21.97$\pm$ 0.01 & 21.18$\pm$ 0.02 \\ 
PC & 408.1 & 645.1 & 21.60$\pm$ 0.01 & 21.21$\pm$ 0.02 \\ 
PC & 678.8 & 225.7 & 21.73$\pm$ 0.01 & 21.36$\pm$ 0.02 \\ 
PC & 298.5 & 365.9 & 21.79$\pm$ 0.01 & 21.25$\pm$ 0.02 \\ 
PC & 193.0 & 397.4 & 22.51$\pm$ 0.01 & 21.28$\pm$ 0.02 \\ 
PC & 459.9 & 595.8 & 22.35$\pm$ 0.01 & 21.46$\pm$ 0.02 \\ 
PC & 261.3 & 331.0 & 22.01$\pm$ 0.01 & 21.49$\pm$ 0.02 \\ 
PC & 492.3 & 298.7 & 22.18$\pm$ 0.01 & 21.46$\pm$ 0.02 \\ 
PC & 726.0 & 108.4 & 21.80$\pm$ 0.02 & 21.62$\pm$ 0.02 \\ 
PC & 412.7 & 405.0 & 22.54$\pm$ 0.01 & 21.57$\pm$ 0.02 \\ 
PC & 400.3 & 433.3 & 23.09$\pm$ 0.01 & 21.08$\pm$ 0.03 \\ 
PC & 421.6 & 631.2 & 23.34$\pm$ 0.02 & 20.78$\pm$ 0.02 \\ 
PC & 599.1 & 682.0 & 22.62$\pm$ 0.01 & 21.70$\pm$ 0.02 \\ 
PC & 504.2 & 607.9 & 22.79$\pm$ 0.01 & 21.83$\pm$ 0.02 \\ 
PC & 659.1 & 317.3 & 22.54$\pm$ 0.01 & 21.93$\pm$ 0.02 \\ 
PC & 391.1 & 467.7 & 23.25$\pm$ 0.01 & 21.54$\pm$ 0.03 \\ 
PC & 376.0 & 129.1 & 23.65$\pm$ 0.02 & 20.96$\pm$ 0.02 \\ 
PC & 275.3 & 155.9 & 23.70$\pm$ 0.02 & 21.00$\pm$ 0.02 \\ 
PC & 494.4 & 141.5 & 23.19$\pm$ 0.01 & 21.92$\pm$ 0.02 \\ 
PC & 337.3 & 413.1 & 23.74$\pm$ 0.02 & 21.18$\pm$ 0.02 \\ 
PC & 106.8 & 295.2 & 22.72$\pm$ 0.02 & 22.08$\pm$ 0.03 \\ 
PC & 643.5 & 701.9 & 23.03$\pm$ 0.01 & 22.57$\pm$ 0.02 \\ 
PC & 537.4 & 723.2 & 22.98$\pm$ 0.02 & 22.57$\pm$ 0.03 \\ 
PC & 524.3 & 387.2 & 23.12$\pm$ 0.02 & 22.41$\pm$ 0.02 \\ 
PC & 362.9 & 725.9 & 24.05$\pm$ 0.02 & 20.95$\pm$ 0.02 \\ 
PC & 380.0 & 517.5 & 23.27$\pm$ 0.01 & 22.62$\pm$ 0.02 \\ 
PC & 164.7 & 545.7 & 23.51$\pm$ 0.02 & 22.60$\pm$ 0.03 \\ 
PC & 235.9 & 573.1 & 23.16$\pm$ 0.01 & 22.77$\pm$ 0.03 \\ 
WF2 & 497.4 & 557.0 & 21.31$\pm$ 0.01 & 18.84$\pm$ 0.07 \\ 
WF2 & 393.2 & 187.3 & 21.81$\pm$ 0.01 & 19.86$\pm$ 0.02 \\ 
WF2 & 525.7 & 776.4 & 21.33$\pm$ 0.02 & 20.29$\pm$ 0.03 \\ 
WF2 & 525.0 & 312.4 & 21.29$\pm$ 0.01 & 20.77$\pm$ 0.02 \\ 
WF2 & 148.8 & 393.7 & 22.31$\pm$ 0.01 & 20.28$\pm$ 0.02 \\ 
WF2 & 488.5 & 731.1 & 22.67$\pm$ 0.01 & 20.32$\pm$ 0.02 \\ 
WF2 & 683.0 & 401.1 & 22.32$\pm$ 0.01 & 21.03$\pm$ 0.02 \\ 
WF2 & 250.3 & 559.1 & 22.96$\pm$ 0.01 & 20.68$\pm$ 0.02 \\ 
WF2 & 130.9 & 486.5 & 22.98$\pm$ 0.01 & 20.59$\pm$ 0.02 \\ 
WF2 & 442.2 & 380.1 & 22.50$\pm$ 0.01 & 21.50$\pm$ 0.02 \\ 
WF2 & 168.5 & 765.4 & 23.34$\pm$ 0.01 & 21.17$\pm$ 0.02 \\ 
WF2 & 263.0 & 729.6 & 22.74$\pm$ 0.01 & 21.86$\pm$ 0.02 \\ 
WF2 & 764.6 & 214.0 & 22.72$\pm$ 0.02 & 21.85$\pm$ 0.02 \\ 
WF2 & 124.6 & 628.2 & 23.47$\pm$ 0.02 & 20.77$\pm$ 0.02 \\ 
WF2 & 533.2 & 344.2 & 22.37$\pm$ 0.01 & 22.05$\pm$ 0.02 \\ 
WF2 & 414.4 & 154.8 & 22.57$\pm$ 0.01 & 21.87$\pm$ 0.02 \\ 
WF2 & 229.6 & 356.1 & 23.22$\pm$ 0.02 & 21.61$\pm$ 0.02 \\ 
WF2 & 497.8 & 309.1 & 23.58$\pm$ 0.02 & 20.51$\pm$ 0.02 \\ 
WF2 &  77.4 & 277.6 & 23.23$\pm$ 0.02 & 21.67$\pm$ 0.02 \\ 
WF2 & 616.0 & 201.5 & 22.82$\pm$ 0.01 & 22.11$\pm$ 0.02 \\ 
WF2 & 366.8 & 525.0 & 23.13$\pm$ 0.01 & 21.84$\pm$ 0.02 \\ 
WF2 & 450.3 & 426.3 & 22.93$\pm$ 0.01 & 22.19$\pm$ 0.03 \\ 
WF2 & 542.5 & 762.1 & 22.64$\pm$ 0.01 & 22.13$\pm$ 0.02 \\ 
WF2 & 638.5 &  96.3 & 23.83$\pm$ 0.02 & 21.04$\pm$ 0.02 \\ 
WF2 & 283.7 & 634.5 & 23.90$\pm$ 0.02 & 21.58$\pm$ 0.02 \\ 
WF2 & 550.5 & 234.8 & 23.13$\pm$ 0.02 & 22.17$\pm$ 0.02 \\ 
WF2 & 367.0 & 718.8 & 23.73$\pm$ 0.02 & 21.74$\pm$ 0.02 \\ 
WF2 & 304.5 & 534.0 & 23.83$\pm$ 0.02 & 21.61$\pm$ 0.02 \\ 
WF2 & 299.3 & 306.8 & 24.24$\pm$ 0.02 & 20.83$\pm$ 0.02 \\ 
WF3 & 141.4 & 721.0 & 21.62$\pm$ 0.01 & 21.13$\pm$ 0.02 \\ 
WF3 & 189.3 & 181.9 & 22.22$\pm$ 0.01 & 20.62$\pm$ 0.02 \\ 
WF3 & 273.9 & 191.0 & 22.12$\pm$ 0.01 & 20.81$\pm$ 0.02 \\ 
WF3 & 360.0 & 576.9 & 22.04$\pm$ 0.01 & 21.19$\pm$ 0.02 \\ 
WF3 & 672.6 & 632.0 & 21.62$\pm$ 0.01 & 21.28$\pm$ 0.02 \\ 
WF3 & 611.8 & 244.4 & 22.84$\pm$ 0.01 & 20.45$\pm$ 0.02 \\ 
WF3 & 464.3 & 452.7 & 22.43$\pm$ 0.01 & 21.61$\pm$ 0.02 \\ 
WF3 & 596.9 & 512.9 & 22.26$\pm$ 0.01 & 21.71$\pm$ 0.02 \\ 
WF3 & 507.5 & 578.6 & 22.81$\pm$ 0.01 & 21.28$\pm$ 0.02 \\ 
WF3 & 249.7 & 189.8 & 23.01$\pm$ 0.01 & 21.08$\pm$ 0.02 \\ 
WF3 & 361.6 & 298.8 & 22.86$\pm$ 0.01 & 21.32$\pm$ 0.02 \\ 
WF3 & 466.0 & 247.2 & 22.93$\pm$ 0.01 & 21.29$\pm$ 0.02 \\ 
WF3 & 251.6 & 290.7 & 22.23$\pm$ 0.01 & 21.86$\pm$ 0.02 \\ 
WF3 & 368.3 & 256.3 & 23.02$\pm$ 0.01 & 21.35$\pm$ 0.02 \\ 
WF3 & 178.7 & 710.5 & 22.24$\pm$ 0.01 & 21.93$\pm$ 0.02 \\ 
WF3 & 370.2 & 650.5 & 22.41$\pm$ 0.01 & 22.21$\pm$ 0.02 \\ 
WF3 & 759.7 & 178.5 & 23.21$\pm$ 0.01 & 20.95$\pm$ 0.02 \\ 
WF3 & 657.5 & 229.4 & 23.23$\pm$ 0.01 & 21.32$\pm$ 0.02 \\ 
WF3 & 491.1 & 319.6 & 23.13$\pm$ 0.01 & 21.49$\pm$ 0.02 \\ 
WF3 & 592.8 &  80.0 & 22.92$\pm$ 0.02 & 22.09$\pm$ 0.03 \\ 
WF3 & 663.5 & 289.6 & 23.44$\pm$ 0.01 & 20.57$\pm$ 0.02 \\ 
WF3 & 243.1 & 740.8 & 23.33$\pm$ 0.01 & 21.26$\pm$ 0.02 \\ 
WF3 & 644.5 & 204.5 & 23.53$\pm$ 0.01 & 20.52$\pm$ 0.02 \\ 
WF3 & 312.6 & 417.3 & 23.59$\pm$ 0.01 & 20.16$\pm$ 0.02 \\ 
WF3 & 337.2 & 699.8 & 23.53$\pm$ 0.01 & 21.65$\pm$ 0.02 \\ 
WF3 & 351.3 & 717.9 & 23.55$\pm$ 0.01 & 21.83$\pm$ 0.02 \\ 
WF3 & 560.3 & 137.1 & 23.28$\pm$ 0.01 & 22.51$\pm$ 0.03 \\ 
WF3 & 568.5 & 341.3 & 23.25$\pm$ 0.01 & 22.31$\pm$ 0.02 \\ 
WF3 & 253.9 & 391.8 & 21.87$\pm$ 0.01 & 20.90$\pm$ 0.02 \\ 
WF3 & 351.7 & 379.5 & 21.55$\pm$ 0.01 & 21.26$\pm$ 0.02 \\ 
WF3 & 618.6 & 371.9 & 21.86$\pm$ 0.01 & 20.97$\pm$ 0.02 \\ 
WF3 & 600.0 & 384.5 & 22.78$\pm$ 0.01 & 20.13$\pm$ 0.02 \\ 
WF3 & 606.4 & 403.8 & 23.03$\pm$ 0.01 & 20.28$\pm$ 0.02 \\ 
WF3 & 551.7 & 455.8 & 22.38$\pm$ 0.01 & 20.92$\pm$ 0.02 \\ 
WF3 & 435.6 & 544.7 & 22.80$\pm$ 0.01 & 20.89$\pm$ 0.02 \\ 
WF3 & 167.4 & 449.3 & 22.57$\pm$ 0.01 & 19.95$\pm$ 0.02 \\ 
WF3 &  73.0 & 386.6 & 21.65$\pm$ 0.02 & 21.38$\pm$ 0.03 \\ 
WF3 & 129.1 & 705.8 & 22.30$\pm$ 0.01 & 20.92$\pm$ 0.02 \\ 
WF3 &  98.1 & 633.0 & 23.42$\pm$ 0.01 & 20.70$\pm$ 0.02 \\ 
WF3 & 471.8 & 596.2 & 23.56$\pm$ 0.02 & 21.16$\pm$ 0.02 \\ 
WF3 & 568.0 & 597.1 & 22.92$\pm$ 0.01 & 21.28$\pm$ 0.02 \\ 
WF3 & 511.7 & 597.6 & 22.60$\pm$ 0.01 & 22.57$\pm$ 0.02 \\ 
WF3 & 634.1 & 591.4 & 22.41$\pm$ 0.01 & 20.26$\pm$ 0.02 \\ 
WF3 & 702.0 & 549.8 & 22.71$\pm$ 0.01 & 21.95$\pm$ 0.02 \\ 
WF3 & 694.4 & 539.0 & 23.22$\pm$ 0.01 & 21.42$\pm$ 0.02 \\ 
WF3 & 201.0 & 770.0 & 22.91$\pm$ 0.01 & 21.49$\pm$ 0.02 \\ 
WF3 & 154.6 & 695.0 & 22.47$\pm$ 0.01 & 21.71$\pm$ 0.02 \\ 
WF3 & 294.5 & 703.8 & 22.72$\pm$ 0.01 & 22.27$\pm$ 0.02 \\ 
WF3 & 481.2 & 697.7 & 22.98$\pm$ 0.01 & 21.31$\pm$ 0.02 \\ 
WF3 & 500.5 & 777.9 & 22.13$\pm$ 0.01 & 21.93$\pm$ 0.02 \\ 
WF3 & 503.5 & 768.7 & 21.82$\pm$ 0.01 & 21.68$\pm$ 0.02 \\ 
WF3 & 726.1 & 718.1 & 22.00$\pm$ 0.01 & 21.80$\pm$ 0.02 \\ 
WF3 & 772.2 & 487.8 & 22.36$\pm$ 0.01 & 21.63$\pm$ 0.02 \\ 
WF3 & 618.2 & 568.7 & 22.47$\pm$ 0.01 & 22.28$\pm$ 0.02 \\ 
WF3 & 611.2 & 431.6 & 22.68$\pm$ 0.01 & 21.85$\pm$ 0.02 \\ 
WF3 & 388.1 & 560.3 & 22.09$\pm$ 0.01 & 21.89$\pm$ 0.02 \\ 
WF3 & 349.8 & 396.0 & 23.33$\pm$ 0.01 & 20.93$\pm$ 0.02 \\ 
WF3 & 466.3 & 265.5 & 22.92$\pm$ 0.01 & 22.90$\pm$ 0.03 \\ 
WF3 & 418.9 & 349.2 & 22.94$\pm$ 0.01 & 22.65$\pm$ 0.03 \\ 
WF3 & 476.0 & 354.3 & 23.27$\pm$ 0.01 & 21.71$\pm$ 0.02 \\ 
WF3 & 578.6 & 367.2 & 23.11$\pm$ 0.01 & 22.11$\pm$ 0.02 \\ 
WF3 & 353.5 & 156.5 & 23.73$\pm$ 0.02 & 21.68$\pm$ 0.02 \\ 
WF3 & 93.9 &  70.3 & 22.24$\pm$ 0.02 & 22.00$\pm$ 0.05 \\ 
WF4 & 514.1 & 201.0 & 22.42$\pm$ 0.01 & 20.61$\pm$ 0.01 \\ 
WF4 & 524.8 & 439.4 & 21.47$\pm$ 0.01 & 20.58$\pm$ 0.02 \\ 
WF4 & 544.7 & 440.4 & 22.20$\pm$ 0.06 & 21.27$\pm$ 0.05 \\ 
WF4 & 469.6 & 370.9 & 21.36$\pm$ 0.01 & 20.52$\pm$ 0.01 \\ 
WF4 & 305.8 & 368.9 & 21.43$\pm$ 0.01 & 20.76$\pm$ 0.02 \\ 
WF4 & 456.4 & 662.0 & 22.05$\pm$ 0.01 & 21.45$\pm$ 0.02 \\ 
WF4 & 501.2 & 149.1 & 21.91$\pm$ 0.01 & 21.30$\pm$ 0.02 \\ 
WF4 & 129.8 & 366.8 & 21.93$\pm$ 0.01 & 21.69$\pm$ 0.02 \\ 
WF4 & 418.8 & 228.7 & 21.60$\pm$ 0.01 & 21.11$\pm$ 0.02 \\ 
WF4 & 702.2 & 139.3 & 22.32$\pm$ 0.01 & 19.68$\pm$ 0.01 \\ 
WF4 & 351.2 & 324.6 & 22.50$\pm$ 0.01 & 20.03$\pm$ 0.01 \\ 
WF4 & 201.6 & 341.5 & 22.77$\pm$ 0.01 & 21.28$\pm$ 0.02 \\ 
WF4 &  90.0 & 535.7 & 22.76$\pm$ 0.01 & 21.47$\pm$ 0.02 \\ 
WF4 & 147.4 & 285.9 & 22.98$\pm$ 0.01 & 20.94$\pm$ 0.02 \\ 
WF4 & 223.7 & 108.2 & 22.20$\pm$ 0.01 & 21.18$\pm$ 0.02 \\ 
WF4 & 444.4 & 784.5 & 23.10$\pm$ 0.02 & 21.04$\pm$ 0.02 \\ 
WF4 & 284.5 & 159.5 & 23.55$\pm$ 0.01 & 19.89$\pm$ 0.01 \\ 
WF4 & 313.5 &  69.6 & 23.44$\pm$ 0.03 & 20.92$\pm$ 0.03 \\ 
WF4 & 661.4 &  86.8 & 23.25$\pm$ 0.02 & 20.91$\pm$ 0.02 \\ 
WF4 &  89.7 & 235.4 & 22.54$\pm$ 0.01 & 21.68$\pm$ 0.02 \\ 
WF4 &  75.6 & 145.1 & 23.93$\pm$ 0.02 & 21.42$\pm$ 0.02 \\ 
WF4 & 384.5 & 215.9 & 22.51$\pm$ 0.01 & 21.80$\pm$ 0.02 \\ 
WF4 & 478.5 & 192.3 & 23.41$\pm$ 0.02 & 20.88$\pm$ 0.02 \\ 
WF4 & 574.5 & 170.7 & 22.15$\pm$ 0.01 & 21.60$\pm$ 0.02 \\ 
WF4 & 735.1 & 152.9 & 23.53$\pm$ 0.02 & 21.35$\pm$ 0.02 \\ 
WF4 & 719.5 &  73.2 & 23.99$\pm$ 0.03 & 22.01$\pm$ 0.03 \\ 
WF4 & 577.4 & 259.0 & 23.01$\pm$ 0.01 & 20.79$\pm$ 0.01 \\ 
WF4 & 419.6 & 327.3 & 23.16$\pm$ 0.02 & 20.12$\pm$ 0.01 \\ 
WF4 & 419.2 & 279.9 & 22.21$\pm$ 0.01 & 21.95$\pm$ 0.02 \\ 
WF4 & 446.1 & 275.4 & 22.15$\pm$ 0.01 & 21.67$\pm$ 0.02 \\ 
WF4 & 325.4 & 266.8 & 23.65$\pm$ 0.02 & 21.32$\pm$ 0.02 \\ 
WF4 & 375.8 & 254.8 & 22.15$\pm$ 0.01 & 21.84$\pm$ 0.02 \\ 
WF4 & 273.9 & 322.2 & 23.36$\pm$ 0.01 & 21.70$\pm$ 0.02 \\ 
WF4 & 250.7 & 387.2 & 23.01$\pm$ 0.01 & 21.78$\pm$ 0.02 \\ 
WF4 & 214.1 & 363.5 & 22.98$\pm$ 0.01 & 22.61$\pm$ 0.03 \\ 
WF4 & 130.5 & 300.4 & 23.59$\pm$ 0.02 & 21.38$\pm$ 0.02 \\ 
WF4 & 129.5 & 219.9 & 22.71$\pm$ 0.01 & 21.52$\pm$ 0.02 \\ 
WF4 & 170.4 & 466.0 & 22.50$\pm$ 0.01 & 21.38$\pm$ 0.02 \\ 
WF4 & 337.9 & 532.2 & 22.88$\pm$ 0.01 & 20.32$\pm$ 0.02 \\ 
WF4 & 297.4 & 476.5 & 23.35$\pm$ 0.02 & 20.96$\pm$ 0.02 \\ 
WF4 & 485.8 & 554.7 & 22.94$\pm$ 0.01 & 22.73$\pm$ 0.02 \\ 
WF4 & 556.3 & 752.4 & 22.83$\pm$ 0.01 & 22.76$\pm$ 0.03 \\ 
WF4 & 484.5 & 730.3 & 23.70$\pm$ 0.02 & 21.20$\pm$ 0.02 \\ 
WF4 & 130.3 & 709.0 & 23.10$\pm$ 0.05 & 20.25$\pm$ 0.02 \\ 
WF4 & 152.4 & 717.3 & 22.01$\pm$ 0.01 & 21.83$\pm$ 0.02 \\ 
WF4 &  96.2 & 752.1 & 22.95$\pm$ 0.02 & 20.90$\pm$ 0.02 \\ 
WF4 & 110.6 & 614.0 & 21.70$\pm$ 0.01 & 21.40$\pm$ 0.02 \\ 
WF4 & 151.4 & 572.7 & 22.88$\pm$ 0.02 & 21.44$\pm$ 0.02 \\ 
\enddata
\tablecomments{$X$ and $Y$ position refer to the coordinate frame of the
  u6dm2101r observation (see Table~\ref{tbl:observations}).}
\end{deluxetable}

\clearpage

\LongTables
\begin{landscape}
\begin{deluxetable}{lccccccc}
\tabletypesize{\scriptsize}
\tablecaption{F555W Photometry for the NGC 5128 Cepheids.\label{tbl:cephphotv}}
\tablewidth{0pt}
\tablehead{
\colhead{} &
\colhead{} &
\colhead{} &
\colhead{} &
\colhead{F555W $\pm \delta$ F555W} &
\colhead{} &
\colhead{} &
\colhead{} 
}
\startdata
JD & C1; P =  5.0d & C2; P =  5.3d & C3; P =  7.0d & C4; P =  7.3d & C5; P =  7.4d & C6; P =  8.2d & C7; P =  8.6d \\ 
   \hline \\ 
2452099.00 & 25.38 $\pm$ 0.14 & 24.36 $\pm$ 0.07 & 24.32 $\pm$ 0.06 & 23.62 $\pm$ 0.04 & 25.07 $\pm$ 0.09 & 25.14 $\pm$ 0.09 & 23.93 $\pm$ 0.05 \\ 
2452105.50 & 24.91 $\pm$ 0.10 & 24.57 $\pm$ 0.07 & 24.25 $\pm$ 0.06 & 23.99 $\pm$ 0.05 & 25.08 $\pm$ 0.09 & 25.13 $\pm$ 0.07 & 23.82 $\pm$ 0.04 \\ 
2452112.50 & 25.55 $\pm$ 0.15 & 24.88 $\pm$ 0.08 & 24.31 $\pm$ 0.07 & 24.01 $\pm$ 0.05 & 25.35 $\pm$ 0.12 & 24.77 $\pm$ 0.05 & 24.38 $\pm$ 0.06 \\ 
2452114.50 & 24.66 $\pm$ 0.08 & 24.42 $\pm$ 0.07 & 24.56 $\pm$ 0.08 & 23.80 $\pm$ 0.05 & 24.69 $\pm$ 0.09 & 25.43 $\pm$ 0.11 & 23.77 $\pm$ 0.06 \\ 
2452116.75 & 25.17 $\pm$ 0.13 & 24.80 $\pm$ 0.08 & 23.83 $\pm$ 0.04 & 24.09 $\pm$ 0.05 & 24.90 $\pm$ 0.09 & 24.67 $\pm$ 0.06 & 23.91 $\pm$ 0.05 \\ 
2452119.25 & 25.01 $\pm$ 0.11 & 24.97 $\pm$ 0.09 & 24.27 $\pm$ 0.06 & 24.21 $\pm$ 0.06 & 25.26 $\pm$ 0.11 & 24.80 $\pm$ 0.06 & 24.27 $\pm$ 0.06 \\ 
2452123.00 & 25.62 $\pm$ 0.15 & 24.69 $\pm$ 0.10 & 23.69 $\pm$ 0.04 & 23.89 $\pm$ 0.04 & 24.68 $\pm$ 0.08 & 25.12 $\pm$ 0.08 & 23.78 $\pm$ 0.05 \\ 
2452125.25 & 24.77 $\pm$ 0.08 & 24.27 $\pm$ 0.06 & 24.11 $\pm$ 0.06 & 24.17 $\pm$ 0.05 & 25.10 $\pm$ 0.11 & 24.51 $\pm$ 0.05 & 23.89 $\pm$ 0.05 \\ 
2452128.75 & 25.32 $\pm$ 0.14 & 24.99 $\pm$ 0.11 & 24.40 $\pm$ 0.08 & 23.71 $\pm$ 0.05 & 24.97 $\pm$ 0.09 & 24.91 $\pm$ 0.07 & 24.36 $\pm$ 0.07 \\ 
2452133.00 & 25.48 $\pm$ 0.16 & 24.69 $\pm$ 0.08 & 24.03 $\pm$ 0.05 & 24.20 $\pm$ 0.07 & 25.01 $\pm$ 0.09 & 24.80 $\pm$ 0.06 & 23.86 $\pm$ 0.04 \\ 
2452137.75 & 25.38 $\pm$ 0.13 & 24.68 $\pm$ 0.08 & 23.84 $\pm$ 0.06 & 23.89 $\pm$ 0.04 & 24.69 $\pm$ 0.09 & 25.00 $\pm$ 0.08 & 24.53 $\pm$ 0.08 \\ 
2452142.00 & 25.14 $\pm$ 0.11 & 24.43 $\pm$ 0.06 & 24.37 $\pm$ 0.07 & 24.07 $\pm$ 0.05 & 25.58 $\pm$ 0.15 & 24.56 $\pm$ 0.06 & 23.93 $\pm$ 0.05 \\ 
   \hline \\
& \multicolumn{7}{c}{F555W $\pm \Delta$ F555W} \\ 
   \hline \\ 
JD & C8; P =  8.8d & C9; P =  9.4d & C10; P = 10.6d & C11; P = 11.0d & C12; P = 11.2d & C13; P = 12.7d & C14; P = 12.7d \\ 
   \hline \\ 
2452099.00 & 24.70 $\pm$ 0.07 & 23.99 $\pm$ 0.05 & 24.31 $\pm$ 0.07 & 23.88 $\pm$ 0.04 & 23.64 $\pm$ 0.05 & 25.14 $\pm$ 0.08 & 24.20 $\pm$ 0.08 \\ 
2452105.50 & 25.18 $\pm$ 0.09 & 24.32 $\pm$ 0.07 & 23.77 $\pm$ 0.04 & 23.63 $\pm$ 0.04 & 23.49 $\pm$ 0.04 & 26.05 $\pm$ 0.18 & 23.55 $\pm$ 0.04 \\ 
2452112.50 & 25.08 $\pm$ 0.09 & 24.12 $\pm$ 0.06 & 23.87 $\pm$ 0.05 & 24.10 $\pm$ 0.06 & 23.81 $\pm$ 0.05 & 25.29 $\pm$ 0.12 & 24.26 $\pm$ 0.08 \\ 
2452114.50 & 25.21 $\pm$ 0.08 & 24.40 $\pm$ 0.06 & 24.02 $\pm$ 0.09 & 24.03 $\pm$ 0.05 & 23.95 $\pm$ 0.06 & 25.31 $\pm$ 0.11 & 24.46 $\pm$ 0.09 \\ 
2452116.75 & 24.58 $\pm$ 0.05 & 24.21 $\pm$ 0.07 & 24.01 $\pm$ 0.05 & 23.67 $\pm$ 0.04 & 23.45 $\pm$ 0.04 & 25.87 $\pm$ 0.16 & 24.05 $\pm$ 0.06 \\ 
2452119.25 & 24.64 $\pm$ 0.06 & 23.92 $\pm$ 0.06 & 24.27 $\pm$ 0.07 & 23.75 $\pm$ 0.04 & 23.41 $\pm$ 0.03 & 25.64 $\pm$ 0.12 & 23.48 $\pm$ 0.04 \\ 
2452123.00 & 25.25 $\pm$ 0.10 & 24.21 $\pm$ 0.06 & 23.83 $\pm$ 0.05 & 24.30 $\pm$ 0.05 & 23.82 $\pm$ 0.05 & 25.03 $\pm$ 0.07 & 23.96 $\pm$ 0.05 \\ 
2452125.25 & 24.82 $\pm$ 0.07 & 24.36 $\pm$ 0.07 & 23.85 $\pm$ 0.05 & 24.05 $\pm$ 0.04 & 23.96 $\pm$ 0.05 & 25.24 $\pm$ 0.10 & 24.26 $\pm$ 0.09 \\ 
2452128.75 & 24.81 $\pm$ 0.09 & -5.03 $\pm$ 0.00 & 24.16 $\pm$ 0.06 & 23.56 $\pm$ 0.03 & 23.30 $\pm$ 0.03 & 25.64 $\pm$ 0.16 & 24.14 $\pm$ 0.06 \\ 
2452133.00 & 25.10 $\pm$ 0.09 & 24.30 $\pm$ 0.07 & 23.88 $\pm$ 0.05 & 24.13 $\pm$ 0.04 & 23.65 $\pm$ 0.05 & 25.33 $\pm$ 0.10 & 23.61 $\pm$ 0.04 \\ 
2452137.75 & 24.92 $\pm$ 0.08 & 23.92 $\pm$ 0.05 & 24.02 $\pm$ 0.06 & 23.60 $\pm$ 0.04 & 23.74 $\pm$ 0.05 & 25.31 $\pm$ 0.10 & 24.15 $\pm$ 0.08 \\ 
2452142.00 & 25.27 $\pm$ 0.11 & 24.23 $\pm$ 0.06 & 24.07 $\pm$ 0.06 & 23.86 $\pm$ 0.04 & 23.46 $\pm$ 0.04 & 25.65 $\pm$ 0.15 & 24.54 $\pm$ 0.14 \\ 
   \hline \\
& \multicolumn{7}{c}{F555W $\pm \Delta$ F555W} \\ 
   \hline \\ 
JD & C15; P = 12.8d & C16; P = 13.9d & C17; P = 13.9d & C18; P = 14.3d & C19; P = 14.9d & C20; P = 15.1d & C21; P = 15.1d \\ 
   \hline \\ 
2452099.00 & 24.79 $\pm$ 0.06 & 23.36 $\pm$ 0.03 & 23.82 $\pm$ 0.06 & 24.44 $\pm$ 0.07 & 23.45 $\pm$ 0.04 & 23.83 $\pm$ 0.04 & 24.11 $\pm$ 0.08 \\ 
2452105.50 & 24.74 $\pm$ 0.06 & 23.39 $\pm$ 0.04 & 23.57 $\pm$ 0.05 & 23.53 $\pm$ 0.05 & 23.57 $\pm$ 0.04 & 24.29 $\pm$ 0.06 & 23.51 $\pm$ 0.04 \\ 
2452112.50 & 24.58 $\pm$ 0.06 & 23.41 $\pm$ 0.05 & 23.77 $\pm$ 0.05 & 24.46 $\pm$ 0.07 & 23.37 $\pm$ 0.04 & 23.68 $\pm$ 0.04 & 24.19 $\pm$ 0.08 \\ 
2452114.50 & 24.25 $\pm$ 0.05 & 22.98 $\pm$ 0.03 & 23.58 $\pm$ 0.04 & 24.46 $\pm$ 0.09 & 23.61 $\pm$ 0.05 & 23.80 $\pm$ 0.05 & 23.97 $\pm$ 0.06 \\ 
2452116.75 & 24.58 $\pm$ 0.05 & 23.16 $\pm$ 0.04 & 23.29 $\pm$ 0.04 & 24.20 $\pm$ 0.06 & 23.87 $\pm$ 0.06 & 24.07 $\pm$ 0.04 & 23.31 $\pm$ 0.04 \\ 
2452119.25 & 24.80 $\pm$ 0.09 & 23.40 $\pm$ 0.04 & 23.56 $\pm$ 0.05 & 23.47 $\pm$ 0.04 & 23.70 $\pm$ 0.05 & 24.36 $\pm$ 0.06 & 23.38 $\pm$ 0.04 \\ 
2452123.00 & 24.92 $\pm$ 0.06 & 23.60 $\pm$ 0.05 & 23.94 $\pm$ 0.05 & 24.03 $\pm$ 0.07 & 22.97 $\pm$ 0.04 & 24.02 $\pm$ 0.04 & 23.80 $\pm$ 0.06 \\ 
2452125.25 & 24.68 $\pm$ 0.08 & 23.45 $\pm$ 0.04 & 24.03 $\pm$ 0.06 & 24.21 $\pm$ 0.07 & 23.23 $\pm$ 0.05 & 23.40 $\pm$ 0.04 & 24.04 $\pm$ 0.07 \\ 
2452128.75 & 24.48 $\pm$ 0.05 & 22.96 $\pm$ 0.03 & 23.38 $\pm$ 0.04 & 24.57 $\pm$ 0.07 & 23.54 $\pm$ 0.04 & 23.77 $\pm$ 0.04 & 24.17 $\pm$ 0.07 \\ 
2452133.00 & 24.97 $\pm$ 0.07 & 23.40 $\pm$ 0.05 & 23.52 $\pm$ 0.04 & 23.80 $\pm$ 0.07 & 23.70 $\pm$ 0.04 & 24.24 $\pm$ 0.04 & 23.21 $\pm$ 0.04 \\ 
2452137.75 & 24.72 $\pm$ 0.07 & 23.67 $\pm$ 0.07 & 24.11 $\pm$ 0.07 & 24.04 $\pm$ 0.06 & 23.04 $\pm$ 0.04 & 23.99 $\pm$ 0.05 & 23.72 $\pm$ 0.05 \\ 
2452142.00 & 24.49 $\pm$ 0.06 & 23.02 $\pm$ 0.03 & 23.69 $\pm$ 0.05 & 24.68 $\pm$ 0.12 & 23.45 $\pm$ 0.04 & 23.54 $\pm$ 0.04 & 24.28 $\pm$ 0.07 \\ 
   \hline \\
& \multicolumn{7}{c}{F555W $\pm \Delta$ F555W} \\ 
   \hline \\ 
JD & C22; P = 15.5d & C23; P = 16.1d & C24; P = 16.5d & C25; P = 16.5d & C26; P = 16.6d & C27; P = 17.3d & C28; P = 17.9d \\ 
   \hline \\ 
2452099.00 & 23.60 $\pm$ 0.07 & 24.67 $\pm$ 0.06 & 24.20 $\pm$ 0.09 & 24.56 $\pm$ 0.06 & 23.72 $\pm$ 0.06 & 24.51 $\pm$ 0.05 & 24.06 $\pm$ 0.06 \\ 
2452105.50 & 23.81 $\pm$ 0.04 & 25.13 $\pm$ 0.10 & 23.47 $\pm$ 0.05 & 25.11 $\pm$ 0.08 & 22.90 $\pm$ 0.04 & 25.37 $\pm$ 0.09 & 23.42 $\pm$ 0.06 \\ 
2452112.50 & 24.00 $\pm$ 0.04 & 25.22 $\pm$ 0.09 & 24.07 $\pm$ 0.06 & 24.32 $\pm$ 0.05 & 23.60 $\pm$ 0.05 & 24.66 $\pm$ 0.06 & 23.81 $\pm$ 0.06 \\ 
2452114.50 & 23.54 $\pm$ 0.05 & 24.90 $\pm$ 0.08 & 24.33 $\pm$ 0.06 & 24.53 $\pm$ 0.05 & 23.72 $\pm$ 0.06 & 24.83 $\pm$ 0.09 & 23.94 $\pm$ 0.05 \\ 
2452116.75 & 23.40 $\pm$ 0.04 & 24.56 $\pm$ 0.07 & 24.25 $\pm$ 0.06 & 24.70 $\pm$ 0.07 & 23.71 $\pm$ 0.06 & 24.89 $\pm$ 0.10 & 24.03 $\pm$ 0.06 \\ 
2452119.25 & 23.68 $\pm$ 0.05 & 24.93 $\pm$ 0.07 & 24.15 $\pm$ 0.07 & 25.07 $\pm$ 0.10 & 23.61 $\pm$ 0.05 & 25.16 $\pm$ 0.13 & 23.92 $\pm$ 0.05 \\ 
2452123.00 & 23.98 $\pm$ 0.05 & 25.24 $\pm$ 0.10 & 23.63 $\pm$ 0.05 & 25.10 $\pm$ 0.09 & 22.95 $\pm$ 0.04 & 25.08 $\pm$ 0.08 & 23.44 $\pm$ 0.04 \\ 
2452125.25 & 24.15 $\pm$ 0.05 & 25.44 $\pm$ 0.12 & 23.80 $\pm$ 0.04 & 24.90 $\pm$ 0.08 & 23.19 $\pm$ 0.05 & 24.24 $\pm$ 0.08 & 23.44 $\pm$ 0.05 \\ 
2452128.75 & 23.90 $\pm$ 0.04 & 25.26 $\pm$ 0.10 & 24.08 $\pm$ 0.05 & 24.28 $\pm$ 0.04 & 23.55 $\pm$ 0.04 & 24.62 $\pm$ 0.08 & 23.77 $\pm$ 0.04 \\ 
2452133.00 & 23.45 $\pm$ 0.03 & 24.65 $\pm$ 0.06 & 24.27 $\pm$ 0.06 & 24.67 $\pm$ 0.07 & 23.83 $\pm$ 0.07 & 24.67 $\pm$ 0.09 & 24.01 $\pm$ 0.06 \\ 
2452137.75 & 23.92 $\pm$ 0.04 & 25.15 $\pm$ 0.11 & 23.57 $\pm$ 0.05 & 25.08 $\pm$ 0.10 & 22.98 $\pm$ 0.04 & 25.03 $\pm$ 0.09 & 23.83 $\pm$ 0.06 \\ 
2452142.00 & 24.20 $\pm$ 0.05 & 25.56 $\pm$ 0.14 & 23.73 $\pm$ 0.04 & 24.79 $\pm$ 0.06 & 23.23 $\pm$ 0.05 & 24.41 $\pm$ 0.06 & 23.39 $\pm$ 0.04 \\ 
   \hline \\
& \multicolumn{7}{c}{F555W $\pm \Delta$ F555W} \\ 
   \hline \\ 
JD & C29; P = 20.5d & C30; P = 20.9d & C31; P = 21.3d & C32; P = 22.2d & C33; P = 22.4d & C34; P = 22.5d & C35; P = 22.6d \\ 
   \hline \\ 
2452099.00 & 24.18 $\pm$ 0.05 & 24.36 $\pm$ 0.05 & 23.99 $\pm$ 0.04 & 23.73 $\pm$ 0.07 & 23.87 $\pm$ 0.06 & 24.34 $\pm$ 0.06 & 25.28 $\pm$ 0.11 \\ 
2452105.50 & 23.89 $\pm$ 0.05 & 24.69 $\pm$ 0.06 & 22.99 $\pm$ 0.03 & 24.11 $\pm$ 0.05 & 23.01 $\pm$ 0.04 & 24.76 $\pm$ 0.08 & -5.06 $\pm$ 0.00 \\ 
2452112.50 & 23.65 $\pm$ 0.04 & 23.80 $\pm$ 0.04 & 23.55 $\pm$ 0.03 & 24.66 $\pm$ 0.07 & 23.43 $\pm$ 0.05 & 24.89 $\pm$ 0.09 & 24.65 $\pm$ 0.07 \\ 
2452114.50 & 23.81 $\pm$ 0.05 & 23.97 $\pm$ 0.04 & 23.73 $\pm$ 0.04 & 24.60 $\pm$ 0.08 & 23.59 $\pm$ 0.06 & 24.72 $\pm$ 0.08 & 24.76 $\pm$ 0.08 \\ 
2452116.75 & 24.06 $\pm$ 0.05 & 24.07 $\pm$ 0.06 & 23.89 $\pm$ 0.04 & 24.44 $\pm$ 0.06 & 23.76 $\pm$ 0.05 & 24.80 $\pm$ 0.08 & 24.72 $\pm$ 0.08 \\ 
2452119.25 & 24.23 $\pm$ 0.05 & 24.20 $\pm$ 0.05 & 23.99 $\pm$ 0.04 & 23.49 $\pm$ 0.04 & 23.75 $\pm$ 0.04 & 24.26 $\pm$ 0.05 & 24.89 $\pm$ 0.07 \\ 
2452123.00 & 24.36 $\pm$ 0.06 & 24.53 $\pm$ 0.05 & 23.86 $\pm$ 0.04 & 23.73 $\pm$ 0.04 & 23.95 $\pm$ 0.05 & 24.44 $\pm$ 0.06 & 25.45 $\pm$ 0.10 \\ 
2452125.25 & 24.27 $\pm$ 0.06 & 24.71 $\pm$ 0.07 & 23.26 $\pm$ 0.03 & 23.92 $\pm$ 0.05 & 23.73 $\pm$ 0.04 & 24.46 $\pm$ 0.09 & 25.43 $\pm$ 0.12 \\ 
2452128.75 & 23.34 $\pm$ 0.03 & 24.61 $\pm$ 0.07 & 23.16 $\pm$ 0.03 & 24.37 $\pm$ 0.09 & 22.49 $\pm$ 0.03 & 24.79 $\pm$ 0.07 & 25.20 $\pm$ 0.09 \\ 
2452133.00 & 23.65 $\pm$ 0.04 & 23.75 $\pm$ 0.04 & 23.48 $\pm$ 0.03 & 24.46 $\pm$ 0.06 & 23.02 $\pm$ 0.06 & 25.00 $\pm$ 0.09 & 24.25 $\pm$ 0.06 \\ 
2452137.75 & 24.22 $\pm$ 0.17 & 24.09 $\pm$ 0.05 & 23.82 $\pm$ 0.04 & 24.61 $\pm$ 0.07 & 23.50 $\pm$ 0.04 & 24.77 $\pm$ 0.09 & 24.88 $\pm$ 0.08 \\ 
2452142.00 & 24.22 $\pm$ 0.05 & 24.41 $\pm$ 0.06 & 23.99 $\pm$ 0.04 & 23.55 $\pm$ 0.04 & 23.74 $\pm$ 0.05 & 24.18 $\pm$ 0.05 & 25.12 $\pm$ 0.08 \\ 
   \hline \\
& \multicolumn{7}{c}{F555W $\pm \Delta$ F555W} \\ 
   \hline \\ 
JD & C36; P = 23.1d & C37; P = 23.1d & C38; P = 24.1d & C39; P = 26.4d & C40; P = 27.1d & C41; P = 27.8d & C42; P = 30.5d \\ 
   \hline \\ 
2452099.00 & 22.97 $\pm$ 0.03 & 22.66 $\pm$ 0.04 & 23.32 $\pm$ 0.03 & 22.96 $\pm$ 0.03 & 23.39 $\pm$ 0.05 & 24.23 $\pm$ 0.05 & 22.87 $\pm$ 0.04 \\ 
2452105.50 & 22.86 $\pm$ 0.03 & 23.19 $\pm$ 0.06 & 23.75 $\pm$ 0.04 & 23.32 $\pm$ 0.03 & 23.04 $\pm$ 0.03 & 23.06 $\pm$ 0.03 & 23.13 $\pm$ 0.04 \\ 
2452112.50 & 23.19 $\pm$ 0.03 & 23.61 $\pm$ 0.07 & 23.70 $\pm$ 0.03 & 23.80 $\pm$ 0.04 & 23.40 $\pm$ 0.03 & 23.64 $\pm$ 0.05 & 23.48 $\pm$ 0.04 \\ 
2452114.50 & 23.35 $\pm$ 0.04 & 23.56 $\pm$ 0.06 & 22.86 $\pm$ 0.04 & 23.72 $\pm$ 0.04 & 23.57 $\pm$ 0.04 & 23.69 $\pm$ 0.04 & 23.61 $\pm$ 0.04 \\ 
2452116.75 & 23.33 $\pm$ 0.03 & 23.45 $\pm$ 0.07 & 22.86 $\pm$ 0.03 & 23.68 $\pm$ 0.04 & 23.59 $\pm$ 0.04 & 23.82 $\pm$ 0.05 & 23.74 $\pm$ 0.05 \\ 
2452119.25 & 23.25 $\pm$ 0.04 & 22.53 $\pm$ 0.05 & 23.03 $\pm$ 0.03 & 23.33 $\pm$ 0.03 & 23.61 $\pm$ 0.04 & 23.91 $\pm$ 0.05 & 23.83 $\pm$ 0.05 \\ 
2452123.00 & 22.75 $\pm$ 0.03 & 22.59 $\pm$ 0.04 & 23.28 $\pm$ 0.03 & 22.85 $\pm$ 0.03 & 23.69 $\pm$ 0.03 & 24.16 $\pm$ 0.06 & 23.52 $\pm$ 0.08 \\ 
2452125.25 & 22.67 $\pm$ 0.03 & 22.90 $\pm$ 0.05 & 23.50 $\pm$ 0.03 & 22.95 $\pm$ 0.03 & 23.58 $\pm$ 0.04 & 24.19 $\pm$ 0.06 & 23.78 $\pm$ 0.05 \\ 
2452128.75 & 22.77 $\pm$ 0.03 & 23.09 $\pm$ 0.19 & 23.72 $\pm$ 0.03 & 23.16 $\pm$ 0.03 & 22.98 $\pm$ 0.03 & 24.19 $\pm$ 0.05 & 23.06 $\pm$ 0.04 \\ 
2452133.00 & 22.89 $\pm$ 0.05 & 23.45 $\pm$ 0.06 & 23.81 $\pm$ 0.04 & 23.43 $\pm$ 0.03 & 23.13 $\pm$ 0.03 & 23.06 $\pm$ 0.03 & 23.00 $\pm$ 0.04 \\ 
2452137.75 & 23.31 $\pm$ 0.04 & 23.20 $\pm$ 0.22 & 23.30 $\pm$ 0.03 & 23.72 $\pm$ 0.05 & 23.33 $\pm$ 0.03 & 23.43 $\pm$ 0.03 & 23.32 $\pm$ 0.05 \\ 
2452142.00 & 23.39 $\pm$ 0.04 & 22.55 $\pm$ 0.04 & 22.94 $\pm$ 0.03 & 23.74 $\pm$ 0.04 & 23.44 $\pm$ 0.03 & 23.69 $\pm$ 0.04 & 23.45 $\pm$ 0.05 \\ 
   \hline \\
& \multicolumn{7}{c}{F555W $\pm \Delta$ F555W} \\ 
   \hline \\ 
JD & C43; P = 34.1d & C44; P = 34.7d & C45; P = 41.2d & C46; P = 42.8d & C47; P = 44.0d & C48; P = 44.0d & C49; P = 44.4d \\ 
   \hline \\ 
2452099.00 & 24.85 $\pm$ 0.08 & 24.68 $\pm$ 0.08 & 23.27 $\pm$ 0.03 & 24.22 $\pm$ 0.09 & 23.04 $\pm$ 0.03 & 23.95 $\pm$ 0.06 & 23.22 $\pm$ 0.04 \\ 
2452105.50 & 24.47 $\pm$ 0.08 & 24.01 $\pm$ 0.05 & 23.50 $\pm$ 0.04 & 24.40 $\pm$ 0.07 & 23.11 $\pm$ 0.04 & 24.13 $\pm$ 0.06 & 23.45 $\pm$ 0.03 \\ 
2452112.50 & 24.49 $\pm$ 0.07 & 24.00 $\pm$ 0.04 & 23.83 $\pm$ 0.04 & 24.64 $\pm$ 0.08 & 23.28 $\pm$ 0.03 & 24.28 $\pm$ 0.06 & 23.60 $\pm$ 0.05 \\ 
2452114.50 & 24.68 $\pm$ 0.08 & 24.10 $\pm$ 0.05 & 23.85 $\pm$ 0.04 & 24.73 $\pm$ 0.08 & 23.29 $\pm$ 0.03 & 24.26 $\pm$ 0.06 & 23.53 $\pm$ 0.04 \\ 
2452116.75 & 24.67 $\pm$ 0.07 & 24.14 $\pm$ 0.08 & 23.96 $\pm$ 0.04 & 24.57 $\pm$ 0.11 & 23.29 $\pm$ 0.04 & 24.34 $\pm$ 0.08 & 23.50 $\pm$ 0.05 \\ 
2452119.25 & 24.82 $\pm$ 0.08 & 24.27 $\pm$ 0.05 & 24.09 $\pm$ 0.05 & 24.79 $\pm$ 0.10 & 23.30 $\pm$ 0.04 & 24.35 $\pm$ 0.08 & 23.23 $\pm$ 0.04 \\ 
2452123.00 & 25.11 $\pm$ 0.10 & 24.37 $\pm$ 0.06 & 24.19 $\pm$ 0.05 & 24.77 $\pm$ 0.09 & 23.40 $\pm$ 0.03 & 24.54 $\pm$ 0.07 & 22.81 $\pm$ 0.03 \\ 
2452125.25 & 25.39 $\pm$ 0.12 & 24.48 $\pm$ 0.06 & 24.04 $\pm$ 0.06 & 24.58 $\pm$ 0.08 & 23.43 $\pm$ 0.03 & 24.37 $\pm$ 0.09 & 22.83 $\pm$ 0.04 \\ 
2452128.75 & 25.44 $\pm$ 0.10 & 24.56 $\pm$ 0.06 & 24.31 $\pm$ 0.06 & 24.16 $\pm$ 0.06 & 23.53 $\pm$ 0.03 & 24.59 $\pm$ 0.10 & 22.75 $\pm$ 0.03 \\ 
2452133.00 & 24.98 $\pm$ 0.09 & 24.74 $\pm$ 0.06 & 23.75 $\pm$ 0.04 & 24.00 $\pm$ 0.05 & 23.56 $\pm$ 0.04 & 24.78 $\pm$ 0.14 & 22.85 $\pm$ 0.03 \\ 
2452137.75 & 24.53 $\pm$ 0.07 & 24.26 $\pm$ 0.05 & 23.29 $\pm$ 0.04 & 24.15 $\pm$ 0.07 & 23.63 $\pm$ 0.06 & 24.85 $\pm$ 0.09 & 23.08 $\pm$ 0.03 \\ 
2452142.00 & 24.53 $\pm$ 0.09 & 23.75 $\pm$ 0.04 & 23.38 $\pm$ 0.04 & 24.23 $\pm$ 0.06 & 23.69 $\pm$ 0.04 & 25.00 $\pm$ 0.11 & 23.17 $\pm$ 0.03 \\ 
   \hline \\
& \multicolumn{7}{c}{F555W $\pm \Delta$ F555W} \\ 
   \hline \\ 
JD & C50; P = 47.0d & C51; P = 48.0d & C52; P = 48.5d & C53; P = 48.5d & C54; P = 50.4d & C55; P = 74.0d & C56; P = 80.0d \\ 
   \hline \\ 
2452099.00 & 24.20 $\pm$ 0.06 & 24.91 $\pm$ 0.09 & 24.43 $\pm$ 0.06 & 22.36 $\pm$ 0.03 & 23.44 $\pm$ 0.04 & 25.42 $\pm$ 0.15 & 24.50 $\pm$ 0.08 \\ 
2452105.50 & 23.76 $\pm$ 0.05 & 24.75 $\pm$ 0.07 & 24.53 $\pm$ 0.06 & 22.83 $\pm$ 0.03 & 23.50 $\pm$ 0.04 & 25.49 $\pm$ 0.15 & 24.55 $\pm$ 0.07 \\ 
2452112.50 & 24.18 $\pm$ 0.05 & 24.83 $\pm$ 0.08 & 24.68 $\pm$ 0.09 & 22.76 $\pm$ 0.04 & 23.72 $\pm$ 0.05 & 25.34 $\pm$ 0.13 & 24.31 $\pm$ 0.07 \\ 
2452114.50 & 24.22 $\pm$ 0.05 & 24.84 $\pm$ 0.08 & 24.89 $\pm$ 0.08 & 22.51 $\pm$ 0.04 & 23.62 $\pm$ 0.04 & 25.39 $\pm$ 0.13 & 24.23 $\pm$ 0.07 \\ 
2452116.75 & 24.33 $\pm$ 0.06 & 24.98 $\pm$ 0.10 & 25.05 $\pm$ 0.08 & 22.00 $\pm$ 0.03 & 23.66 $\pm$ 0.04 & 25.07 $\pm$ 0.10 & 24.28 $\pm$ 0.06 \\ 
2452119.25 & 24.35 $\pm$ 0.05 & 24.92 $\pm$ 0.10 & 25.31 $\pm$ 0.10 & 21.94 $\pm$ 0.03 & 23.72 $\pm$ 0.04 & 25.13 $\pm$ 0.11 & 24.23 $\pm$ 0.06 \\ 
2452123.00 & 24.38 $\pm$ 0.06 & 25.16 $\pm$ 0.11 & 25.52 $\pm$ 0.12 & 22.03 $\pm$ 0.03 & 23.68 $\pm$ 0.05 & 24.73 $\pm$ 0.10 & 24.06 $\pm$ 0.06 \\ 
2452125.25 & 24.48 $\pm$ 0.06 & 25.10 $\pm$ 0.10 & 25.75 $\pm$ 0.16 & 22.04 $\pm$ 0.02 & 23.79 $\pm$ 0.04 & 24.81 $\pm$ 0.09 & 24.08 $\pm$ 0.05 \\ 
2452128.75 & 24.51 $\pm$ 0.06 & 25.21 $\pm$ 0.10 & 25.77 $\pm$ 0.15 & 22.19 $\pm$ 0.03 & 23.69 $\pm$ 0.05 & 24.61 $\pm$ 0.07 & 23.87 $\pm$ 0.05 \\ 
2452133.00 & 24.69 $\pm$ 0.07 & 25.50 $\pm$ 0.13 & 25.94 $\pm$ 0.15 & 22.27 $\pm$ 0.03 & 23.67 $\pm$ 0.05 & 24.57 $\pm$ 0.07 & 23.84 $\pm$ 0.05 \\ 
2452137.75 & 24.58 $\pm$ 0.06 & 25.51 $\pm$ 0.13 & 25.67 $\pm$ 0.13 & 22.40 $\pm$ 0.02 & 23.56 $\pm$ 0.05 & 24.70 $\pm$ 0.09 & 23.95 $\pm$ 0.05 \\ 
2452142.00 & 24.66 $\pm$ 0.08 & 25.65 $\pm$ 0.15 & 24.74 $\pm$ 0.09 & 22.55 $\pm$ 0.04 & 23.31 $\pm$ 0.08 & 24.70 $\pm$ 0.09 & 23.84 $\pm$ 0.04 \\ 
\enddata
\end{deluxetable}
\clearpage
\end{landscape}

\LongTables
\begin{landscape}
\begin{deluxetable}{lccccccc}
\tabletypesize{\scriptsize}
\tablecaption{F814W Photometry for the NGC 5128 Cepheids.\label{tbl:cephphoti}}
\tablewidth{0pt}
\tablehead{
\colhead{} &
\colhead{} &
\colhead{} &
\colhead{} &
\colhead{F814W $\pm \delta$ F814W} &
\colhead{} &
\colhead{} &
\colhead{} 
}
\startdata
JD & C1; P =  5.0d & C2; P =  5.3d & C3; P =  7.0d & C4; P =  7.3d & C5; P =  7.4d & C6; P =  8.2d & C7; P =  8.6d \\ 
   \hline \\ 
2452099.00 & 24.20 $\pm$ 0.11 & 23.42 $\pm$ 0.07 & 23.59 $\pm$ 0.08 & 23.06 $\pm$ 0.05 & 24.09 $\pm$ 0.12 & 23.95 $\pm$ 0.08 & 23.00 $\pm$ 0.06 \\ 
2452112.75 & 24.17 $\pm$ 0.12 & 23.51 $\pm$ 0.08 & 23.66 $\pm$ 0.09 & 23.29 $\pm$ 0.06 & 24.04 $\pm$ 0.13 & 23.60 $\pm$ 0.07 & 23.26 $\pm$ 0.08 \\ 
2452117.00 & 24.09 $\pm$ 0.13 & 23.48 $\pm$ 0.07 & 23.36 $\pm$ 0.06 & 23.33 $\pm$ 0.07 & 23.89 $\pm$ 0.11 & 23.64 $\pm$ 0.06 & 23.08 $\pm$ 0.06 \\ 
2452123.25 & 24.43 $\pm$ 0.16 & 23.61 $\pm$ 0.08 & 23.17 $\pm$ 0.10 & 23.15 $\pm$ 0.05 & 23.83 $\pm$ 0.10 & 23.84 $\pm$ 0.08 & 22.39 $\pm$ 0.09 \\ 
2452129.00 & 24.58 $\pm$ 0.17 & 23.87 $\pm$ 0.09 & 23.60 $\pm$ 0.06 & 23.10 $\pm$ 0.06 & 24.25 $\pm$ 0.27 & 23.64 $\pm$ 0.07 & 23.42 $\pm$ 0.05 \\ 
2452137.75 & 24.28 $\pm$ 0.14 & 23.47 $\pm$ 0.08 & 23.32 $\pm$ 0.06 & 23.13 $\pm$ 0.06 & 23.82 $\pm$ 0.11 & 23.76 $\pm$ 0.07 & 23.48 $\pm$ 0.09 \\ 
   \hline \\ 
& \multicolumn{7}{c}{F814W $\pm \Delta$ F814W} \\ 
   \hline \\ 
JD & C8; P =  8.8d & C9; P =  9.4d & C10; P = 10.6d & C11; P = 11.0d & C12; P = 11.2d & C13; P = 12.7d & C14; P = 12.7d \\ 
   \hline \\ 
2452099.00 & 23.36 $\pm$ 0.07 & 22.80 $\pm$ 0.06 & 23.33 $\pm$ 0.09 & 22.92 $\pm$ 0.01 & 22.74 $\pm$ 0.01 & 23.40 $\pm$ 0.06 & 22.98 $\pm$ 0.01 \\ 
2452112.75 & 23.46 $\pm$ 0.06 & 23.03 $\pm$ 0.06 & 22.84 $\pm$ 0.08 & 23.33 $\pm$ 0.06 & 22.76 $\pm$ 0.07 & 23.55 $\pm$ 0.08 & 23.12 $\pm$ 0.07 \\ 
2452117.00 & 23.27 $\pm$ 0.08 & 22.99 $\pm$ 0.08 & 22.97 $\pm$ 0.06 & 22.94 $\pm$ 0.01 & 22.71 $\pm$ 0.01 & 23.93 $\pm$ 0.10 & 23.13 $\pm$ 0.01 \\ 
2452123.25 & 23.55 $\pm$ 0.09 & 23.00 $\pm$ 0.01 & 22.94 $\pm$ 0.05 & 23.31 $\pm$ 0.06 & 22.82 $\pm$ 0.01 & 23.34 $\pm$ 0.07 & 22.99 $\pm$ 0.06 \\ 
2452129.00 & 23.37 $\pm$ 0.07 & -4.89 $\pm$ 0.01 & 23.21 $\pm$ 0.06 & 22.80 $\pm$ 0.01 & 22.56 $\pm$ 0.01 & 23.73 $\pm$ 0.07 & 22.94 $\pm$ 0.01 \\ 
2452137.75 & 23.34 $\pm$ 0.07 & 22.92 $\pm$ 0.01 & 22.87 $\pm$ 0.07 & 23.00 $\pm$ 0.01 & 22.88 $\pm$ 0.07 & 23.41 $\pm$ 0.06 & 23.08 $\pm$ 0.07 \\ 
   \hline \\ 
& \multicolumn{7}{c}{F814W $\pm \Delta$ F814W} \\ 
   \hline \\ 
JD & C15; P = 12.8d & C16; P = 13.9d & C17; P = 13.9d & C18; P = 14.3d & C19; P = 14.9d & C20; P = 15.1d & C21; P = 15.1d \\ 
   \hline \\ 
2452099.00 & 23.44 $\pm$ 0.08 & 22.52 $\pm$ 0.01 & 22.72 $\pm$ 0.01 & 23.15 $\pm$ 0.01 & 22.48 $\pm$ 0.01 & 22.75 $\pm$ 0.01 & 23.07 $\pm$ 0.10 \\ 
2452112.75 & 23.23 $\pm$ 0.01 & 22.38 $\pm$ 0.01 & 22.73 $\pm$ 0.01 & 23.11 $\pm$ 0.06 & 22.47 $\pm$ 0.01 & 22.66 $\pm$ 0.01 & 23.11 $\pm$ 0.09 \\ 
2452117.00 & 23.13 $\pm$ 0.06 & 22.37 $\pm$ 0.01 & 22.51 $\pm$ 0.01 & 23.03 $\pm$ 0.01 & 22.84 $\pm$ 0.01 & 23.02 $\pm$ 0.01 & 22.60 $\pm$ 0.06 \\ 
2452123.25 & 23.45 $\pm$ 0.06 & 22.57 $\pm$ 0.06 & 22.62 $\pm$ 0.01 & 22.83 $\pm$ 0.01 & 22.25 $\pm$ 0.01 & 23.02 $\pm$ 0.01 & 22.94 $\pm$ 0.08 \\ 
2452129.00 & 23.11 $\pm$ 0.01 & 22.26 $\pm$ 0.01 & 22.48 $\pm$ 0.01 & 23.23 $\pm$ 0.07 & 22.02 $\pm$ 0.09 & 22.65 $\pm$ 0.01 & 22.92 $\pm$ 0.06 \\ 
2452137.75 & 23.33 $\pm$ 0.07 & 22.71 $\pm$ 0.01 & 22.97 $\pm$ 0.06 & 22.82 $\pm$ 0.01 & 22.27 $\pm$ 0.01 & 23.03 $\pm$ 0.01 & 22.79 $\pm$ 0.08 \\ 
   \hline \\ 
& \multicolumn{7}{c}{F814W $\pm \Delta$ F814W} \\ 
   \hline \\ 
JD & C22; P = 15.5d & C23; P = 16.1d & C24; P = 16.5d & C25; P = 16.5d & C26; P = 16.6d & C27; P = 17.3d & C28; P = 17.9d \\ 
   \hline \\ 
2452099.00 & 22.61 $\pm$ 0.01 & 23.06 $\pm$ 0.06 & 23.16 $\pm$ 0.07 & 22.91 $\pm$ 0.01 & 22.61 $\pm$ 0.08 & 23.21 $\pm$ 0.01 & 22.88 $\pm$ 0.07 \\ 
2452112.75 & 22.80 $\pm$ 0.01 & 23.23 $\pm$ 0.07 & 22.91 $\pm$ 0.06 & 22.78 $\pm$ 0.01 & 22.44 $\pm$ 0.08 & 23.17 $\pm$ 0.01 & 22.65 $\pm$ 0.01 \\ 
2452117.00 & 22.22 $\pm$ 0.01 & 22.94 $\pm$ 0.01 & 23.07 $\pm$ 0.07 & 22.90 $\pm$ 0.01 & 22.77 $\pm$ 0.18 & 23.29 $\pm$ 0.06 & 22.85 $\pm$ 0.07 \\ 
2452123.25 & 22.75 $\pm$ 0.01 & 23.29 $\pm$ 0.07 & 22.59 $\pm$ 0.01 & 23.21 $\pm$ 0.01 & 22.08 $\pm$ 0.01 & 23.56 $\pm$ 0.06 & 22.45 $\pm$ 0.01 \\ 
2452129.00 & 22.81 $\pm$ 0.01 & 23.35 $\pm$ 0.06 & 22.96 $\pm$ 0.06 & 22.81 $\pm$ 0.01 & 22.48 $\pm$ 0.01 & 23.12 $\pm$ 0.01 & 22.65 $\pm$ 0.01 \\ 
2452137.75 & 22.70 $\pm$ 0.01 & 23.19 $\pm$ 0.01 & 22.57 $\pm$ 0.01 & 23.24 $\pm$ 0.01 & 22.19 $\pm$ 0.06 & 23.55 $\pm$ 0.06 & 22.83 $\pm$ 0.06 \\ 
   \hline \\ 
& \multicolumn{7}{c}{F814W $\pm \Delta$ F814W} \\ 
   \hline \\ 
JD & C29; P = 20.5d & C30; P = 20.9d & C31; P = 21.3d & C32; P = 22.2d & C33; P = 22.4d & C34; P = 22.5d & C35; P = 22.6d \\ 
   \hline \\ 
2452099.00 & 22.53 $\pm$ 0.09 & 22.90 $\pm$ 0.01 & 22.79 $\pm$ 0.01 & 22.47 $\pm$ 0.01 & 22.68 $\pm$ 0.01 & 22.72 $\pm$ 0.06 & 23.12 $\pm$ 0.01 \\ 
2452112.75 & 22.38 $\pm$ 0.01 & 22.63 $\pm$ 0.01 & 22.40 $\pm$ 0.01 & 22.98 $\pm$ 0.01 & 22.12 $\pm$ 0.01 & 23.10 $\pm$ 0.06 & 22.77 $\pm$ 0.01 \\ 
2452117.00 & 22.49 $\pm$ 0.01 & 22.75 $\pm$ 0.01 & 22.42 $\pm$ 0.06 & 22.93 $\pm$ 0.01 & 22.41 $\pm$ 0.01 & 22.97 $\pm$ 0.05 & 22.89 $\pm$ 0.07 \\ 
2452123.25 & 22.88 $\pm$ 0.07 & 23.00 $\pm$ 0.01 & 22.78 $\pm$ 0.01 & 22.45 $\pm$ 0.01 & 22.59 $\pm$ 0.01 & 22.83 $\pm$ 0.05 & 23.16 $\pm$ 0.06 \\ 
2452129.00 & 22.21 $\pm$ 0.01 & 23.14 $\pm$ 0.06 & 22.24 $\pm$ 0.01 & 22.70 $\pm$ 0.01 & 21.93 $\pm$ 0.01 & 23.09 $\pm$ 0.06 & 23.23 $\pm$ 0.07 \\ 
2452137.75 & 22.47 $\pm$ 0.08 & 22.74 $\pm$ 0.01 & 22.65 $\pm$ 0.01 & 22.99 $\pm$ 0.01 & 22.24 $\pm$ 0.01 & 22.97 $\pm$ 0.05 & 22.73 $\pm$ 0.06 \\ 
   \hline \\ 
& \multicolumn{7}{c}{F814W $\pm \Delta$ F814W} \\ 
   \hline \\ 
JD & C36; P = 23.1d & C37; P = 23.1d & C38; P = 24.1d & C39; P = 26.4d & C40; P = 27.1d & C41; P = 27.8d & C42; P = 30.5d \\ 
   \hline \\ 
2452099.00 & 22.13 $\pm$ 0.01 & 21.40 $\pm$ 0.07 & 22.19 $\pm$ 0.01 & 22.01 $\pm$ 0.01 & 22.28 $\pm$ 0.07 & 22.78 $\pm$ 0.01 & 21.90 $\pm$ 0.01 \\ 
2452112.75 & 22.20 $\pm$ 0.01 & 22.27 $\pm$ 0.07 & 22.61 $\pm$ 0.01 & 22.49 $\pm$ 0.01 & 22.33 $\pm$ 0.01 & 22.35 $\pm$ 0.01 & 22.16 $\pm$ 0.01 \\ 
2452117.00 & 22.42 $\pm$ 0.01 & 22.18 $\pm$ 0.07 & 21.99 $\pm$ 0.01 & 22.55 $\pm$ 0.01 & 22.44 $\pm$ 0.01 & 22.36 $\pm$ 0.01 & 22.41 $\pm$ 0.01 \\ 
2452123.25 & 21.92 $\pm$ 0.01 & 21.71 $\pm$ 0.01 & 22.19 $\pm$ 0.01 & 21.98 $\pm$ 0.01 & 22.60 $\pm$ 0.01 & 22.59 $\pm$ 0.01 & 22.52 $\pm$ 0.01 \\ 
2452129.00 & 21.98 $\pm$ 0.01 & 21.61 $\pm$ 0.08 & 22.52 $\pm$ 0.01 & 22.09 $\pm$ 0.01 & 22.09 $\pm$ 0.01 & 22.78 $\pm$ 0.01 & 22.01 $\pm$ 0.01 \\ 
2452137.75 & 22.26 $\pm$ 0.01 & 22.15 $\pm$ 0.10 & 22.33 $\pm$ 0.01 & 22.49 $\pm$ 0.01 & 22.27 $\pm$ 0.01 & 22.13 $\pm$ 0.01 & 22.06 $\pm$ 0.01 \\ 
   \hline \\ 
& \multicolumn{7}{c}{F814W $\pm \Delta$ F814W} \\ 
   \hline \\ 
JD & C43; P = 34.1d & C44; P = 34.7d & C45; P = 41.2d & C46; P = 42.8d & C47; P = 44.0d & C48; P = 44.0d & C49; P = 44.4d \\ 
   \hline \\ 
2452099.00 & 24.02 $\pm$ 0.11 & 22.60 $\pm$ 0.01 & 21.72 $\pm$ 0.01 & 22.24 $\pm$ 0.01 & 21.51 $\pm$ 0.05 & 22.31 $\pm$ 0.05 & 21.69 $\pm$ 0.05 \\ 
2452112.75 & 23.88 $\pm$ 0.12 & 22.15 $\pm$ 0.01 & 21.86 $\pm$ 0.01 & 22.51 $\pm$ 0.01 & 21.49 $\pm$ 0.05 & 22.38 $\pm$ 0.05 & 22.06 $\pm$ 0.05 \\ 
2452117.00 & 24.07 $\pm$ 0.12 & 22.20 $\pm$ 0.01 & 22.06 $\pm$ 0.01 & 22.55 $\pm$ 0.01 & 21.62 $\pm$ 0.05 & 22.41 $\pm$ 0.05 & 22.03 $\pm$ 0.05 \\ 
2452123.25 & 24.43 $\pm$ 0.23 & 22.34 $\pm$ 0.01 & 22.20 $\pm$ 0.01 & 22.57 $\pm$ 0.01 & 21.63 $\pm$ 0.05 & 22.45 $\pm$ 0.05 & 21.58 $\pm$ 0.05 \\ 
2452129.00 & 24.34 $\pm$ 0.15 & 22.52 $\pm$ 0.01 & 22.31 $\pm$ 0.01 & 22.18 $\pm$ 0.01 & 21.75 $\pm$ 0.05 & 22.50 $\pm$ 0.05 & 21.53 $\pm$ 0.05 \\ 
2452137.75 & 23.78 $\pm$ 0.11 & 22.14 $\pm$ 0.01 & 21.73 $\pm$ 0.01 & 22.15 $\pm$ 0.01 & 21.77 $\pm$ 0.05 & 22.59 $\pm$ 0.05 & 21.63 $\pm$ 0.05 \\ 
   \hline \\ 
& \multicolumn{7}{c}{F814W $\pm \Delta$ F814W} \\ 
   \hline \\ 
JD & C50; P = 47.0d & C51; P = 48.0d & C52; P = 48.5d & C53; P = 48.5d & C54; P = 50.4d & C55; P = 74.0d & C56; P = 80.0d \\ 
   \hline \\ 
2452099.00 & 23.19 $\pm$ 0.06 & 22.91 $\pm$ 0.05 & 23.74 $\pm$ 0.08 & 21.50 $\pm$ 0.01 & 22.57 $\pm$ 0.05 & 22.93 $\pm$ 0.06 & 23.19 $\pm$ 0.06 \\ 
2452112.75 & 23.17 $\pm$ 0.05 & 22.80 $\pm$ 0.05 & 23.87 $\pm$ 0.08 & 21.65 $\pm$ 0.01 & 22.81 $\pm$ 0.08 & 22.90 $\pm$ 0.06 & 23.20 $\pm$ 0.07 \\ 
2452117.00 & 23.29 $\pm$ 0.06 & 22.87 $\pm$ 0.05 & 23.86 $\pm$ 0.09 & 21.27 $\pm$ 0.01 & 22.69 $\pm$ 0.06 & 22.94 $\pm$ 0.05 & 23.05 $\pm$ 0.05 \\ 
2452123.25 & 23.47 $\pm$ 0.06 & 22.96 $\pm$ 0.05 & 24.35 $\pm$ 0.12 & 21.22 $\pm$ 0.01 & 22.78 $\pm$ 0.06 & 22.76 $\pm$ 0.06 & 22.83 $\pm$ 0.05 \\ 
2452129.00 & 23.61 $\pm$ 0.08 & 23.03 $\pm$ 0.05 & 24.46 $\pm$ 0.14 & 21.24 $\pm$ 0.01 & 22.88 $\pm$ 0.05 & 22.50 $\pm$ 0.07 & 22.66 $\pm$ 0.06 \\ 
2452137.75 & 23.60 $\pm$ 0.09 & 23.14 $\pm$ 0.05 & 24.55 $\pm$ 0.15 & 21.37 $\pm$ 0.01 & 22.78 $\pm$ 0.05 & 22.66 $\pm$ 0.05 & 22.78 $\pm$ 0.05 \\ 
\enddata
\end{deluxetable}
\clearpage
\end{landscape}

\begin{deluxetable}{lrrccrccccl}
\tabletypesize{\scriptsize}
\tablecaption{Suspected Variable Stars in NGC 5128.\label{tbl:var}}
\tablewidth{0pt}
\tablehead{
\colhead{ID} &
\colhead{$X$} &
\colhead{$Y$} &
\colhead{RA} &
\colhead{Dec} &
\colhead{Period} &
\colhead{$V^{Int}$} &
\colhead{$I^{Int}$} &
\colhead{$V^{ph}$} &
\colhead{$I^{ph}$} &
\colhead{Chip}\\
\colhead{} &
\colhead{(pixel)} &
\colhead{(pixel)} &
\colhead{(h:m:s)} &
\colhead{(\deg:\min:\sec)} &
\colhead{(days)} &
\colhead{(mag)} &
\colhead{(mag)} &
\colhead{(mag)} &
\colhead{(mag)} &
\colhead{}
}
\startdata
V1 & 477.0 & 384.2 & 13:25:17.853 & $-$43:00:11.30 &  \phantom{$\geq$}5.2 & 25.76 & 24.45 & 25.80 & 24.37 & PC\\
V2 & 244.4 & 592.1 & 13:25:20.198 & $-$42:59:29.75 &  \phantom{$\geq$}5.3 & 23.62 & 22.76 & 23.63 & 22.76 & WF4\\
V3 & 633.1 &  95.7 & 13:25:15.042 & $-$42:59:03.10 &  \phantom{$\geq$}5.6 & 24.40 & 23.14 & 24.41 & 23.14 & WF4\\
V4 & 241.4 & 647.5 & 13:25:12.689 & $-$42:59:07.93 &  \phantom{$\geq$}5.8 & 24.32 & 23.55 & 24.30 & 23.59 & WF3\\
V5 & 383.6 & 563.3 & 13:25:19.665 & $-$42:59:16.90 &  \phantom{$\geq$}6.2 & 24.39 & 22.95 & 24.34 & 22.94 & WF4\\
V6 & 387.4 & 314.7 & 13:25:17.458 & $-$42:59:22.04 &  \phantom{$\geq$}6.3 & 24.26 & 23.05 & 24.19 & 23.03 & WF4\\
V7 & 433.1 & 411.9 & 13:25:18.227 & $-$42:59:15.45 &  \phantom{$\geq$}6.3 & 24.17 & 23.23 & 24.20 & 23.25 & WF4\\
V8 & 124.3 & 112.0 & 13:25:15.183 & $-$43:00:10.38 &  \phantom{$\geq$}6.4 & 24.39 & 23.09 & 24.40 & 23.12 & WF2\\
V9 & 274.2 & 709.5 & 13:25:21.170 & $-$42:59:24.26 &  \phantom{$\geq$}6.7 & 24.29 & 23.88 & 24.24 & 23.92 & WF4\\
V10 & 636.8 & 342.9 & 13:25:18.457 & $-$43:00:07.82 &  \phantom{$\geq$}6.8 & 25.36 & 24.02 & 25.45 & 24.07 & PC\\
V11 & 423.9 & 632.9 & 13:25:11.113 & $-$42:59:13.50 &  \phantom{$\geq$}6.8 & 24.20 & 23.30 & 24.22 & 23.33 & WF3\\
V12 & 532.8 & 218.3 & 13:25:16.316 & $-$42:59:10.07 &  \phantom{$\geq$}6.8 & 24.64 & 23.64 & 24.63 & 23.67 & WF4\\
V13 & 522.7 & 269.0 & 13:25:17.928 & $-$43:00:05.74 &  \phantom{$\geq$}6.9 & 25.14 & 23.97 & 25.16 & 23.96 & PC\\
V14 & 336.4 & 481.9 & 13:25:19.041 & $-$42:59:23.29 &  \phantom{$\geq$}7.0 & 23.77 & 22.98 & 23.83 & 23.01 & WF4\\
V15 & 195.8 & 383.2 & 13:25:13.647 & $-$42:59:32.38 &  \phantom{$\geq$}7.2 & 24.51 & 22.16 & 24.52 & 22.16 & WF3\\
V16 & 416.2 & 204.5 & 13:25:16.427 & $-$42:59:21.67 &  \phantom{$\geq$}7.2 & 23.82 & 23.60 & 23.82 & 23.57 & WF4\\
V17 & 454.9 & 262.4 & 13:25:16.861 & $-$42:59:16.64 &  \phantom{$\geq$}7.2 & 24.34 & 23.15 & 24.32 & 23.11 & WF4\\
V18 & 357.2 & 426.7 & 13:25:18.510 & $-$42:59:22.49 &  \phantom{$\geq$}7.3 & 24.06 & 23.06 & 24.10 & 23.11 & WF4\\
V19 & 334.9 & 506.7 & 13:25:19.263 & $-$42:59:22.88 &  \phantom{$\geq$}7.3 & 24.05 & 23.02 & 24.09 & 23.03 & WF4\\
V20 & 602.2 & 549.3 & 13:25:09.717 & $-$42:59:25.67 &  \phantom{$\geq$}7.4 & 24.73 & 23.59 & 24.79 & 23.62 & WF3\\
V21 & 410.0 & 120.6 & 13:25:15.724 & $-$43:00:38.06 &  \phantom{$\geq$}7.4 & 24.30 & 23.11 & 24.27 & 23.09 & WF2\\
V22 & 596.0 &  93.0 & 13:25:15.091 & $-$42:59:06.74 &  \phantom{$\geq$}7.5 & 24.28 & 23.50 & 24.32 & 23.54 & WF4\\
V23 & 593.0 & 142.6 & 13:25:15.531 & $-$42:59:05.93 &  \phantom{$\geq$}7.6 & 24.26 & 23.45 & 24.21 & 23.38 & WF4\\
V24 & 376.3 & 575.6 & 13:25:19.788 & $-$42:59:17.34 &  \phantom{$\geq$}7.9 & 24.10 & 23.13 & 24.10 & 23.12 & WF4\\
V25 & 313.9 & 658.5 & 13:25:20.643 & $-$42:59:21.56 &  \phantom{$\geq$}8.2 & 23.94 & 22.94 & 23.88 & 22.91 & WF4\\
V26 & 476.6 & 518.7 & 13:25:19.084 & $-$42:59:08.88 &  \phantom{$\geq$}8.4 & 24.20 & 23.38 & 24.19 & 23.34 & WF4\\
V27 & 704.7 & 663.6 & 13:25:08.586 & $-$42:59:17.05 &  \phantom{$\geq$}8.5 & 24.53 & 23.45 & 24.51 & 23.44 & WF3\\
V28 & 158.1 & 643.4 & 13:25:13.429 & $-$42:59:06.44 &  \phantom{$\geq$}8.8 & 23.76 & 23.12 & 23.70 & 23.10 & WF3\\
V29 & 314.6 & 344.2 & 13:25:17.866 & $-$42:59:28.45 &  \phantom{$\geq$}8.8 & 24.62 & 23.52 & 24.65 & 23.52 & WF4\\
V30 & 372.6 & 406.1 & 13:25:18.297 & $-$42:59:21.45 &  \phantom{$\geq$}9.1 & 24.18 & 23.81 & 24.23 & 23.84 & WF4\\
V31 & 356.3 & 414.3 & 13:25:18.402 & $-$42:59:22.85 &  \phantom{$\geq$}9.4 & 24.32 & 22.99 & 24.28 & 22.96 & WF4\\
V32 & 272.7 & 548.1 & 13:25:12.623 & $-$42:59:18.20 &  \phantom{$\geq$}9.5 & 23.96 & 23.05 & 23.95 & 23.05 & WF3\\
V33 & 118.6 & 613.0 & 13:25:13.840 & $-$42:59:08.47 &  \phantom{$\geq$}9.7 & 24.58 & 23.81 & 24.56 & 23.79 & WF3\\
V34 & 465.6 & 310.4 & 13:25:11.421 & $-$42:59:45.62 &  \phantom{$\geq$}9.8 & 25.74 & 23.74 & 25.72 & 23.72 & WF3\\
V35 & 207.9 & 381.6 & 13:25:12.993 & $-$43:00:24.68 & \phantom{$\geq$}10.4 & 25.08 & 23.30 & 25.08 & 23.31 & WF2\\
V36 & 271.3 & 133.8 & 13:25:16.786 & $-$43:00:02.39 & \phantom{$\geq$}10.7 & 25.01 & 23.74 & 25.04 & 23.75 & PC\\
V37 & 376.1 & 310.5 & 13:25:17.443 & $-$42:59:23.22 & \phantom{$\geq$}11.6 & 24.07 & 23.17 & 24.03 & 23.14 & WF4\\
V38 &  92.3 & 630.8 & 13:25:14.031 & $-$42:59:06.17 & \phantom{$\geq$}11.7 & 23.83 & 22.67 & 23.82 & 22.66 & WF3\\
V39 & 338.2 & 435.0 & 13:25:18.622 & $-$42:59:24.15 & \phantom{$\geq$}12.7 & 23.77 & 22.97 & 23.74 & 22.96 & WF4\\
V40 & 418.1 & 379.9 & 13:25:17.973 & $-$42:59:17.61 & \phantom{$\geq$}13.7 & 23.79 & 22.76 & 23.84 & 22.82 & WF4\\
V41 &  99.8 & 445.4 & 13:25:14.359 & $-$42:59:24.17 & \phantom{$\geq$}13.9 & 24.81 & 23.01 & 24.87 & 22.94 & WF3\\
V42 & 333.2 & 141.0 & 13:25:12.941 & $-$42:59:58.92 & \phantom{$\geq$}14.6 & 25.33 & 23.24 & 25.37 & 23.25 & WF3\\
V43 & 324.8 & 515.3 & 13:25:19.359 & $-$42:59:23.67 & \phantom{$\geq$}15.6 & 23.69 & 22.58 & 23.69 & 22.59 & WF4\\
V44 & 246.3 & 556.4 & 13:25:19.880 & $-$42:59:30.36 & \phantom{$\geq$}15.6 & 23.56 & 22.60 & 23.52 & 22.58 & WF4\\
V45 & 540.4 & 578.9 & 13:25:18.292 & $-$43:00:19.25 & \phantom{$\geq$}17.8 & 25.71 & 23.31 & 25.67 & 23.28 & PC\\
V46 & 249.3 & 556.0 & 13:25:19.870 & $-$42:59:30.08 & \phantom{$\geq$}19.5 & 23.83 & 22.83 & 23.81 & 22.78 & WF4\\
V47 & 670.1 & 482.2 & 13:25:18.722 & $-$43:00:13.63 & \phantom{$\geq$}20.4 & 25.15 & 23.28 & 25.22 & 23.33 & PC\\
V48 & 190.6 & 579.7 & 13:25:20.195 & $-$42:59:35.23 & \phantom{$\geq$}20.8 & 23.56 & 22.51 & 23.44 & 22.47 & WF4\\
V49 & 197.6 & 741.9 & 13:25:09.809 & $-$43:00:32.22 & \phantom{$\geq$}28.0 & 25.16 & 23.76 & 25.17 & 23.77 & WF2\\
V50 & 140.8 & 339.5 & 13:25:16.455 & $-$43:00:12.82 & \phantom{$\geq$}29.2 & 25.06 & 24.13 & 25.07 & 24.14 & PC\\
V51 &  69.0 & 679.1 & 13:25:14.131 & $-$42:59:01.04 & \phantom{$\geq$}29.2 & 23.04 & 21.93 & 23.03 & 21.92 & WF3\\
V52 & 767.8 & 495.4 & 13:25:13.191 & $-$43:01:21.27 & \phantom{$\geq$}29.7 & 24.64 & 23.58 & 24.65 & 23.58 & WF2\\
V53 & 156.2 & 309.5 & 13:25:17.878 & $-$42:59:44.53 & $\geq$30.0 & 25.22 & 22.79 & 25.27 & 22.82 & WF4\\
V54 & 718.2 & 688.8 & 13:25:08.418 & $-$42:59:14.95 & $\geq$32.0 & 25.34 & 24.15 & 25.34 & 24.14 & WF3\\
V55 & 595.9 & 474.0 & 13:25:13.011 & $-$43:01:04.31 & $\geq$44.0 & 25.48 & 23.66 & 25.47 & 23.67 & WF2\\
V56 & 599.0 & 714.2 & 13:25:09.407 & $-$42:59:09.75 & $\geq$44.0 & 24.64 & 23.61 & 24.64 & 23.63 & WF3\\
V57 & 340.2 & 701.6 & 13:25:10.462 & $-$43:00:44.99 & $\geq$45.0 & 24.28 & 22.19 & 24.28 & 22.19 & WF2\\
V58 & 110.5 & 677.8 & 13:25:13.773 & $-$42:59:02.08 & $\geq$45.0 & 24.25 & 22.85 & 24.27 & 22.87 & WF3\\
V59 & 360.5 & 749.0 & 13:25:21.343 & $-$42:59:15.04 & $\geq$45.0 & 25.08 & 23.62 & 25.09 & 23.58 & WF4\\
V60 & 289.6 & 576.4 & 13:25:12.415 & $-$42:59:15.86 & $\geq$48.0 & 23.66 & 23.17 & 23.66 & 23.16 & WF3\\
V61 & 522.7 & 691.8 & 13:25:10.940 & $-$43:01:02.33 & $\geq$49.6 & 24.17 & 23.10 & 24.18 & 23.11 & WF2\\
V62 & 387.0 & 319.3 & 13:25:12.095 & $-$42:59:42.95 & \phantom{$\geq$}49.6 & 25.13 & 22.88 & 25.11 & 22.88 & WF3\\
V63 & 550.7 & 215.7 & 13:25:16.258 & $-$42:59:08.39 & $\geq$50.0 & 24.36 & 22.54 & 24.38 & 22.55 & WF4\\
V64 & 581.4 & 696.8 & 13:25:11.023 & $-$43:01:08.09 & $\geq$55.0 & 25.73 & 23.49 & 25.68 & 23.46 & WF2\\
V65 & 327.7 & 198.9 & 13:25:12.870 & $-$42:59:53.21 & $\geq$60.0 & 25.75 & 23.14 & 25.86 & 23.17 & WF3\\
V66 & 300.0 & 402.2 & 13:25:12.689 & $-$42:59:32.93 & $\geq$70.0 & 26.03 & 23.02 & 26.08 & 23.07 & WF3\\
V67 & 294.2 & 145.8 & 13:25:15.254 & $-$43:00:27.48 & $\geq$80.0 & 23.23 & 23.30 & 23.14 & 23.28 & WF2\\
V68 & 166.1 & 276.3 & 13:25:14.131 & $-$42:59:42.02 & $\geq$80.0 & 25.59 & 23.15 & 25.48 & 23.10 & WF3\\
V69 & 297.6 & 589.2 & 13:25:20.066 & $-$42:59:24.67 & $\geq$80.0 & 23.39 & 23.02 & 23.29 & 22.99 & WF4\\
V70 & 345.3 & 176.8 & 13:25:16.326 & $-$42:59:29.15 & $\geq$80.0 & 24.36 & 22.11 & 24.35 & 22.13 & WF4\\
\enddata
\tablecomments{Comments on the individual Variables: 
V1: found in region of high background and likely contaminated; the
light curve is noisy.
V2: located five pixels from a very bright star; the resulting bias in
the photometry might be responsible for the low amplitude of the light
curve.
V3: found in a crowded region.
V4: it appears to be a blend from visual inspection.
V5, V6, V7: all found in a crowded region.
V8: it appears to be a blend based on visual inspection, possibly
resulting in a light curve with low amplitude.
V9: found in a crowded region; the resulting bias in
the photometry might be responsible for the low amplitude of the light
curve.
V10 and V11: the light variations are not entirely convincing from a
visual inspection, the light curve has small amplitude. 
V12: found in a crowded region.
V13: located two pixels away from a bright star.
V14: found in a crowded region.
V15: the light variations are not entirely convincing from a
visual inspection.
V16: found in a crowded region; the resulting bias in
the photometry might be responsible for the low amplitude of the light
curve.
V17 to V21: found in a crowded region; the light curve of V20 is symmetric.
V22: located close to the diffraction spike of a bright star; the light
curve is noisy.
V23: it appears to be a blend on visual inspection.
V24: located ten pixels away from a bright star; the resulting bias in
the photometry might be responsible for the symmetric nature of the
light curve.
V25: found in a crowded region.
V26: the light variations are not entirely convincing from a
visual inspection.
V27: the light curve appears to be rather symmetric. 
V28 to V34: found in a crowded region; the light curves are symmetric
and/or noisy. 
V35: the light curve appears to be rather symmetric. 
V36 and V37: found in crowded region; the light curve of V36 is noisy.
V38: it appears to be a blend on visual inspection. 
V39 to V44: found in crowded region; the light curves are noisy and/or symmetric.
V45: the light curve is noisy. 
V46: found in crowded region. 
V47: it appears to be a blend on visual inspection. 
V48 and V49: found in crowded region; the light curves are noisy.
V50: the light curve is noisy.  
V51: it appears to be a blend on visual inspection. 
V52: found very close to the edge of the chip; the photometry might be
contaminated, possibly resulting in the flat minimum of the light curve.
V53 to V56: found in crowded region; some light curves are noisy or symmetric.
V57: the light variations are not entirely convincing from a
visual inspection.
V58: found in crowded region; the light curve has small amplitude. 
V59: the light variations are not entirely convincing from a
visual inspection.
V60: found in crowded region; the light curve has small amplitude.
V61: located four pixels away from a bright star; the photometry might be
contaminated, possibly resulting in the small amplitude observed for the light curve.
V62: it appears to be a blend on visual inspection. 
V63: found in a crowded region, the light curve has slow ascent.
V64 and V65: the light variations are not entirely convincing from a
visual inspection.
V66: found in a crowded region.
V67: the light curve has very sparse phase coverage, the nature of the
variations could not be established.
V68 to V70: found in crowded region.}
\end{deluxetable}

\clearpage

NOTE: Figures 11 to 13 can be found on the version of the paper available at http://astrowww.phys.uvic.ca/~lff/publications.html
				   
\end{document}